\documentclass[12pt,a4paper]{article}
\pdfoutput=1
\usepackage[utf8]{inputenc}
\usepackage[british]{babel}

\usepackage{authblk}
\usepackage{amsmath,amssymb,bm,multirow,bbold,slashed}
\usepackage{graphicx}
\usepackage{latexsym}
\usepackage{mathrsfs}
\usepackage{amsfonts}
\usepackage{float}
\usepackage{longtable}
\usepackage{bbm}
\usepackage{cite}
\usepackage{color}
\usepackage{lscape}

\usepackage{hyperref}
\usepackage{hypcap}

\allowdisplaybreaks
\makeatletter
\providecommand*{\diff}%
 {\@ifnextchar^{\DIfF}{\DIfF^{}}}
\def\DIfF^#1{%
 \mathop{\mathrm{\mathstrut d}}%
  \nolimits^{#1}\gobblespace}
\def\gobblespace{%
  \futurelet\diffarg\opspace}
\def\opspace{%
  \let\DiffSpace\!%
  \ifx\diffarg(%
   \let\DiffSpace\relax
   \else
   \ifx\diffarg[%
	\let\DiffSpace\relax
   \else
	\ifx\diffarg\{%
	 \let\DiffSpace\relax
	\fi\fi\fi\DiffSpace}

\hypersetup{%
colorlinks,citecolor=blue,urlcolor=blue,linkcolor=blue,%
hypertexnames=false,%
pdftitle={Lepton-flavour violation in hadronic tau decays and mu--tau conversion in nuclei},%
pdfauthor={T. Husek, K. Monsalvez-Pozo, J. Portoles}%
}

\title{\bf\boldmath Lepton-flavour violation in hadronic tau decays and $\mu$--$\tau$ conversion in nuclei}
\author[]{\sc\small Tom\'a\v{s} Husek\thanks{Email:~Tomas.Husek@ific.uv.es}}
\author[]{\sc\small Kevin Mons\'alvez-Pozo\thanks{Email:~Kevin.Monsalvez@ific.uv.es}}
\author[]{\sc\small Jorge~Portol\'es\thanks{Email:~Jorge.Portoles@ific.uv.es}}
\affil[]{\small Instituto de F\'isica Corpuscular, CSIC -- Universitat de Val\`encia,\\ Apt.\ Correus 22085, E-46071 Val\`encia, Spain}
\date{}

\begin{document}


\maketitle

\begin{abstract}
\noindent
Within the Standard Model Effective Field Theory framework, with operators up to dimension 6, we perform a model-independent analysis of the lepton-flavour-violating processes involving tau leptons. Namely, we study hadronic tau decays and $\ell$--$\tau$ conversion in nuclei, with $\ell = e,\mu$. Based on available experimental limits, we establish constraints on the Wilson coefficients of the operators contributing to these processes. Our work paves the way to extract the related information from Belle II and foreseen future experiments.
%
%
\end{abstract}


\section{Introduction}\label{s:1}

Although both light quark and lepton families are triple replicated in nature, their description in the Standard Model (SM) of particles has a significative difference: While quark families mix, giving a rich flavour-physics phenomenology, lepton families do not. This fact is due to the neutrino-mass degeneracy (they are all massless in the SM), that produces a set of conservation laws for the lepton flavours, namely, the Lagrangian is invariant under global $L \equiv \text{U}(1)_e \times \text{U}(1)_{\mu} \times \text{U}(1)_{\tau}$ rotations of the lepton fields. However, the present situation, that supersedes and extends the SM setting, involves massive non-degenerate neutrinos that mix their flavours while $\text{U}(1)_L$, the diagonal subgroup of $L$, is kept as a good global symmetry (conservation of lepton number) as far as neutrinos are Dirac \cite{Abe:2011fz,Abe:2012jj,An:2012eh,Ahn:2012nd,Adamson:2011qu} (Majorana neutrinos would imply lepton-number violation).
This new situation gives, within the extended SM, lepton-flavour violation also in processes involving electrically charged leptons, but their rates are tiny, far below the experimental reach \cite{Cheng:1977nv,Petcov:1976ff,Bilenky:1977du,Marciano:1977wx,Lee:1977tib}, for instance $\Gamma(\mu \rightarrow e \gamma) / \Gamma(\mu \rightarrow e \nu \bar{\nu}) < 10^{-40}$.
\par 
It is natural to query, once neutrinos are known to be non-degenerate, if dynamics not included in the SM can produce charged-lepton-flavour-violating (CLFV) processes. In fact, there is no known reason why this symmetry should be sustained, and many models beyond SM can violate it. As a consequence, we would have a much more involved dynamics of leptons, similar to the one of quarks, and displaying a more symmetric and egalitarian electroweak (EW) interaction. Indeed, we already have relevant experimental hints that point out to a non-trivial lepton dynamics, as is the case of the apparent violation of universality around the third family (allocating the tau lepton and its neutrino) in some decays of the B mesons \cite{Lees:2013uzd,Aaij:2014ora,Aaij:2015yra,Aaij:2017deq,Aaij:2017vbb,Aaij:2019wad}. Even earlier, similar hints appeared in $W \rightarrow \ell \nu_{\ell}$ \cite{Filipuzzi:2012mg}, although this tension seems to have been released by a recent experimental determination by ATLAS \cite{Aad:2020ayz}.
\par 
However, we are interested here in new-physics dynamics that generates CLFV processes involving the tau lepton. Although there has been done plenty of research in the study of CLFV processes (see, for instance, the reviews \cite{Raidal:2008jk,deGouvea:2013zba,Calibbi:2017uvl}), most of these involve only the lightest first two families. The tau family has novel features as it adds to the analogous processes that appear with $\mu$ and $e$ the decays into hadrons. This, together with the suspicion of universality violation in the third family, puts the focus on the tau lepton. While the study of CLFV tau decays into leptons has a wide bibliography, the one that considers tau decays into hadrons --- precisely the new feature of the third family --- has been less studied. We find, for instance, the model-independent approach of Refs.~\cite{Black:2002wh,Celis:2013xja,Celis:2014asa}, supersymmetric settings \cite{Brignole:2004ah,Arganda:2008jj,Herrero:2009tm} and little-Higgs models \cite{Lami:2016vrs}.
\par 
Our purpose is to perform a model-independent analysis of CLFV tau processes that involve hadrons. Our framework will be that of the Standard Model Effective Field Theory (SMEFT) \cite{Buchmuller:1985jz} where to the SM Lagrangian we add higher D-dimensional operators ${\cal O}_i^{{\text{(D)}}}$ that involve the same particle spectrum as the SM and are invariant under $\text{SU}(3)_\text{C} \times \text{SU}(2)_\text{L} \times \text{U}(1)_\text{Y}$,
\begin{equation}
\label{eq:smeft}
{\cal L_{{\text{SMEFT}}}}
={\cal L}_{{\text{SM}}}
+\sum_{D>4} \left( \frac{1}{\Lambda^{{\text{D}}-4}} \sum_i \, C_i^{{\text{(D)}}} \, {\cal O}^{{\text{(D)}}}_i \right),\qquad
\left[{\cal O}^{{\text{(D)}}}_i \right]
=\left[E^{{\text{D}}}\right],
\end{equation}
where $\Lambda$ is the scale that drives the new dynamics. With this setting, the dimensionless couplings of different operators --- the Wilson coefficients $C_i^{{\text{(D)}}}$ --- should be naturally ${\cal O}(1)$. The SMEFT Lagrangian --- written in terms of  fields at the electroweak scale and given by operators invariant under the electroweak gauge symmetry group --- represents an effective field theory at the EW scale that collects the dynamics produced by new physics. Any extension of the SM should be reduced to the SMEFT at the electroweak scale. Hence, it is a model-independent framework that ignores the origin or dynamics of the ultraviolet completion at $E \gg M_W$. The first contribution to CLFV processes can be generated by $D=6$ operators, and we will keep only those. For definiteness, we will use the basis and notation consistently with Ref.~\cite{Grzadkowski:2010es}. It is necessary to point out that our procedure is not valid if the origin of CLFV lies at $E \lesssim M_W$, in which case the SMEFT Lagrangian (\ref{eq:smeft}) does not serve as an appropriate framework.
\par 
The experimental search for LFV decays of the tau into hadrons has been carried out by BaBar and Belle experiments \cite{Amhis:2019ckw},
and it is foreseen that it will have a strong push with the results from Belle II \cite{Kou:2018nap}. Hence, it is timely to perform a thorough analysis of those processes in a model-independent framework in order to provide a key tool for model builders. We will be interested in the decays $\tau^- \rightarrow \ell^- P$, $\tau^- \rightarrow \ell^- P_1P_2$ and $\tau^- \rightarrow \ell^- V$, where $P$ stands for any pseudoscalar meson, $V$ for a vector resonance and $\ell = \mu, e$. These processes are simple enough to fit comfortably into our framework, but rich enough to offer a general view of the landscape of operators. We also wish to consider $\ell$--$\tau$ conversion in the presence of nuclei: $\mu^- (e^-)+ \mathcal{N}(A,Z) \rightarrow \tau^- \,  X$, i.e.\ with a fixed-target of atomic and mass numbers $Z$ and $A$, respectively. This latter process has received minor attention due to the complexities of its experimental setting; see, however, Refs.~\cite{Gninenko:2001id,Sher:2003vi,Abada:2016vzu,Takeuchi:2017btl,Gninenko:2018num} and references therein. Its feasibility at NA64 has been pointed out in Ref.~\cite{Gninenko:2018num}. Moreover, future foreseen fixed-target experiments such as the muon collider \cite{Delahaye:2013jla}, the electron-ion collider (EIC) \cite{Deshpande:2013paa}, 
the ILC \cite{Baer:2013cma} or circular colliders as LHeC \cite{Acar:2016rde} could consider to look for this conversion.
\par 
We will take into account the processes above and, using $\texttt{HEPfit}$ \cite{deBlas:2016ojx,deBlas:2019okz}, we perform an overall analysis of the participating $D=6$ Wilson coefficients in Eq.~(\ref{eq:smeft}) with the present experimental bounds of the CLFV processes. As we are expecting that the latter get a big improvement in the near future, the relevance of our work is more to provide a tool for the foreseen new results rather than the present numerical status.
\par 
In the next section, we specify our notation and the detailed account of the processes that we consider. In Section \ref{s:3}, we perform the fit and provide numerical results and their analyses. Section \ref{s:4} will collect our conclusions. Several appendices help to ease up and complement the main text.

\section{The SMEFT Lagrangian: lepton-flavour-violating processes}\label{s:2}

The SMEFT Lagrangian (\ref{eq:smeft}) provides the relevant effective theory that describes the physical processes at the electroweak scale. The spectrum contained in its operators and the symmetry of the latter are those of the SM; the relevant scale $\Lambda$ reflects the energy at which new physics appears. Clearly, $\Lambda$ depends on the new dynamics that we wish to describe, i.e.\ it does not need to be the same for LFV or violation of lepton number, for instance. In the following, we will use $\Lambda_{{\text{CLFV}}}$ to denote the scale of interest of the present work.
\par 
The processes we wish to analyse are those involving the third family of leptons. We choose this system since, as commented in the Introduction, it has been involved in several cases that could be related to some deviations from the SM dynamics.
Accordingly, our SMEFT operators are still invariant under global $\text{U}(1)_e \times \text{U}(1)_\mu$ rotations that do not involve the tau lepton anymore. The large mass of the tau lepton allows its CLFV decays into hadrons, opening a wide set of modes that can be looked for in dedicated experiments. Moreover, $\ell$--$\tau$ conversion ($\ell = \mu,e$) in the presence of nuclei has been scarcely considered in the bibliography. The obvious relevance of studying both type of processes at the same time is that they involve, generically, the same SMEFT operators. In addition, we will assume that the Wilson coefficients involving the muon and the electron are the same, i.e.\ we presume universality in the two lighter lepton families. This seems endorsed, up to now, by the experimental results. 
\par 
There is only one $D=5$ operator in the SMEFT Lagrangian \cite{Weinberg:1979sa}, but it violates lepton number and is of no interest for our research here. The list of $D=6$ operators is rather large \cite{Buchmuller:1985jz} and contains the first relevant operators that implement CLFV processes. We will limit our study to those and we will use, essentially, the notation from Ref.~\cite{Grzadkowski:2010es}. The operators contributing to processes under consideration here are collected in Table~\ref{tab:1}.
\begin{table}[!t]
\capstart
\begin{center}
\renewcommand{\arraystretch}{1.5}
\begin{tabular}{|c|c||c|c|}
\hline
\multicolumn{1}{|c}{$\Lambda^2 \, \times$ Coupling} &
\multicolumn{1}{|c||}{Operator}  &
\multicolumn{1}{c}{$\Lambda^2 \, \times$ Coupling} &
\multicolumn{1}{|c|}{Operator}   \\
\hline
\hline
 $C_{LQ}^{(1)}$ & $\left( \bar{L}_p \gamma_{\mu} L_r \right) \left( \bar{Q}_s \gamma^{\mu} Q_t \right)$ & $C_{e \varphi}$  &
 $\left( \varphi^{\dagger} \varphi \right) \left( \bar{L}_p e_r \varphi \right)$   \\
 $C_{LQ}^{(3)}$ & $\left( \bar{L}_p \gamma_{\mu} \sigma^I L_r \right) \left( \bar{Q}_s \gamma^{\mu} \sigma^I Q_t \right)$ &
 $C_{\varphi e}$  &
 $\left( \varphi^{\dagger}  i \overset{\leftrightarrow}{D}_{\mu} \varphi \right) \left( e_p \gamma^{\mu} e_r  \right)$   \\
 $C_{eu}$ & $\left( \bar{e}_p \gamma_{\mu} e_r \right) \left( \bar{u}_s \gamma^{\mu} u_t \right)$ & $C_{\varphi L}^{(1)}$ &
 $\left( \varphi^{\dagger} i \overset{\leftrightarrow}{D}_{\mu} \varphi \right) \left( \bar{L}_p \gamma^{\mu} L_r \right)$ \\
 $C_{ed}$ &  $\left( \bar{e}_p \gamma_{\mu} e_r \right) \left( \bar{d}_s \gamma^{\mu} d_t \right)$ & $C_{\varphi L}^{(3)}$ &
 $\left( \varphi^{\dagger} i \overset{\leftrightarrow}{D}_{I\mu} \varphi \right) \left( \bar{L}_p \sigma_I \gamma^{\mu} L_r \right)$ \\
 $C_{Lu}$ & $\left( \bar{L}_p \gamma_{\mu} L_r \right) \left( \bar{u}_s \gamma^{\mu} u_t \right)$ & $C_{eW}$ & $ \left( \bar{L}_p \sigma^{\mu \nu} e_r \right) \sigma_I \varphi W^I_{\mu \nu}$ \\
 $C_{Ld}$ &  $\left( \bar{L}_p \gamma_{\mu} L_r \right) \left( \bar{d}_s \gamma^{\mu} d_t \right)$ & $C_{eB}$ & $ \left( \bar{L}_p \sigma^{\mu \nu} e_r \right) \varphi B_{\mu \nu}$ \\ 
 $C_{Qe}$ &  $\left( \bar{Q}_p \gamma_{\mu} Q_r \right) \left( \bar{e}_s \gamma^{\mu} e_t \right)$ & & \\
 $C_{LedQ}$ & $\left( \bar{L}^j_p e_r \right) \left( \bar{d}_s Q^j_t \right)$ & & \\ 
 $C_{LeQu}^{(1)}$ & $\left( \bar{L}_p^j e_r \right) \varepsilon_{jk}  \left( \bar{Q}_s^k u_t \right)$ & & \\
 $C_{LeQu}^{(3)}$ & $\left( \bar{L}_p^j \sigma_{\mu \nu} e_r \right)  \varepsilon_{jk}  \left( \bar{Q}_s^k \sigma^{\mu \nu} u_t \right)$ & & \\
\hline
\multicolumn{4}{c}{}
\end{tabular}
\end{center}
\vspace*{-1cm}
\caption{\label{tab:1}
$D=6$ operators appearing in the Lagrangian (\ref{eq:smeft}) and contributing to the CLFV processes that we study in this work. The four-fermion operators appear on the left-hand side, while those involving the Higgs doublet $\varphi$ and the gauge bosons are on the right. The notation (up to small apparent changes), is the one from Ref.~\cite{Grzadkowski:2010es}. For the family indices we use $p$, $r$, $s$ and $t$, while $j$ and $k$ are isospin indices. For $I=1,2,3$, $\sigma_I$ are the Pauli matrices, with $\varepsilon = i \sigma_2$, and $\sigma^{\mu\nu}\equiv\frac i2[\gamma^\mu,\gamma^\nu]$. $\Lambda$ is then the scale where the new dynamics arises. The operators share the same notation with the associated couplings, substituting simply $C \rightarrow {\cal O}$, i.e.\ ${\cal O}_{LQ}^{(1)}$ and so on.}
\end{table}

\subsection{Hadronic tau decays} \label{ss:2.1}

\begin{figure}[!t]
\capstart
\begin{center}
      \includegraphics[scale=0.75]{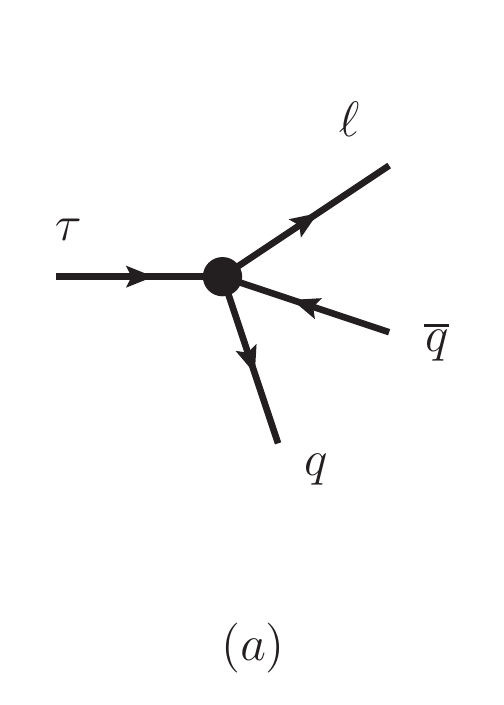}
      \hspace*{1.5cm}
      \includegraphics[scale=0.75]{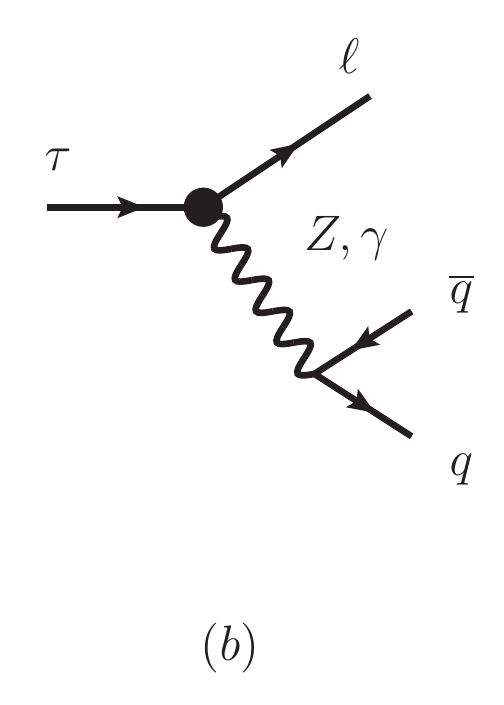}
      \hspace*{1.5cm}
      \includegraphics[scale=0.75]{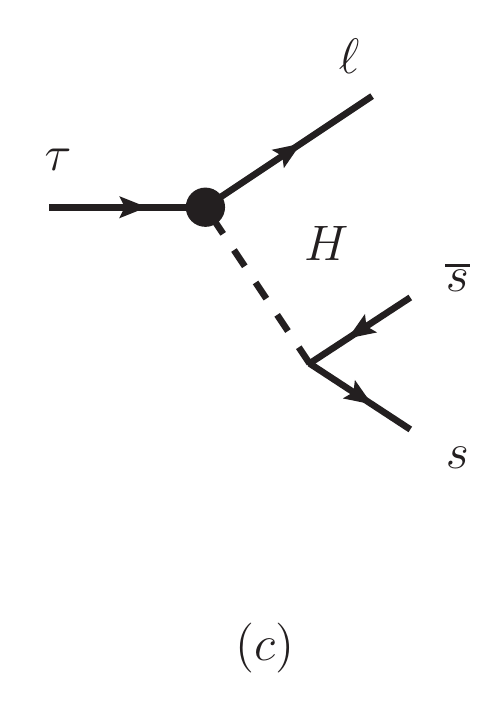}
\caption[]{\label{fig:1} Different contributions of the SMEFT Lagrangian to $\tau \rightarrow \ell \, \overline{q} q$, with $\ell = e,\mu$. The dot indicates the CLFV vertex. We consider $m_u=m_d=0$, but $m_s \neq 0$: The contribution of the Higgs in (c) thus only exists for the production of $\overline{s} s$.}
\end{center}
\end{figure}

\begin{figure}[!t]
\capstart
\begin{center}
\includegraphics[scale=0.75]{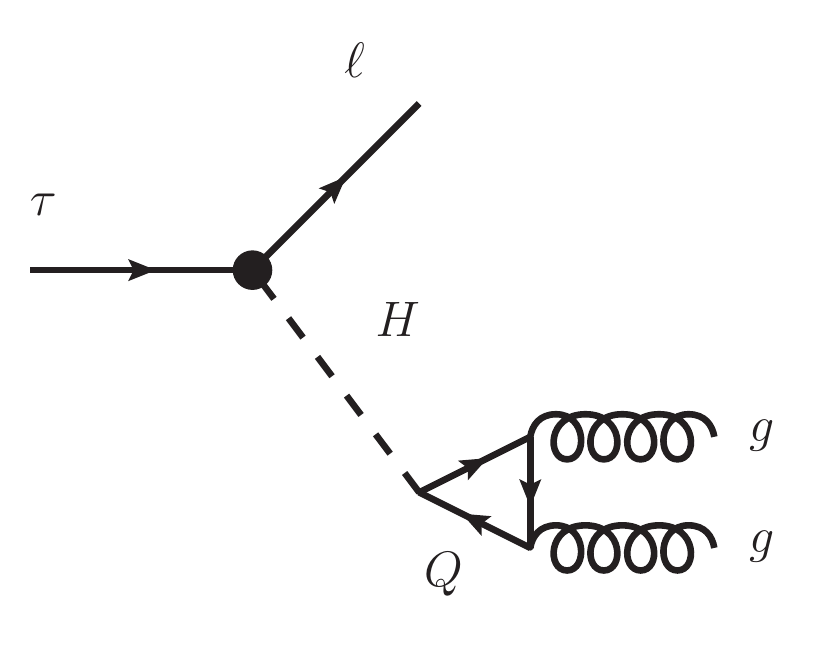}
\caption[]{\label{fig:2} Dominant scalar contribution to $\tau \rightarrow \ell \overline{P}P$, with $\ell = e,\mu$ and $P=\pi, K$. The dot indicates the CLFV vertex. The $Q$ in the loop stands for a heavy quark, namely $Q=c,b,t$.}
\end{center}
\end{figure}

We will study the CLFV decays $\tau^- \rightarrow \ell^- \{P,P_1P_2,V\}$, with $\ell = \mu, e$, and $P$ and $V$ standing for pseudoscalar and vector fields with light-quark content (namely $u$, $d$ and $s$), respectively.  In order to determine the widths of the hadronic tau decays using the $D=6$ operators in Table~\ref{tab:1}, the procedure has two steps:
\begin{itemize}
\item[1)] determine the perturbative amplitudes at parton level,
\item[2)] hadronize those partons into mesons.
\end{itemize}
The key role in the perturbative contribution has been customarily given to $\tau \rightarrow \ell q \overline{q}$ shown in Fig.~\ref{fig:1}. However, in Ref.~\cite{Celis:2013xja} it was pointed out that the scalar contribution via the diagram shown in Fig.~\ref{fig:2} should also be considered --- as the hadronization of gluons into mesons is not small at these energies. In the second step, we will proceed to hadronize the $\overline{q} \Gamma q$ currents (with $\Gamma = \{1, \gamma_5, \gamma_{\mu}, \gamma_{\mu} \gamma_5, \sigma_{\mu \nu}, \sigma_{\mu \nu} \gamma_5 \}$) into observable pseudoscalar and vector mesons, and analogously, for the gluons from Fig.~\ref{fig:2}. Let us now discuss separately the two points stated above.

\subsubsection{Perturbative amplitudes} \label{sss:2.1.1}

The SMEFT framework provides the processes $\tau \rightarrow (\ell + \text{hadrons})$ already at the tree level. Operators in Table~\ref{tab:1} generate two kinds of contributions:
\begin{itemize}
\item[i)] those yielding a two-quark current $\tau \rightarrow \ell \overline{q} q$ (shown in Fig.~\ref{fig:1}), either provided by local vertices as in (a), or by gauge-boson (b) or Higgs (c) exchanges; in the latter case, we will consider massless up and down quarks, but $m_s \neq 0$, and, accordingly, the diagram (c) will only contribute to the production of $\overline{s} s$,
\item[ii)] the scalar two-gluon contribution $\tau \rightarrow \ell g g$ (shown in Fig.~\ref{fig:2}), with heavy quarks in the finite fermion loop, that was also considered previously in Ref.~\cite{Celis:2013xja}; here it was concluded that, in spite of the loop suppression, this is the dominant Higgs contribution to these processes.
\end{itemize}
The results of the tree-level amplitudes related to diagrams from Fig.~\ref{fig:1} and generated by LFV operators from Table~\ref{tab:1} are collected in Section~\ref{apps:1} of Appendix~\ref{app:1}; the loop contribution from Fig.~\ref{fig:2} is given in Section \ref{apps:2}. In the latter amplitude, we have written (for hadronization purposes) the two-gluon final state in terms of the trace of the energy--momentum tensor \cite{Donoghue:1990xh}, as explained in Subsection \ref{apps:4} of Appendix~\ref{app:2}.
Accordingly, the general perturbative amplitude that describes the CLFV tau decays into hadrons is given by
\begin{equation} \label{eq:tottau}
{\cal M}_{\tau} = {\cal M}_{{\text{tree}}} + {\cal M}_{\tau gg} \, ,
\end{equation}
with ${\cal M}_{{\text{tree}}}$ and ${\cal M}_{\tau gg}$ defined in Eqs.~(\ref{eq:fourst}) and (\ref{eq:emuu}), respectively.

\subsubsection{Hadronization} \label{sss:2.1.2}

Our results for the tau decay amplitude ${\cal M}_{\tau}$ are given, for the tree-level contributions, in terms of light-quark bilinears, and, for the gluon case, by the energy--momentum tensor. The final states we take into account involve pseudoscalar mesons and vector resonances. Therefore, we need to hadronize the quark bilinears and the energy--momentum tensor. The chiral perturbation theory ($\chi$PT) \cite{Gasser:1983yg,Gasser:1984gg} provides a model-independent scheme for this procedure. Unfortunately, this framework only provides reliable results (typically) for $E \ll 1 \, \text{GeV}$, while the mass of the tau lepton is much larger: The energy region that happens to be populated by hadron resonances becomes kinematically available and relevant for the final state. A complementary scenario consistent with the constraints of the chiral symmetry is given by the resonance chiral theory (R$\chi$T) \cite{Ecker:1988te,Ecker:1989yg,Cirigliano:2006hb}. It provides us with a phenomenological Lagrangian, driven also by the chiral symmetry, that includes not only the light pseudoscalar mesons, but also the lightest $\text{U}(3)$ nonets of resonances that remain in the large-number-of-colours framework (i.e.\ when $N_\text{C} \rightarrow \infty$). In addition, external currents with appropriate quantum numbers allow to hadronize the relevant quark bilinears. 
\par 
We consider first the hadronization of the latter. We identify the scalar, pseudoscalar, vector, axial-vector and tensor currents:
\begin{alignat}{3}
\label{eq:quarkb}
S^i &= -\overline{q} \, \lambda^i \, q \,,\qquad\qquad\qquad\;\;
P^i &&= \overline{q} \, i \gamma_5 \, \lambda^i \, q \,,\notag\\
V^i_{\mu} & = \overline{q} \, \gamma_{\mu} \frac{\lambda^i}{2} \, q \,,\qquad\qquad\qquad
A^i_{\mu} &&= \overline{q} \, \gamma_{\mu} \gamma_5 \, \frac{\lambda^i}{2} \, q \,,\\
T^{i \, \mu \nu} & = \overline{q} \, \sigma^{\mu \nu} \, \frac{\lambda^i}{2} \, q \,,\qquad\qquad\;\;
T^{i \, \mu \nu}_5 &&= \overline{q} \, \sigma^{\mu \nu} \, \gamma_5 \, \frac{\lambda^i}{2} \, q \,,\notag
\end{alignat}
where $\lambda^i$, $i=0,\dots,8$ are the Gell-Mann matrices and $\gamma_5\equiv i\gamma^0\gamma^1\gamma^2\gamma^3$; the chiral currents are defined as is customary: $J_\text{L}^{i} = (S^i - i P^i)/2$, $J_\text{R}^i = (S^i + i P^i)/2$, $J_\text{L}^{i \, \mu} = (V^{i \, \mu} - A^{i \, \mu})/2$,  $J_\text{R}^{i \, \mu} = (V^{i \, \mu} + A^{i \, \mu})/2$, 
$T_\text{L}^{i \, \mu \nu} = (T^{i \, \mu \nu} - T_5^{i \, \mu \nu})/2$, $T_\text{R}^{i \, \mu \nu} = (T^{i \, \mu \nu} + T_5^{i \, \mu \nu})/2$.
\par 
Within the $\chi$PT and R$\chi$T frameworks, one attaches external currents to the massless QCD Lagrangian ${\cal L}_\text{QCD}^0$, allowing to determine the relevant QCD currents:
\begin{equation} \label{eq:QCDext}
{\cal L}_\text{QCD}^\text{ext} = {\cal L}_\text{QCD}^0 + \overline{\psi} \gamma_{\mu}\left( v^{\mu} + a^{\mu} \gamma_5 \right) \psi - \overline{\psi} \left( s - i p \gamma_5 \right) \psi + \overline{\psi} \sigma_{\mu \nu} \, \overline{t}^{\mu \nu} \psi \,. 
\end{equation}
The auxiliary fields $v_{\mu} = v_{\mu}^i \lambda^i/2$, $a_{\mu} = a_{\mu}^i \lambda^i/2$, $s= s_i \lambda^i$, $p = p_i \lambda^i$ and $\overline{t}^{\mu \nu} = \overline{t}_{\mu \nu}^i \lambda^i/2$ are Hermitian matrices in the flavour space.%
\footnote{For practical purposes, in the tensor case we define the $t_{\mu \nu}$ and $t_{\mu \nu}^{\dagger}$ external fields by
$\bar\psi \sigma_{\mu \nu} \overline{t}^{\mu \nu} \psi = \bar\psi_L \sigma^{\mu \nu} t_{\mu \nu}^{\dagger} \psi_R + \bar{\psi}_R \sigma^{\mu \nu} t_{\mu \nu} \psi_L$ \cite{Cata:2007ns}.}
Hence, within the R$\chi$T framework and once the relevant Lagrangian (see Appendix~\ref{app:3}) is fixed, the quark bilinear hadronization is given by the functional derivatives with respect to the external fields of the R$\chi$T action. The SMEFT operators of our interest are:
\begin{alignat}{4}
\label{eq:functder}
S^i &= \frac{\partial {\cal L}_\text{R$\chi$T}}{\partial s^i} \bigg|_{j=0} \,, \qquad \qquad
P^i = \frac{\partial {\cal L}_\text{R$\chi$T}}{\partial p^i} \bigg|_{j=0} \,, \qquad \qquad \notag \\
V_{\mu}^i &= \frac{\partial {\cal L}_\text{R$\chi$T}}{\partial v_{\mu}^i} \bigg|_{j=0} \,, \qquad \qquad
A_{\mu}^i = \frac{\partial {\cal L}_\text{R$\chi$T}}{\partial a_{\mu}^i} \bigg|_{j=0} \,, \qquad \qquad \\
& \qquad \qquad T_{\mu \nu}^i = \frac{\partial {\cal L}_\text{R$\chi$T}}{\partial (t_{\mu \nu}^{i \, \dagger} + t_{\mu \nu}^i)} \bigg|_{j=0} \,, \qquad \notag
\end{alignat}
where the R$\chi$T Lagrangian is the one in Eq.~(\ref{eq:ourrl}). The `$j=0$' notation indicates that after the derivatives are calculated, all external currents are set to zero. We obtain
\begin{equation}
\label{eq:rescur}
\begin{split}
S^i & = - \, B_0 \, \langle \, \lambda^i \, \Phi^2\, \rangle \, + \, 8 \, \frac{B_0}{F^2} \, L_5^\text{SD} \, \langle \, \lambda^i \, \partial_{\mu} \, \Phi \, \partial^{\mu} \, \Phi \, \rangle \, + \, 4 \, B_0 \, c_m \, \langle \, \lambda^i \, S \, \rangle \, \\ & + \,  \, 4 \, \frac{B_0}{F^2} \, \frac{d_m^2}{M_P^2} \, \left[ \, 2 \, \langle \, \lambda^i \, \Phi \, M \, \Phi \, \rangle \, + \, \langle \, \lambda^i \, \left\{ M, \Phi^2 \right\} \, \rangle \, \right] \,  +  \, 4 \, B_0 \, \gamma \, g^{\mu \nu} \, \langle  \, \lambda^i \, T_{\mu \nu} \, \rangle \, , \\
P^i & = \sqrt{2} \, B_0 \, F \, \langle \, \lambda^i \, \Phi \, \rangle \, - \, 4 \, \sqrt{2} \, \frac{B_0}{F} \, \frac{d_m^2}{M_P^2} \, \langle \, \lambda^i \, \left\{ M , \Phi \right\} \, \rangle \, , \\
V_{\mu}^i & = \frac{i}{2}  \, \langle \, \lambda^i \, \left[ \, (\partial_{\mu} \Phi) \, \Phi \, - \, \Phi \, \partial_{\mu} \Phi \, \right] \, \rangle \, - \, \frac{F_V}{\sqrt{2}} \, \langle \, \lambda^i \, \partial^{\nu} V_{\nu \mu} \, \rangle \, , \\
A_{\mu}^i & = - \, \frac{F}{\sqrt{2}} \, \langle \, \lambda^i \, \partial_{\mu} \Phi \, \rangle \, , \\
T_{\mu \nu}^i & = - \, i \, \frac{\Lambda_2^\text{SD}}{2 \, F^2} \, \langle \, \lambda^i \, \left[ \, (\partial_{\mu} \Phi) \, \partial_{\nu} \Phi \, - \, (\partial_{\nu} \Phi) \, \partial_{\mu} \Phi \, \right] \, \rangle \, + \, \frac{T_V}{2} \, \langle \, \lambda^i \, V_{\mu \nu} \, \rangle \, ,
\end{split}
\end{equation}
with $\Phi$ being the octet of Goldstone fields, and
\begin{equation} \label{eq:massg}
M = \left( \begin{array}{ccc}
m_{\pi}^2 & & \\
& m_{\pi}^2 & \\
& & 2 m_K^2 - m_{\pi}^2
\end{array} \right) . 
\end{equation}
These are not the final results for the hadronization of the quark bilinears. The resonance chiral Lagrangian provides additional interactions that allow to introduce resonance contributions to the currents above. The mesonic final states will be two pseudoscalars $P_1P_2$, one pseudoscalar $P$, or one vector resonance $V$. The final results are:
\begin{alignat}{3}
\label{eq:finalh}
\left[ \, i \, \overline{q}_i \, \gamma_5 \, q_j \, \right. 
&\rightarrow \,\left. P \, \right]
&&\simeq \quad 2 \,  B_0 \, F \, \Omega_P^{(1)}(ij) \, + \,  2 \, \frac{B_0}{F} \, \frac{d_m^2}{M_P^2} \, m_K^2 \, \Omega_P^{(2)}(ij) \, ,\nonumber \\
\left[ \, \overline{q}_i \, \gamma_{\mu} \, \gamma_5 \, q_j \, \right. 
&\rightarrow \left. \, P \, \right]
&&\simeq \quad - i \, 2 \, F \, \Omega_A^{(1)}(ij) \, p_{\mu}\, ,\nonumber \\
\left[ \, \overline{q}_i \, \gamma_{\mu} \, q_j \, \right. 
&\rightarrow \left. \, V \, \right]
&&\simeq \quad - 2 \, F_V \, M_V \, \Omega_V^{(1)}(ij) \, \varepsilon_{\mu} \,  , \nonumber \\
\left[ \,  \overline{q}_i \, \sigma_{\mu \nu}  \,  q_j \,\right.  
&\rightarrow \left. \, V  \, \right]
&&\simeq \quad i \, 2 \, \frac{T_V}{M_V} \, \Omega_T^{(1)}(ij) \left( \, p_{\mu} \, \varepsilon_{\nu} \, - \, p_{\nu} \, \varepsilon_{\mu} \, \right) \, , \nonumber \\
\left[ \, \overline{q}_i \, q_j  \right.  
& \rightarrow \left.  P_1 \, P_2  \, \right]
&&\simeq 2 \, B_0\, \Omega_S^{(1)}(ij) \left[ \, 1  +  4 \, \frac{L_5^\text{SD}}{F^2}  \, \left(  s  -m_1^2 -  m_2^2  \right)  \right]  + \, 2 \, \frac{B_0}{F^2} \, \frac{d_m^2}{M_P^2} \, m_K^2 \, \Omega_S^{(2)}(ij)  \, \nonumber \\
& && + \frac{B_0}{F^2} \, c_m \, \sum_{S} \frac{\Omega_{S}^{(3)}(ij)}{s \, - \, M_S^2} \, \left[ \, c_d \, \Omega_{S}^{(4)} \, \left( \, s \, - \, m_1^2 \, - \, m_2^2 \, \right) \, + \, 2 \, c_m \, m_K^2 \, \Omega_{S}^{(5)} \, \right]  \, \nonumber \\
& && + \frac{1}{3} \, \frac{B_0}{F^2} \, \gamma \,\sum_{T} \, \frac{\Omega_{T}^{(2)}(ij)}{M_T^4} \,\left\{ \, g_T \, \Omega_{T}^{(3)} \, \left[ \, (m_1^2-m_2^2)^2\, + \, M_T^2 \, (m_1^2+m_2^2) \, \right. \right. \nonumber \\
& && - \left.\left. s \, (M_T^2+s) \, \right]
+ \, 2 \, (2 M_T^2+ s) \, \left[ \, \beta \, \Omega_{T}^{(4)} (m_1^2+m_2^2-s) \,
- \, 2 \, \gamma \, m_K^2 \, \Omega_{T}^{(5)} \, \right] \right\}, \nonumber \\
\left[ \,  \overline{q}_i \, \gamma_{\mu} \, q_j  \right. 
&\rightarrow \left.  P_1 \, P_2 \, \right]
&&\simeq \left[ \, 2 \, \Omega_V^{(2)}(ij)\, + \, 
 \sqrt{2} \, \frac{F_V \, G_V}{F^2} \, \sum_{V} \, \frac{s}{M_V^2 \, - \, s} \,  \Omega_{V}^{(1)}(ij) \, \Omega_{V}^{(3)} \, \right] (\, p_1\, -\, p_2\, )_{\mu} \nonumber \\
& && + \left[ \, \sqrt{2} \, \frac{F_V \, G_V}{F^2} \, (m_2^2 - m_1^2) \, \sum_{V} \, \frac{\Omega_{V}^{(1)}(ij) \, \Omega_{V}^{(3)}}{M_V^2 \, - \, s} \, \right] (\, p_1 \, + \, p_2 \,)_{\mu} \,, \nonumber \\
\left[ \,  \overline{q}_i \, \sigma^{\mu \nu}  \,  q_j \right. 
&\rightarrow \left.  P_1 \,P_2  \, \right] &&\simeq \frac{i}{F^2} \left[ \, - \Lambda_2^\text{SD} \, \Omega_T^{(6)}(ij) \, 
+  \, 2 \, \sqrt{2} \, G_V \, T_V \, \sum_{V} \, \frac{\Omega_{T}^{(1)}(ij) \, \Omega_{V}^{(3)}}{M_V^2 \, - \, s} \, \right] \left( \, p_1^{\mu} \, p_2^{\nu} \,- \, p_1^{\nu} \, p_2^{\mu} \, \right).
\end{alignat}
Note that the remaining currents do not contribute to these final states. The dimensionless parameters $\Omega$ identify the different intermediate and final states and are listed in Appendix~\ref{app:4}. These results within the R$\chi$T correspond to a model of the Large-$N_\text{C}$ limit, assuming that only the lightest multiplet of intermediate resonances (that survive in the $N_\text{C} \rightarrow \infty$ limit) contribute to the dynamics. The widths of resonances appear at the subleading order ($\mathcal{O}(1/N_\text{C}$)), and are therefore not present in the expressions above. However, this is not satisfactory from the phenomenological point of view and we thus implement the widths in the corresponding poles: $[M_R^2 - s] \to [M_R^2 - s - i M_R \Gamma_R]$. Moreover, a constant width reasonably represents only narrow resonances (typically when $\Gamma_R / M_R  \lesssim 0.1$) and in such a case it is a good approximation. The analytical construction of momentum-dependent widths is only known for dominant two-body decays \cite{GomezDumm:2000fz}.
\par 
The hadronization of the vector current into two pseudoscalars of equal mass%
\footnote{For two pseudoscalars of different mass the matrix element has, in addition, another form factor.}
is driven by the well-known vector form factor, that has been thoroughly studied in the literature. In fact, the expression in Eq.~(\ref{eq:finalh}) is just the starting point of a more thorough model-independent construction based both on the R$\chi$T and the use of dispersion relations \cite{Guerrero:1997ku,Pich:2001pj,Dumm:2013zh} or Pad\'e approximants \cite{Masjuan:2008fv}. The vector form factor for the pseudoscalar $P$ is defined as
\begin{equation} \label{eq:fvvp}
\langle P(p_1) \overline{P}(p_2) | V_{\mu}^\text{em} | 0 \rangle = ( p_1 - p_2 )_{\mu} F_\text{V}^P(q^2)\,, 
\end{equation}
where $q=p_1 + p_2$, and where the electromagnetic current is defined by
\begin{equation} \label{eq:vmuem}
 V_{\mu}^\text{em} = \sum_{Q_q} Q_q \bar{q} \, \gamma_{\mu} \, q =  V_{\mu}^3 + \frac{1}{\sqrt{3}} \, V_{\mu}^8 \, ,
\end{equation}
with $Q_q$ standing for the electric charge of the quark $q=u,d,s$. Hence, we notice that, for instance, we can hadronize the $\bar{u} \gamma_{\mu} u$ current through
\begin{equation} \label{eq:uualter}
\langle P(p_1) \overline{P}(p_2) | \bar{u} \gamma_{\mu} u | 0 \rangle = 
(p_1 - p_2)_{\mu} F_\text{V}^P(q^2) +  \frac{2}{\sqrt{6}} \, \langle P(p_1) \overline{P}(p_2) | V_{\mu}^0 | 0 \rangle \,.
\end{equation} 
Analogous expressions could be obtained for the $\bar{d} \gamma_{\mu} d$ and $\bar{s} \gamma_{\mu} s$ currents, where the $V_{\mu}^3$ current appears in addition on the r.h.s. We only apply this procedure to the hadronization of the $u$-quark current in order to point out the possibility for further improvements. The singlet vector current is determined using the R$\chi$T framework, giving
\begin{equation} \label{eq:vsinglet}
V_{\mu}^0 = F_\text{V} \left[\, \sin \theta_V \, \partial^{\nu} \phi_{\nu \mu} - \cos \theta_V \, \partial^{\nu} \omega_{\nu \mu} \, \right] ,
\end{equation}
which further contributes to the hadronization of two pseudoscalars. For our vector form factor $F_\text{V}^P(q^2)$, we have employed the result from Appendix~B of Ref.~\cite{Arganda:2008jj}, modifying the hadronization for the final states $\pi^+ \pi^-$, $K^+ K^-$ and $\overline{K}^0 K^0$.
\par 
We turn now to the hadronization of the two gluons from the diagram in Fig.~\ref{fig:2}. The amplitude represented by this diagram contributes only to two pseudoscalars in the final state, and is given in Eq.~(\ref{eq:emuu}) of Appendix~\ref{apps:2} in terms of the matrix element $\theta_P(q^2)$ of the energy--momentum tensor (see Subsection~\ref{apps:4} in Appendix~\ref{app:2}). We could perform an evaluation of the matrix element within the R$\chi$T, but final-state interactions (FSI) in the two-pseudoscalar final state are ruled by the $I=J=0$ scattering phase shift. This implies that FSI are important and have to be considered, for instance, using dispersion relations. This was already pointed out in Ref.~\cite{Donoghue:1990xh} for $\pi \pi$ and $\overline{K} K$, and has been recently reevaluated for the two-pion case in Ref.~\cite{Celis:2013xja}. We will only consider the $\pi \pi$ final state and we will use the results of the latter reference: Incidentally, they have also improved the matrix element for the two-pion final state of the mass term $m_s \bar{s} s$ (see Eq.~\ref{eq:tauamp3}), which we have also used.

\subsection{\texorpdfstring{\boldmath $\ell$--$\tau$}{l--tau} conversion in nuclei} \label{ss:2.2}

The conversion among flavours of charged leptons in the presence of a nucleus is a well-motivated way to study CLFV phenomena that has been pursued already in the past, namely for the $\mu$--$e$ conversion in nuclei with the strongest limit set by Sindrum II \cite{Bertl:2006up}:
\begin{equation}
\mathcal{B}_{\mu e}^{Au}=\frac{\Gamma(\mu^{-}\, \text{Au} \, \rightarrow \, e^{-}\, \text{Au})}{\Gamma_{\text{capture}}(\mu^{-}\, \text{Au} )}<7\times 10^{-13}\, , \quad 90\% \, \text{C.L.}
\end{equation}
The $\mu$--$e$ experiments are of a different nature than those concerning the $\tau$ lepton: Typically, these experiments are performed at low energy and the muon becomes bounded before decaying in orbit or being captured by the nucleus.

For $\mu$--$\tau$, the conversion is expected to occur by deep inelastic scattering (DIS) of the lepton off the nucleus, thus these experiments are based on a fixed-target nucleus hit by an incoming lepton beam of a given flavour $\ell$. If the energy of the beam leptons is high enough, they will penetrate the hadronic structure of the nucleons within the nucleus and interact with its constituents, the partons, i.e.\ quarks and gluons. Due to the fact that lepton flavour is conserved within the SM (which also holds at the tree level for the charged-lepton sector in its minimally extended version), a change of the flavour of the incoming charged lepton as a result of the interaction with nuclei is forbidden. Therefore, any measurable signal of a process of this kind would suggest new physics.

Our aim here is to perform a model-independent analysis within the SMEFT framework of $\ell$--$\tau$ conversion in nuclei $(\ell=\mu,e)$. Regarding the product of the interaction, we consider a $\tau$ lepton plus any hadronic content of no particular relevance to us, i.e.\ we are only interested in the inclusive process $\ell+\mathcal{N}(A,Z)\to\tau+X$, where we do not have any information about $X$.

Since the interacting parton lives in the hadronic environment of the nucleus, its dynamics is heavily influenced by low-energy non-perturbative QCD effects. However, we can make use of QCD factorization theorems to separate the non-perturbative behaviour --- encoded in the so-called parton distribution functions (PDFs) --- from the part we can compute perturbatively. Once the perturbative calculation is done, we calculate the convolution of the result with the PDFs to obtain the total cross section of the process.

\subsubsection{Perturbative amplitudes}
\label{sss:2.2.1}

\begin{figure}[!tb]
\capstart
\begin{center}
    \includegraphics[scale=0.75]{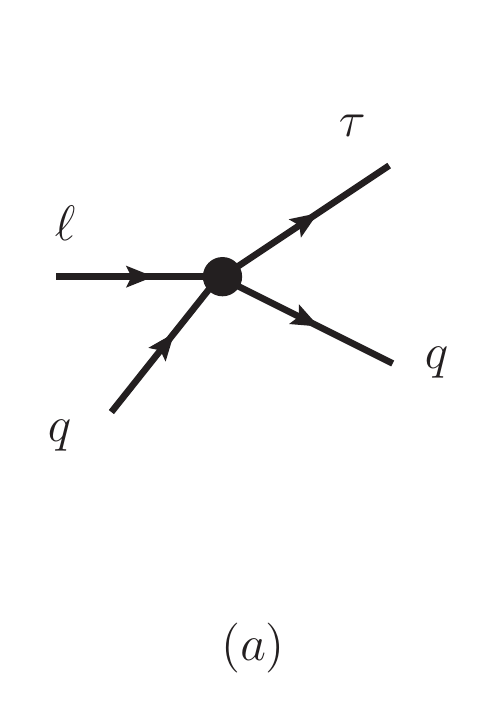}
    \hspace*{1.5cm}
    \includegraphics[scale=0.75]{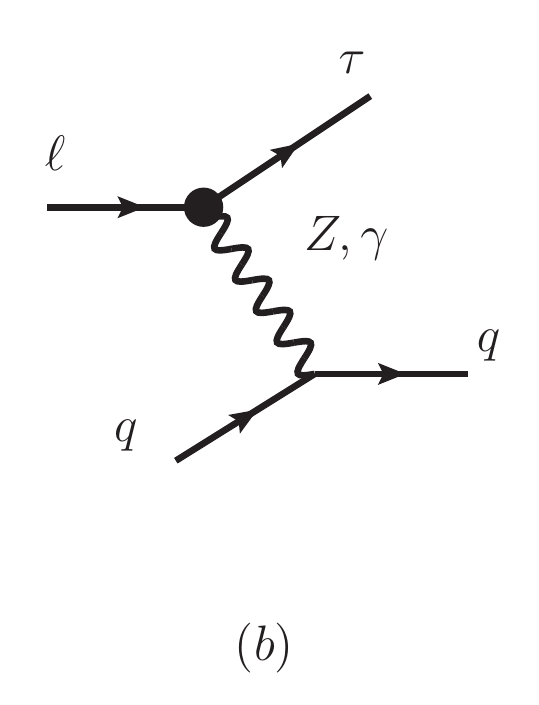}
    \hspace*{1.5cm}
    \includegraphics[scale=0.75]{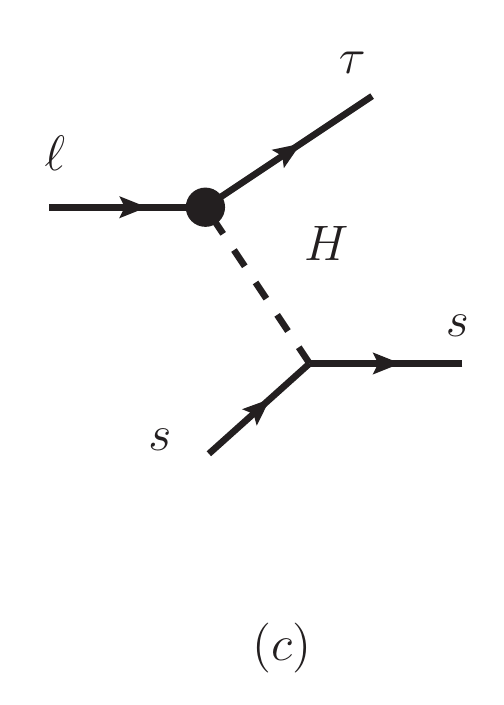}
\caption[]{\label{fig:3} Different contributions of the SMEFT Lagrangian to $\ell q \rightarrow \tau q$ for $\ell= e,\mu$ and $q=u,d,s$. The dot indicates the CLFV vertex. We consider $m_u=m_d=0$ but $m_s \neq 0$. Hence, the contribution of the Higgs in (c) only exists for $q=s$.}
\end{center}
\end{figure}

\begin{figure}[!tb]
\capstart
\begin{center}
  \includegraphics[scale=0.75]{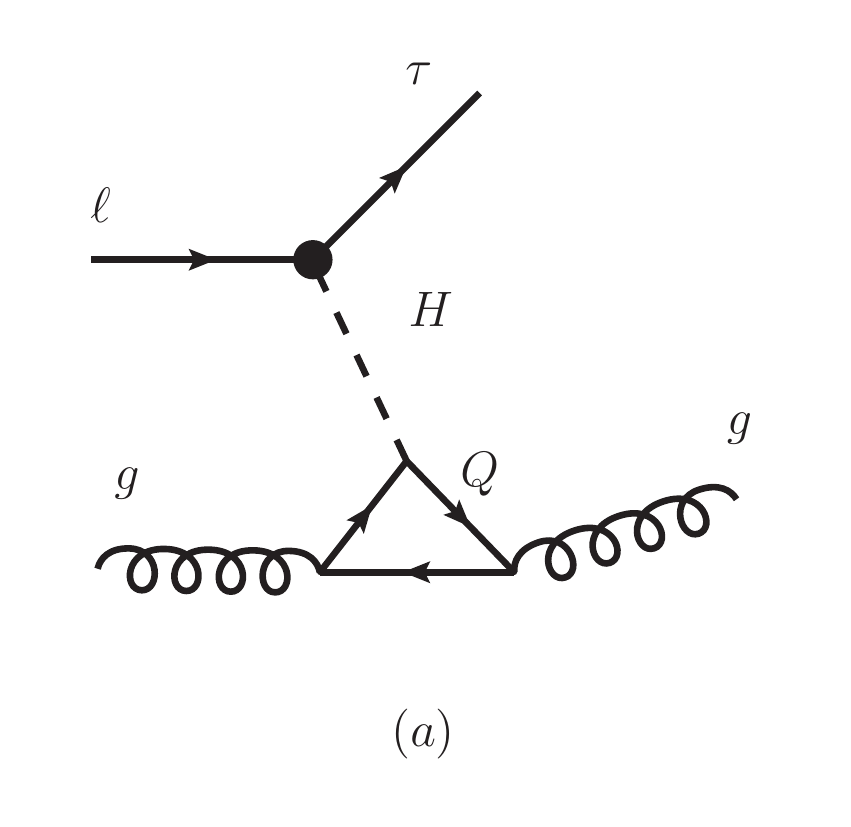}
  \hspace*{2cm}
  \includegraphics[scale=0.75]{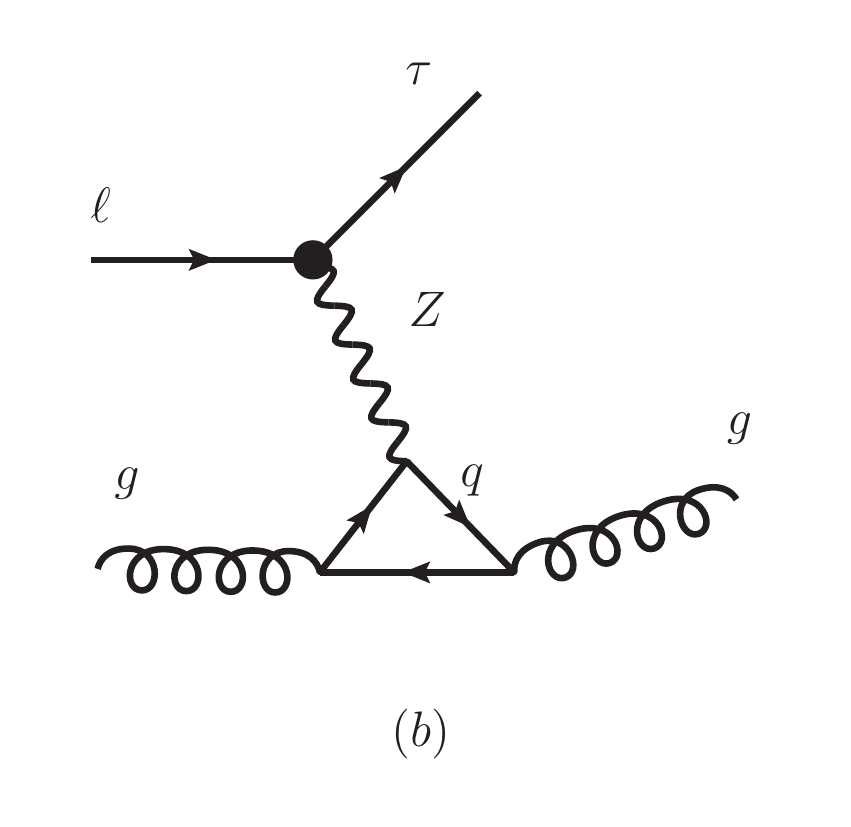}
\caption[]{\label{fig:4} Higgs and $Z$ contribution to $\ell g \rightarrow \tau g$, with $\ell = e,\mu$. The dot indicates the CLFV vertex. $Q$ is a heavy quark, namely $Q=c,b,t$, and $q=u,d,c,s,t,b$.}
\end{center}
\end{figure}

The perturbative cross sections involved in this process are computed using the SMEFT operators listed in Table \ref{tab:1}. These yield three different leading contributions:
\begin{itemize}
\item[i)] the process $\ell q \rightarrow \tau q^{(\prime)}$ (see Fig.~\ref{fig:3}), represented in terms of local vertices (a), the gauge-boson (b), and Higgs (c) exchange. We consider massless up and down quarks, while $m_{s}\neq 0$: The diagram (c) thus only contributes to the production of $\bar{s}s$,
\item[ii)] the same process as i), but with antiquarks: $\ell \bar{q} \rightarrow \tau \bar{q}^{(\prime)}$. This leads to different cross sections of the process and also the non-perturbative behaviour of anti-quarks inside the nucleons is not the same as of their opposite-charged partners,
\item[iii)] the process $\ell g \rightarrow \tau g$ (see Fig.~\ref{fig:4}), represented by the Higgs (a) or $Z$-gauge-boson (b) exchange, and a quark triangle loop.
\end{itemize}
Note that for the processes of type i) and ii), $\ell q (\bar{q}) \rightarrow \tau q^{(\prime)} (\bar{q}^{(\prime)})$, we may also allow for quark-flavour change. Therefore, we also take into account quark currents such as $\bar{c}u, \bar{b}s$, \dots, which contribute to the total amplitude only via contact-interaction contribution, i.e.\ via the process depicted in Fig.~\ref{fig:3}(a). Nevertheless, we consider the same Wilson coefficients for all quark flavours, thus assuming minimal flavour violation in the quark sector (driven only by the CKM matrix). Allowing quarks to change flavour during the interaction allows us to consider a wider variety of final states for the hadronic $\tau$ decays, as well as it leads to an increased cross section of the process of $\ell$--$\tau$ conversion in nuclei. This implies a richer phenomenology and stronger constraints on the Wilson coefficients. These flavour-changing neutral currents (FCNCs) are forbidden at the tree level in the SM due to the GIM mechanism~\cite{PhysRevD.2.1285}, while there is no reason to assume that such a mechanism is also relevant for the beyond-Standard Model (BSM) physics at higher energy scales. We will thus study both cases: 1) CLFV with FCNC, and 2) CLFV only. However, for the first scenario, one could expect the CLFV and FCNC phenomena to be mediated by different type of new physics, since there might be no reason for both phenomena to be related to each other. Describing both interactions within the SMEFT framework, this would mean different energy scales $\Lambda$. In spite of these considerations, as we are studying both scenarios, we will be assuming throughout this work only one energy scale driving all new-physics interactions, and thus implicitly considering the same energy scale for both phenomena.

The results for the tree-level amplitudes shown in Fig.~\ref{fig:3} and generated by LFV operators from Table~\ref{tab:1} are given in Section~\ref{apps:1} of Appendix~\ref{app:1}. 
Regarding the contribution iii), for the Higgs exchange with heavy quarks in the finite fermion loop (as it happened in its hadronic $\tau$ decay counterpart; see Fig.~\ref{fig:2}) it was shown in Ref.~\cite{Takeuchi:2017btl} that (in spite of the loop suppression) this is the dominant Higgs contribution to the $\ell$--$\tau$ conversion in nuclei. We also conclude that the diagram in Fig.~\ref{fig:4}(b) is not negligible compared to the Higgs exchange: Actually, they are of the same order of magnitude.
All the relevant amplitudes for process iii) are collected in Section~\ref{apps:3} of Appendix~\ref{app:1}, while for a thorough discussion regarding the contribution shown in Fig.~\ref{fig:4}(b), an interested reader is referred to Section~\ref{apps:5} of Appendix~\ref{app:2}.
Accordingly, the general perturbative amplitudes that describe the CLFV $\mu (e)$--$\tau$ conversion in nuclei are given in terms of the amplitudes from Appendix~\ref{app:1}, using
\begin{itemize}
\item[i)]
\begin{equation}
\mathcal{M}_{qq} = \, {\cal M}_{{\text{loc}}} + {\cal M}_{Z} + {\cal M}_{\gamma} + {\cal M}_H \, , 
\end{equation}
\item[ii)]
\begin{equation}
\mathcal{M}_{\bar{q}\bar{q}} = \, {\cal M}_{{\text{loc}}}^{\prime} + {\cal M}_{Z}^{\prime} + {\cal M}_{\gamma}^{\prime} + {\cal M}_H^{\prime} \, , 
\end{equation}
\item[iii)]
\begin{equation}
\mathcal{M}_{gg} = \, \mathcal{M}_{H l} + \mathcal{M}_{Z l}\, ,
\end{equation}
\end{itemize}
where $\mathcal{M}^{\prime}$s stand for the same amplitudes as those from Appendix~\ref{app:1}, but for antiquarks.

\subsubsection{Non-Perturbativity: Nuclear parton distribution functions}
\label{sss:2.2.2}

Nuclei are bound systems where the low-energy non-perturbative effects of QCD among their constituents are non-negligible. Therefore, to address this problem properly, we make use of quantities that describe these long-distance effects: the parton distribution functions. 
By means of the QCD factorization theorems, the total cross section can be computed as a convolution of the non-perturbative PDFs ($f$) and the perturbative cross sections ($\hat{\sigma}$) calculated using the amplitudes of the previous section (Section~\ref{sss:2.2.1}):
\begin{equation}
\sigma_{\ell-\tau}=\hat{\sigma}\otimes f \, .
\end{equation}

The total cross section is an observable quantity. However, since the perturbative cross sections are computed within perturbation theory, our inability to calculate them at every order of the perturbative expansion generates a non-physical dependence on the energy scale which propagates into the PDFs; this also means that they depend on the renormalization scheme. The above-mentioned scale is usually taken as $Q^{2}=-q^{2}$, with $q^{2}$ being the transferred momentum of the system; $Q^{2}$ is also often called the characteristic scale of the process. Furthermore, it is customary to characterize the PDFs through the Lorentz invariant quantity $\xi$, the fraction of the nucleus momentum carried by the interacting parton.
Consequently, we express the perturbative cross section as well as the PDFs in terms of the two discussed invariant quantities
\begin{equation}
\sigma_{\ell-\tau}=\hat{\sigma}(\xi,Q^{2})\otimes f(\xi, Q^{2}) \, ,
\end{equation}
where the total cross section still depends on the Wilson coefficients $C_{i}$ and the BSM energy scale $\Lambda_{{\text{CLFV}}}$.

Whereas the dependence of the PDFs on the momentum fraction $\xi$ is completely non-perturbative and has to be extracted from the data, their evolution in terms of $Q^{2}$ is achieved by using the DGLAP evolution equations: Once the PDFs are determined at a given scale $Q^{2}_{0}$, we can calculate it at any other scale $Q^{2}$.
There are several groups performing this global QCD analysis using state-of-the-art perturbative theoretical computations to obtain the best PDFs given the current data; for an overview of the field, see Ref.~\cite{Rojo:2019uip} and references therein. Since, in our case, we are dealing with heavy nuclei instead of free nucleons, nuclear binding effects alter significantly the non-perturbative behaviour of the constituents at different $\xi$ regimes, as it was first discovered in Ref.~\cite{Aubert:1983xm}. All these effects are included in the nuclear parton distribution functions (nPDFs), which we find more suitable to describe the $\ell$--$\tau$ conversion in nuclei: We use the nCTEQ15-np fit of the nPDFs provided by the group around the nCTEQ15 project~\cite{Kovarik:2015cma}, incorporated within the ManeParse Mathematica package~\cite{Clark:2016jgm}.

\subsubsection{Total cross section}
\label{sss:2.2.3}

The convolution of the perturbative and non-perturbative pieces is a rather complicated topic due to higher-order QCD corrections, target mass corrections, etc. However, when including next-to-leading-order (NLO) QCD corrections, it can be shown within the QCD-improved parton model \cite{Leader:1996hm} that the modifications can be absorbed into the PDFs, while keeping the perturbative cross sections at tree level. This is the leading-order (LO) QCD formalism (or twist-2 factorization) which we follow in this work: Our SMEFT perturbative cross sections are calculated at tree level, while the nCTEQ15 nPDFs that we use are computed at NLO~\cite{Kovarik:2015cma}. We would like to point out that the twist-2 factorization is appropriate in the limit of massless partons~\cite{Owens:1992hd}. This may be the case for the $u$, $d$ and $s$ quarks, however, note that larger uncertainties are expected when considering the quark currents of diagrams of type (a) in Fig.~\ref{fig:3} of processes i) and ii) with massive $c$ and $b$ quarks. 

The perturbative unpolarized differential cross sections for the processes from Section~\ref{sss:2.2.1} (contributing to $\ell$--$\tau$ conversion in nuclei) in terms of the invariants $\xi$ and $Q^{2}$ and computed within the SMEFT framework (using the operators from Table~\ref{tab:1}) are
\begin{align}
\label{eq:pertampl1}
\frac{\diff\hat{\sigma} (\ell \, q_{i}(\xi P) \rightarrow \tau \, q_{j})}{\diff\xi \diff Q^{2}}&=\frac{1}{16 \pi\lambda(s(\xi),m_{\ell}^{2},m_{i}^{2})}\,\overline{|\mathcal{M}_{qq}(\xi,Q^{2})|^{2}}\,,\\
\label{eq:pertampl2}
\frac{\diff\hat{\sigma} (\ell \, \bar{q}_{i}(\xi P) \rightarrow \tau \, \bar{q}_{j})}{\diff\xi \diff Q^{2}}&=\frac{1}{16 \pi\lambda(s(\xi),m_{\ell}^{2},m_{i}^{2})}\,\overline{|\mathcal{M}_{\bar{q}\bar{q}}(\xi,Q^{2})|^{2}}\,,\\
\label{eq:pertampl3}
\frac{\diff\hat{\sigma} (\ell \, g(\xi P) \rightarrow \tau \, g)}{\diff\xi \diff Q^{2}}&=\frac{1}{16 \pi\lambda(s(\xi),m_{\ell}^{2},m_{i}^{2})}\,\overline{|\mathcal{M}_{gg}(\xi,Q^{2})|^{2}}\,,
\end{align}
with the momentum of the interacting parton $p_{i}=\xi P$ being a fraction of the nucleus total momentum $P$; thus, we consider $m_{i}^{2}=\xi^{2}M^{2}$ since only the nucleus mass $M$ is physical. $\lambda(a,b,c)\equiv(a+b-c)^{2}-4  a  b$ is the K\"all\'en's triangle function.
Finally, using the LO QCD formalism, the total differential cross section reads
\begin{equation}
\label{eq:totcrosssect}
\begin{split}
\sigma(\ell\, \mathcal{N}(P)\rightarrow \, \tau \, X)
&=\sum_{i,j} \int_{\xi_\text{min}}^{1} \int_{Q^{2}_{-}(\xi)}^{Q^{2}_{+}(\xi)}\diff\xi \diff Q^{2}\, \left\lbrace \frac{\diff\hat{\sigma} (\ell \, q_{i}(\xi P) \rightarrow \tau \, q_{j})}{\diff\xi \diff Q^{2}}\,f_{q_{i}}(\xi, Q^{2})\right.\\
&+\left.\frac{\diff\hat{\sigma} (\ell \, \bar{q}_{i}(\xi P) \rightarrow \tau \, \bar{q}_{j})}{\diff\xi \diff Q^{2}}\,f_{\bar{q}_{i}}(\xi, Q^{2})+\frac{\diff\hat{\sigma} (\ell \, g(\xi P) \rightarrow \tau \, g)}{\diff\xi \diff Q^{2}}\,f_{g}(\xi, Q^{2})\right\rbrace \,.
\end{split}
\end{equation}
The integration limits are given in Appendix~\ref{app:5}.

\section{Numerical results} \label{s:3}

In this section, we present the main features and results of our numerical analysis performed on the SMEFT $D=6$ operators generating CLFV $\tau$-involved processes: hadronic $\tau$ decays and $\ell$--$\tau$ conversion in nuclei. In the first part, we introduce the HEPfit tool \cite{deBlas:2019okz} employed in the analysis and its statistical framework. We also present the existing or expected experimental limits on these processes. In the second part, we present the results of the fits for each process class (tau decays or conversion in nuclei) individually as well as the combined analysis, making always the distinction between the 1) CLFV with FCNC, and 2) CLFV only cases.

\subsection{Set-up}
\label{ss:3.1}

The effects of new physics on the physical observables are parametrized within the SMEFT framework by the Wilson coefficients (WCs) $C_i$ and the energy scale where the new degrees of freedom live; in our case we denote this scale as $\Lambda_{{\text{CLFV}}}$. In general, every observable with a specific experimental bound will depend on several WCs. Consequently, in our work, we have a set of observables related to CLFV phenomena, each depending on several WCs and $\Lambda_{{\text{CLFV}}}$. Our goal is to translate the available information on the former into relations and constraints for the latter. Actually, we will be fitting (as is usual) the ratio $C/\Lambda_\text{CLFV}^2$. This is achieved with help of HEPfit, an open-source tool embedded with a Bayesian statistical framework that uses a Markov chain Monte Carlo (MCMC) routine. In this way the complete WC parameter space can be sampled. As output, we obtain allowed values for the WCs at different confidence levels, as well as the correlations among all of them. Since we are working within a Bayesian framework, the priors we set for the WCs (i.e.\ their initial probability distributions we choose) will be of primordial importance: We use flat distributions for the WCs since we do not have any reason to favour some values over others.

Due to the fact that the studied CLFV processes depend on more than one WC, we could not know which WC (or WCs) is (are) behind one possible experimental signal.
Measurements of other CLFV processes would be required in order to answer what kind of new physics is responsible for these observations. This is where the main importance of our general numerical analysis lies. Without additional information, a naive analysis of the sensitivity of the observables on individual Wilson coefficients would lead to overestimated (too strong) bounds on the latter: If the actual new physics contributes through more than one operator, this sensitivity gets diluted due to the correlations among different WCs.

\subsubsection{Experimental bounds}
\label{sss:3.1.1}
The best experimental results on CLFV hadronic $\tau$ decays (as upper limits on the widths) have been given mainly by Belle and BaBar \cite{Amhis:2019ckw}. Possible final states considered in this work are
\begin{align*}
&\tau\to\ell P: & P&=\pi^{0},K^{0}, \eta,\eta^{\prime}\,,\\
&\tau\to\ell P_1P_2: & P_1P_2&=\pi^{+}\pi^{-},K^{0}\bar{K}^{0},K^{+}K^{-},\pi^{+}K^{-},K^{+}\pi^{-}\,,\\
&\tau\to\ell V: & V&=\rho^{0}(770),\omega(782),\phi(1020),K^{*0}(892),\bar{K}^{*0}(892)\,.
\end{align*}
Note that in Appendix~\ref{app:4} we give the R$\chi$T results for the hadronization of the quark currents into two pseudoscalars for more states than those listed here. This is because of the lack of experimental data on those decays; there are still many processes that have not been searched for. The expressions for the decay widths and our definition of the width of tau decays into hadron resonances are given in Appendix~\ref{app:8}.
This analysis is expected to be improved when new data appear and better bounds are set, as it should be the case with Belle II \cite{Aushev:2010bq}: They claim an improvement on the sensitivity by at least one order of magnitude. Hence, we will also consider the Belle II expected limits.

Regarding $\ell$--$\tau$ conversion in nuclei, there are no experimental limits yet. However, for the numerical analysis and intending to show the relevance of this process for CLFV searches, we will consider the most conservative expected sensitivity of the NA64 experiment \cite{Gninenko:2018num}. This can be further translated in terms of the limits on the physical observables of our interest as
\begin{equation}
\label{eq:ratiomutau}
R_{\ell \, \tau}=\frac{\sigma(\ell\, \mathcal{N}\rightarrow \, \tau \, X)}{\sigma(\ell\, \mathcal{N}\rightarrow \, \ell \, X)}\sim 10^{-13}-10^{-12}\,.
\end{equation}
Here, the numerator is given by Eq.~\eqref{eq:totcrosssect} and the denominator is the dominant contribution to the inclusive $\ell + \mathcal{N}$ process: the lepton bremsstrahlung on nuclei, that we take from Eq.~(21) of Ref.~\cite{Gninenko:2018num}. We will use two specific nuclei, namely Fe(56,26) and Pb(208,82). Following the prospects of this experiment, we consider for the energy of the incident lepton beam $E_{e}=100$\,GeV and $E_{\mu}=150$\,GeV for electrons and muons, respectively.

\subsubsection{Wilson coefficients}

In what follows, we slightly modify the basis of $D=6$ operators contributing to CLFV shown in Table~\ref{tab:1}, so that it suits better our study. We comment on some (hopefully well-motivated) modifications of several WCs, as well as their running.

First, we find that the operators ${\cal O}_{\varphi L}^{(1)}$ and ${\cal O}_{\varphi L}^{(3)}$ lead to the same contribution to CLFV $\tau$-involved processes. Therefore, our analysis is not sensitive to associated WCs separately, but only to their combination, namely
\begin{equation}
C_{\varphi L}^{(1) \, \prime} \equiv C_{\varphi L}^{(1)}+C_{\varphi L}^{(3)} \, .
\end{equation}  
Likewise, for the non-FCNC case, once the analysis is performed, it turns out that we cannot distinguish between this redefined $C_{\varphi L}^{(1) \, \prime}$ and $C_{L Q}^{(3)}$. This forces us to consider only the following combination as independent:
\begin{equation}
\label{eq:CLQ3noFCNC}
C_{L Q}^{(3) \, \prime}\equiv \ C_{L Q}^{(3)} + C_{\varphi L}^{(1)}+C_{\varphi L}^{(3)} \, .
\end{equation}
Similarly, the contributions stemming from ${\cal O}_{e B}$ and ${\cal O}_{e W}$ are equal up to factors of $c_\text{W}\equiv \cos \theta_\text{W}$ and $s_\text{W}\equiv \sin \theta_\text{W}$, with $\theta_\text{W}$ being the weak angle. We are thus again not sensitive to these two WCs, but only to their combination. Moreover, both operators contribute through a photon and $Z$ exchange. Hence, to disentangle these contributions, we can do a `rotation' of both WCs and define their particular combinations $C_{\gamma}$ and $C_{Z}$ as
\begin{equation}
\begin{pmatrix}
C_{\gamma} \\
C_{Z}
\end{pmatrix}=
\begin{pmatrix}
c_\text{W} & -s_\text{W}\\
s_\text{W} & c_\text{W}
\end{pmatrix}
\begin{pmatrix}
C_{eB} \\
C_{eW}
\end{pmatrix}.
\end{equation}
We will then put constrains on $C_{\gamma}$ and $C_{Z}$ instead of $C_{e B}$ and $C_{e W}$.

Second, as we discussed in Section~\ref{sss:2.2.2}, our inability to calculate the physical observables to all orders in the perturbation theory produces an artificial dependence of the WCs on the energy scale. Therefore, in order to compare the constraints set on the WCs coming from different processes at different energies, we should apply the renormalization-group equations (RGEs) to run all the WCs to the same energy scale. We will only consider the QCD running since it is (by far) dominant, and perform the analysis at the scale of $\tau$ decays. Likewise, since, in effective field theories, the scale dependence of WCs is closely related to the scale dependence of associated currents of the fundamental theory, we should only worry about the scalar and tensorial quark currents present in the four-fermion operators of Table~\ref{tab:1}: Vectorial currents do not run in QCD and so neither do their WCs.

For the scalar quark currents, both for $\bar{q}_{i}q_{i}$ and $\bar{q}_{i}q_{j}$, we arrive at scale-independent WCs $C_{LedQ}^{\, \prime}$ and $C_{LeQu}^{(1) \prime}$ using the following redefinitions:
\begin{equation}
\label{eq:scalar_redefinition}
\begin{split}
 C_{LedQ}=   \frac{m_{i}}{m_{\tau}} \, C_{LedQ }^{\, \prime}   \, , \qquad \qquad
C_{LeQu}^{(1)}=  \frac{m_{i}}{m_{\tau}}\, C_{LeQu}^{(1) \prime}  \, .
\end{split}
\end{equation}
This allows us to remove the scale dependence of $B_{0}$ in $\tau$ decays through the $\chi$PT relation $2B_{0}M_{q}\simeq M$, with $M_{q}=\text{diag}(m_{u},m_{d},m_{s})$ being the diagonal matrix of the light quark masses and $M$ the pseudoscalar physical-mass matrix defined in Eq.~\eqref{eq:massg}.
The running of the tensorial WC is given by
\begin{equation}
C_{LeQu}^{(3)}(m_{\tau})=Z(m_{\tau},\mu_{\ell-\tau})\,C_{LeQu}^{(3)}(\mu_{\ell-\tau})\,,
\end{equation}
where $C_{LeQu}^{(3)}(m_{\tau})$ is the WC at the scale of hadronic $\tau$ decays. The $Z$-factor is given by
\begin{equation}
Z(m_{\tau},\mu_{ \ell-\tau})=\bigg[ \frac{\alpha_{s}^{4}(m_{\tau})}{\alpha_{s}^{4}(m_{b}) }\bigg]^{-\frac{12}{75}}\bigg[ \frac{\alpha_{s}^{5}(m_{b})}{\alpha_{s}^{5}(\mu_{\ell-\tau}) }\bigg]^{-\frac{12}{69}}\, .
\end{equation}

Finally, the set of 15 independent WCs considered in our general (with FCNC) analysis reads
\begin{equation}
\Big\{C_{LQ}^{(1)},C_{LQ}^{(3)},C_{eu},C_{ed},C_{Lu},C_{Ld},C_{Qe},C_{LedQ}^{\, \prime},C_{LeQu}^{(1) \, \prime},C_{LeQu}^{(3)}(m_{\tau}),C_{\varphi L}^{(1) \, \prime},C_{\varphi e},C_{\gamma},C_{Z},C_{e \varphi}\Big\}\,.
\end{equation}
For the non-FCNC scenario, one needs to trade $C_{\varphi L}^{(1) \, \prime}$ and $C_{LQ}^{(3)}$ for their combination $C_{LQ}^{(3)\,\prime}$ from Eq.~\eqref{eq:CLQ3noFCNC}.

\subsection{Results}
Here, we present the main results obtained from the numerical analysis in several different scenarios. First, we address the case of hadronic $\tau$ decays both for the existing Belle and expected Belle II limits. Second, we focus on $\ell$--$\tau$ conversion in nuclei. Finally, we show the results of the combined analysis.

\subsubsection{Hadronic \texorpdfstring{$\tau$}{tau} decays}
\label{sss:3.2.1}
The observables used in this analysis are the branching ratios of $\tau$ decays into an electron or a muon and the hadronic final states given in Section~\ref{sss:3.1.1}. In total, there are 14 observables for each final-state-lepton flavour. Needless to mention, these observables are not equally sensitive to all WCs and the experimental limits are not equally strong either. This then leads to different constraints on the ratios $C/\Lambda_{{\text{CLFV}}}^{2}$ and correlations among them.

\begin{figure}[!tb]
\capstart
\begin{center}
\includegraphics[scale=0.9]{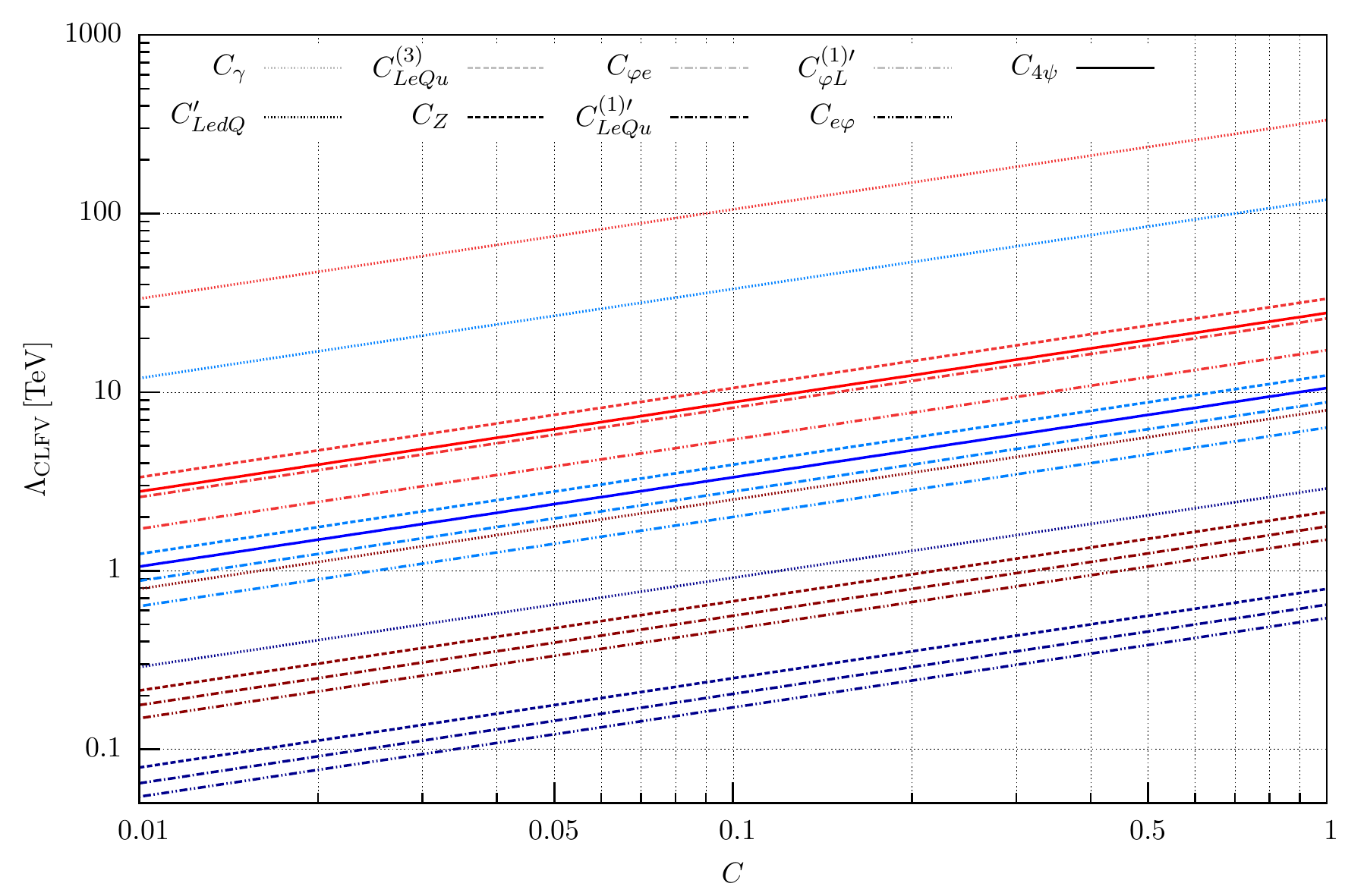}
\caption[]{\label{fig:5} Constraints on $\Lambda_\text{CLFV}$ with respect to the values of WCs, based on the current Belle (shades of blue) and expected Belle II (shades of red) limits, given at the 99.8\,\% confidence level.
The four-fermion WCs are represented altogether as $C_{4\psi}$.
For a given set of limits (distinguished by blue and red color shades), the lighter shades correspond to the WCs listed in the first row of the key (omitting now for a moment the common four-fermion WC $C_{4\psi}$) and the darker shades correspond to the WCs listed in the second row.
To make the use of the plot even simpler, for a given set of bounds (red or blue lines), the WCs (again up to $C_{4\psi}$) are listed in the key in the same order as they appear in the plot, the light-shaded lines being always above the dark-shaded ones.
Similarly, for a given WC, the red line appears always above the blue line, corresponding to stronger limits expected from Belle II.}
\end{center}
\end{figure}

After using the current limits from Belle (see Fig.~\ref{fig:5}), the least constrained WCs are --- due to the small quark masses involved --- the scalar (Higgs) $C_{e \varphi}$ and (up-type-quark) $C_{LeQu}^{(1) \, \prime}$.
These are followed by the `rotated' $C_{Z}$ and the other scalar WC $C_{LedQ}^{\, \prime}$. The $C_{\varphi L}^{(1) \, \prime}$ and $C_{\varphi e}$, both contributing via an intermediate-$Z$-exchange diagram, as well as the 4-fermion vectorial WCs are practically equally constrained: Here, the down-type-quark  WCs are constrained slightly stronger. The constraint on the tensorial $C_{LeQu}^{(3)}$ is then slightly stronger than on the 4-fermion ones. Finally, the strongest constraint is on the `rotated' $C_{\gamma}$.
Let us now comment on how the situation changes while including/excluding the FCNCs.
First, the non-FCNC case results in an incapability of disentangling the contribution of $C_{L Q}^{(3)}$ and $C_{\varphi L}^{(1) \, \prime}$, so we are only sensitive to their combination \eqref{eq:CLQ3noFCNC}. Second, FCNCs only happen through 4-fermion operators. One would then expect these to be less constrained in the non-FCNC case. However, the lost of sensitivity on $C_{L Q}^{(3)}$ and $C_{\varphi L}^{(1) \, \prime}$ separately results in lower correlations among the redefined $C_{LQ}^{(3) \, \prime}$, $C_{LQ}^{(1)}$ and $C_{Lu}$; see also Figs.~\ref{fig:G1} and~\ref{fig:G2} of Appendix~\ref{app:6} for direct comparison. As compared to the FCNC case, this in turn leads to a slightly stronger constraint on $C_{LQ}^{(3) \, \prime}$, equal constraints for both $C_{LQ}^{(1)}$ and $C_{Lu}$, and slightly weaker ones for the rest of the 4-fermion WCs. The latter effect is enhanced even further for the cases of $C_{eu}$, $C_{ed}$ and $C_{Qe}$ due to the increase of the correlations among these WCs. Note also that, due to the strong correlation between $C_{LedQ}^{ \, \prime}$ and $C_{e \varphi}$, and the lower constraint on the former due to the previous argument, the constraint on the latter is also reduced in the non-FCNC case.

Considering WCs are of order 1, the current Belle limits are probing energy scales up to $\approx$~120\,TeV: This holds for the best case related to $C_{\gamma}$, while the scale of 1\,TeV is not reached for the least constrained WCs. This situation is expected to improve with Belle II by approximately a factor of 3.
Examining the expected Belle II limits, the analysis results in the same pattern of constraints as in the previous case of Belle limits (see again Fig.~\ref{fig:5}).  In Table~\ref{tab:Belle_BelleII} we give (including FCNCs) the energy scales probed both by Belle and Belle~II for WCs of order 1.
\begin{table}[!htb]
\capstart
\begin{center}
\renewcommand{\arraystretch}{1.5}
\begin{tabular}{||c|c|c|c|c|c|}
\cline{1-6}
 \multicolumn{6}{||c|}{Bounds on $\Lambda_{{\text{CLFV}}}$\,[TeV] }  \\
\hline
 WC & Belle & Belle II & WC &  Belle & Belle II\\
\hline
\hline
 $C_{LQ}^{(1)}$ & $\gtrsim 8.5 $  & $\gtrsim 26 $  &  $C_{LeQu}^{(1) \, \prime}$ & $\gtrsim 0.65 $  & $\gtrsim 1.8 $  \\
\hline
 $C_{LQ}^{(3)}$ & $\gtrsim 7.5 $   & $\gtrsim 21 $ &  $C_{LeQu}^{(3)}$ & $\gtrsim 12$ & $\gtrsim 33 $ \\
\hline
 $C_{eu}$ &  $\gtrsim 7.7 $  & $\gtrsim 22 $ & $C_{\varphi L}^{(1) \, \prime}$ & $\gtrsim 6.3 $  & $\gtrsim 17 $  \\
\hline
$C_{ed}$, $C_{Ld}$ & $\gtrsim 10 $   & $\gtrsim 26 $  & $C_{\varphi e}$ & $\gtrsim 8.8 $  & $\gtrsim 26 $  \\
\hline
$C_{Lu}$ & $\gtrsim 6.5 $  & $\gtrsim 20 $   & $C_{\gamma}$ & $\gtrsim 120$  & $\gtrsim 330$ \\
\hline
$C_{Qe}$ & $\gtrsim 11 $  & $\gtrsim 28 $  & $C_{Z}$ & $\gtrsim 0.79$  & $\gtrsim 2.1 $ \\
\hline
$C_{LedQ}^{\, \prime}$ & $\gtrsim 2.9 $  & $\gtrsim 7.9 $  &  $C_{e \varphi }$ & $\gtrsim 0.54$  & $\gtrsim 1.5 $ \\
\hline
\end{tabular}
\end{center}
\vspace*{-0.5cm}
\caption{\label{tab:Belle_BelleII} Bounds on the new-physics energy scale mediating CLFV phenomena ($\Lambda_{{\text{CLFV}}}$) in tau decays. Here, we consider $C\approx1$. The results are based on Belle and Belle II limits, given at the 99.8\,\% confidence level.}
\end{table}

\begin{figure}[tb]
\capstart
\begin{center}
\includegraphics[scale=0.9]{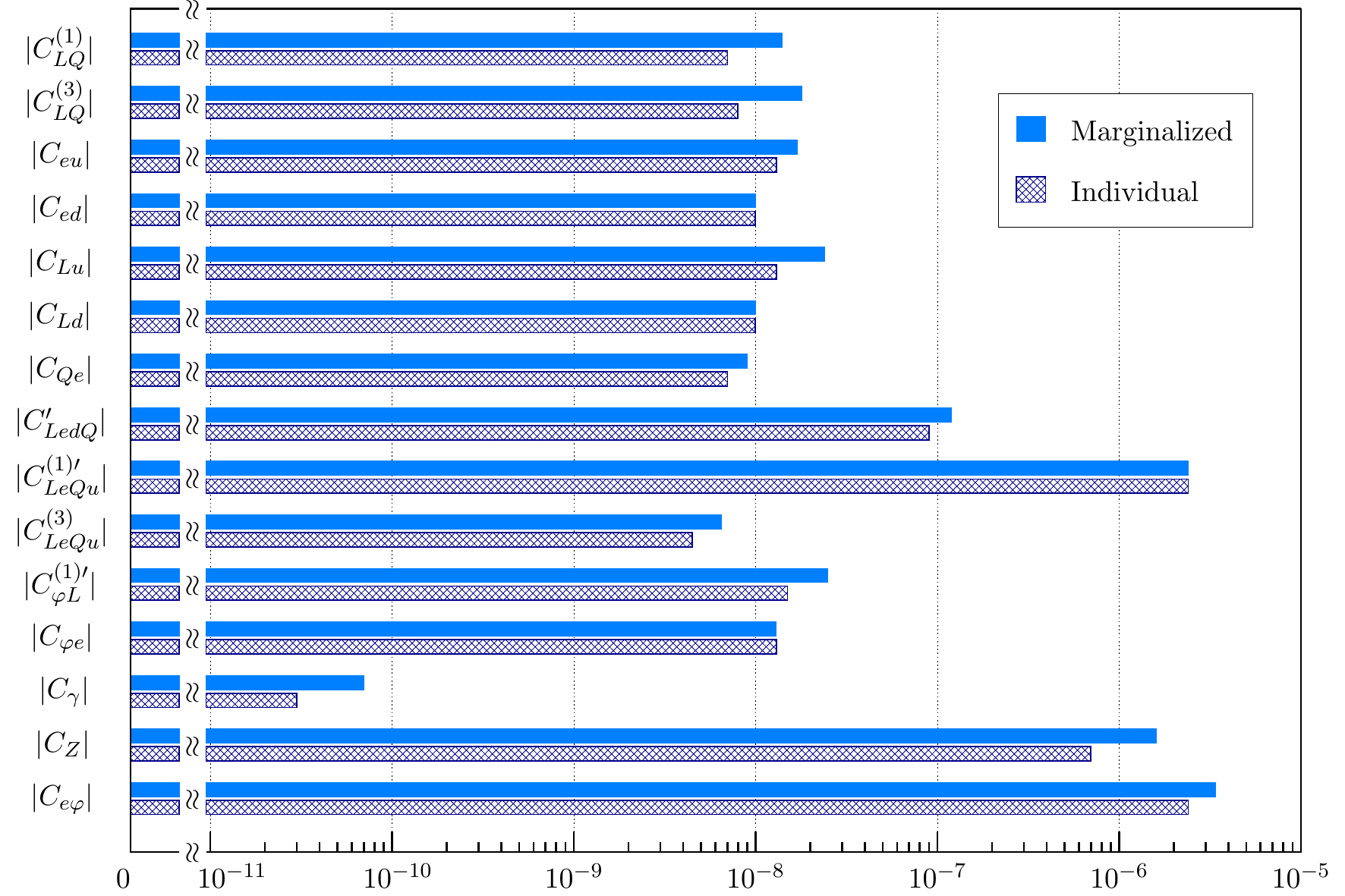}
\caption[]{\label{fig:6} Constraints on $C/ \Lambda_{{\text{CLFV}}}^{2}$\,[GeV$^{-2}$] based on the current Belle limits, stemming from the marginalized and individual analyses for tau decays, given at the 99.8\% confidence level.}
\end{center}
\end{figure}

The single-parameter analyses (i.e.\ when only one WC is kept nonzero) typically provide stronger bounds as compared to the marginalized approach, where all parameters are varied simultaneously. As it was pointed out at the beginning of Section~\ref{ss:3.1}, since more than one $D=6$ operator could be involved in each CLFV process, single-parameter analyses could be missing some important information --- for instance, the possible correlations among some of the parameters. The bounds on these parameters would then be too strong. In Fig.~\ref{fig:6} we give both the individual and marginalized bounds based on the Belle and Belle II limits; we refer the reader to Appendix~\ref{app:6} for correlation matrices.  

\subsubsection{\texorpdfstring{\boldmath $\ell$--$\tau$}{l--tau} conversion in nuclei}

Considering $\ell$--$\tau$ conversion in nuclei only, we perform the fit by taking into account four observables: two for each lepton (electron and muon) and other two for different nuclei, Fe(56,26) and Pb(208,82). The normalization channel (the bremsstrahlung cross section) in the observable under consideration \eqref{eq:ratiomutau} is much larger for electrons than for muons, as it is for lead compared to iron. This means that the results will be mainly driven by the $\mu$--$\tau$ conversion in Fe(56,26) and to a lesser degree in Pb(208,82), as it, indeed, turns out to be the case. Accordingly, this fit behaves as a single-parameter analysis. The correlation matrix (see Fig.~\ref{fig:G3} of Appendix~\ref{app:6}) is thus almost diagonal, except for the $C_{LeQu}^{(1) \, \prime}$--$C_{LeQu}^{(3)}$ correlation of $\rho \approx 0.5$, and for the case $C_{LQ}^{(3)}$--$C_{\varphi L}^{(1) \, \prime}$ for which we find $\rho \approx 0.66$: Even though we are able to constrain the WCs from the latter pair separately (as opposed to the non-FCNC case), their correlation is still strong. 

\begin{figure}[tb]
\capstart
\begin{center}
\includegraphics[scale=0.9]{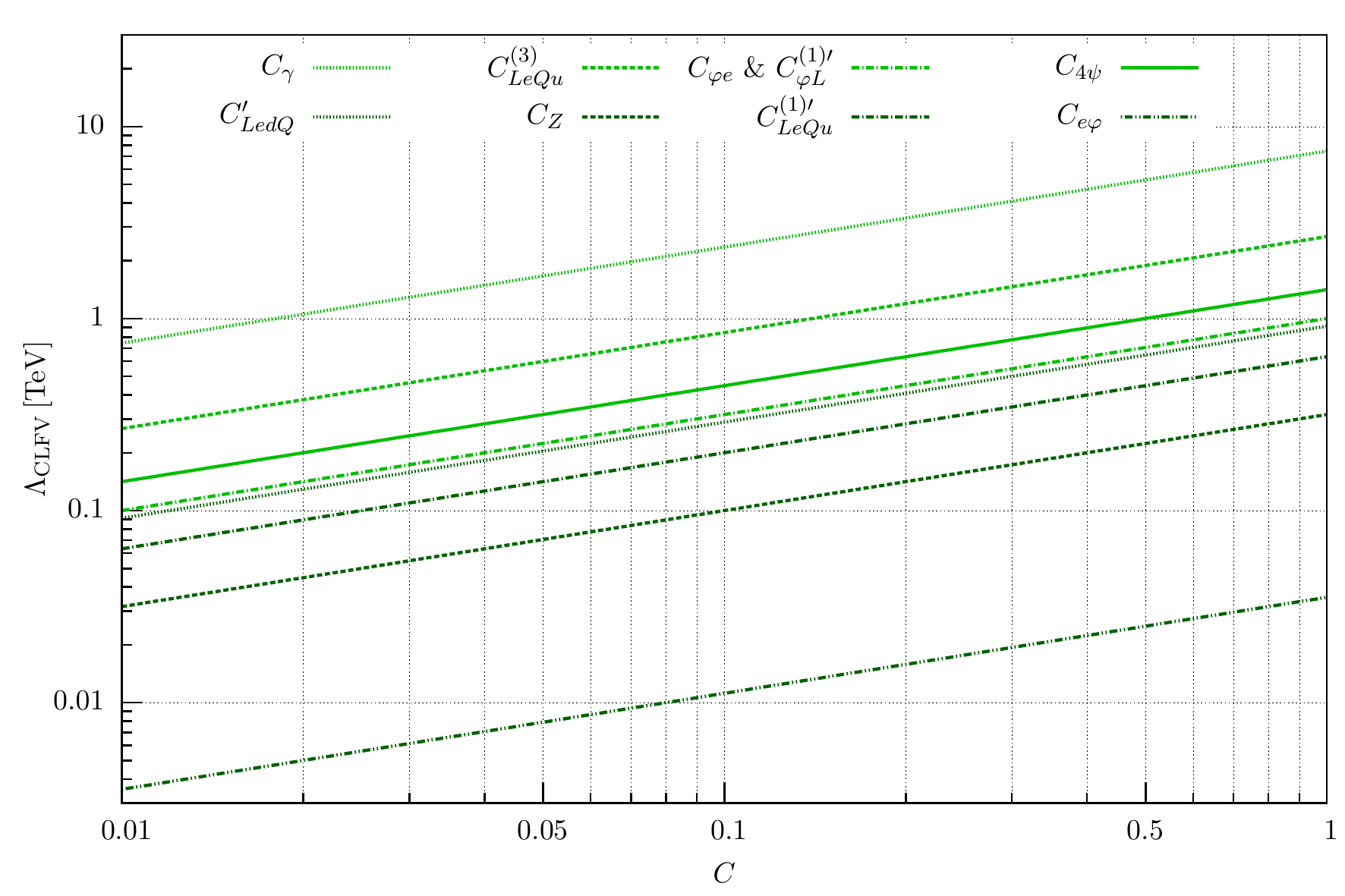}
\caption[]{\label{fig:7} Constraints on $\Lambda_{{\text{CLFV}}}$ with respect to the values of WCs from $\mu$-$\tau$ conversion in Fe(56,26), based on the expected sensitivity of the NA64 experiment, given at the 99.8\,\% confidence level.}
\end{center}
\end{figure}

The pattern of constraints is shown in Fig.~\ref{fig:7}. The weakest constraints are again for the Higgs WC $C_{e \varphi}$, followed by the `rotated' $C_{Z}$ one order of magnitude away. After this comes the scalar $C_{LeQu}^{(1) \, \prime}$, and the remaining WCs then follow the same pattern as for hadronic $\tau$ decays. Note that the order of $C_{Z}$ and $C_{LeQu}^{(1) \, \prime}$ has been inverted compared to the previous case. This is due to the fact (since we are considering FCNCs) that for $\ell$--$\tau$ conversion it is possible to have an outgoing charm quark after the effective interaction has taken place. Hence, due to the redefinition of Eq.~\eqref{eq:scalar_redefinition}, the related matrix element is enhanced by the mass of the charm quark.

Based on the expected sensitivity of the NA64 experiment, it would be possible to probe energy scales from $\approx$~30\,GeV for $C_{e \varphi}$ up to $\approx$~7.5\,TeV for $C_{\gamma}$, as it is shown in Table~\ref{tab:etau_mutau}. There, we give also the numbers for the $e$--$\tau$ case separately. As we said above, the numerical analysis is dominated mainly by the $\mu$--$\tau$ conversion, which means that the constraints obtained by considering the $e$--$\tau$ case only are much worse.
This implies that the quantity $R_{\mu\tau}$ related to the $\mu$--$\tau$ conversion in nuclei is the one more sensitive to new physics in this case.
\begin{table}[htb]
\capstart
\begin{center}
\renewcommand{\arraystretch}{1.5}
\begin{tabular}{||c|c|c|c|c|c|}
\cline{1-6}
 \multicolumn{6}{||c|}{Bounds on $\Lambda_{{\text{CLFV}}}$\,[TeV] }  \\
\hline
 WC & $e$--$\tau$ & $\mu$--$\tau$ & WC &  $e$--$\tau$ & $\mu$--$\tau$  \\
\hline
\hline
 $C_{LQ}^{(1)}$ & $\gtrsim 0.13 $  & $\gtrsim 1.7 $  &  $C_{LedQ}$ & $\gtrsim 0.06$  & $\gtrsim 0.9  $  \\
\hline
 $C_{LQ}^{(3)}$ & $\gtrsim 0.11 $   & $\gtrsim 1.5 $ &  $C_{LeQu}^{(1)}$ & $\gtrsim 0.05$ & $\gtrsim 0.6 $ \\
\hline
 $C_{eu}$ &  $\gtrsim 0.11 $  & $\gtrsim 1.4 $ & $C_{LeQu}^{(3)}$ & $\gtrsim 0.2 $  & $\gtrsim 2.7 $  \\
\hline
$C_{ed}$ & $\gtrsim 0.11 $   & $\gtrsim 1.4 $  & $C_{\varphi e}, C_{\varphi L}^{(1)}$ & $\gtrsim 0.08 $  & $\gtrsim 1 $  \\
\hline
$C_{Lu}$ & $\gtrsim 0.09 $  & $\gtrsim 1.1 $   & $C_{\gamma}$ & $\gtrsim 0.6$  & $\gtrsim 7.5 $ \\
\hline
$C_{Ld}$ & $\gtrsim 0.09 $  & $\gtrsim 1.2 $  & $C_{Z}$ & $\gtrsim 0.02$  & $\gtrsim 0.3 $ \\
\hline
$ C_{Qe} $ & $\gtrsim 0.1 $  & $\gtrsim 1.4 $  &  $C_{e \varphi }$ & $\gtrsim 0.003$  & $\gtrsim 0.04 $ \\
\hline
\end{tabular}
\end{center}
\vspace*{-0.5cm}
\caption{\label{tab:etau_mutau} Bounds on the new-physics energy scale mediating CLFV phenomena ($\Lambda_{{\text{CLFV}}}$), both for $e$--$\tau$ and $\mu$--$\tau$ conversion in Fe(56,26). Here, we consider $C \approx 1$. The results are based on the expected sensitivity of the NA64 experiment, given at the 99.8\,\% confidence level.}
\end{table}

Comparing the non-FCNC scenario with respect to the FCNC case for the $\ell$--$\tau$ conversion, the main differences are as follows. First, the incapability to disentangle, as for $\tau$ decays, the contributions from $C_{LQ}^{(3)}$ and $C_{\varphi L}^{(1) \, \prime}$ (due to their strong correlation) forces us again to consider the redefinition \eqref{eq:CLQ3noFCNC}: We can thus be sensitive to both operators independently only when FCNCs are included. Second, all 4-fermion WCs are less constrained, the largest difference occurring for $C_{LedQ}^{\, \prime}$ (two orders of magnitude weaker constraint regarding the ratio $C/  \Lambda_{{\text{CLFV}}}^{2}$) and $C_{LeQu}^{(1) \, \prime}$ (for which this analysis is actually not sensitive at all). The previous correlation among the latter and $C_{LeQu}^{(3)}$ is trivially lost, because of the redefinition \eqref{eq:scalar_redefinition} together with the vanishing up-quark mass, which we consider throughout the work. The constraints on the remaining WCs stay practically the same and the correlation matrix is rather diagonal.

\subsubsection{Combined analysis}

From the discussion and results of the previous sections, it is straightforward to see that the results of the combined analysis --- where we consider 28 hadronic $\tau$ decay channels and 4 cross sections of $\ell$--$\tau$ conversion in nuclei --- are dominated by the current Belle or expected Belle II limits.
We may try different ratios \eqref{eq:ratiomutau} for $\ell$--$\tau$ conversion in nuclei in order to see at which point these processes become competitive with the hadronic $\tau$ decays. We find that already with $R_{\ell \, \tau}\sim 10^{-13}$ the scalar $C_{LeQu}^{(1) \, \prime}$ receives a stronger constraint from the $\mu$--$\tau$ conversion due to its large sensitivity to the charm-quark mass (when considering FCNCs and the redefinition of Eq.~\eqref{eq:scalar_redefinition}). Nevertheless, it is not until we reach $R_{\ell \, \tau}\sim 10^{-15}$ that $\mu$--$\tau$ plays a significant role in the analysis: Most of the correlations among the WCs are then removed or diluted, which allows for slightly stronger constraints on the WCs compared to the ones Belle alone provides.
This implies that, in case that several LFV hadronic $\tau$ decays would be observed by Belle II, $\mu$--$\tau$ conversion in nuclei may have the last word in unveiling what $D=6$ operator(s) is/are behind the new-physics mechanism responsible for this manifestation of charged-lepton-flavour violation.

\section{Conclusions} \label{s:4} 

We have presented a model-independent numerical analysis of the SMEFT dimension-6 operators related to CLFV $\tau$-involved processes: We used the current Belle and the expected Belle II limits on hadronic $\tau$ decays, as well as a more exotic process, the $\ell$--$\tau$ conversion in nuclei, still not tried but feasible at the NA64 experiment at CERN. We have used HEPfit to perform the statistical part of the analysis.
\par
After BaBar and Belle experiments, tau decays started to be considered complementary to processes involving electrons and muons in the search for CLFV. That capability will be enforced even more with the expected results of the Belle II experiment. Here, we have studied the LFV decays of the tau lepton into hadrons by explaining in detail the procedure of hadronization. The wide range of possible final states provides 14 specific observables to include in our analysis. Our results show that the WC of the operator ${\cal O}_{\gamma} = c_W {\cal O}_{eB} - s_W {\cal O}_{eW}$ is the most constrained one providing, for $C_{\gamma} \approx 1$, a bound of $\Lambda_{{\text{CLFV}}} > 120 \, \mbox{TeV}$ (based on Belle data) or $\Lambda_{{\text{CLFV}}} > 330 \, \mbox{TeV}$ (foreseen by Belle II).
\par
In comparison with the $\mu$--$e$ conversion in nuclei widely studied in the bibliography, $\mu$--$\tau$ conversion has not attracted much attention, mainly due to the fact that its possible experimental determination has non-trivial complexities. However, in our opinion, the $\mu$--$\tau$ conversion is again a complementary tool in the endeavour of looking for CLFV since it obeys different dynamics compared to that of the $\mu$--$e$ conversion and, accordingly, it could provide an independent setting. In addition, its feasibility at NA64 at CERN should be strongly considered, although other fixed-target experiments (ILC, EIC, etc.) also offer good expectations. In our study, we have taken into account both the $e$--$\tau$ and $\mu$--$\tau$ conversion in Fe(56,26) and Pb(208,82) and we have concluded that $\mu$--$\tau$ conversion in Fe(56,26) imposes the strongest constraints.
In the latter case, the ${\cal O}_{\gamma}$ operator is again the most constrained, but giving only $\Lambda_{{\text{CLFV}}} > 7.5 \, \mbox{TeV}$ for $C_{\gamma} \approx 1$. We conclude that the current expected sensitivity, for instance, of the NA64 experiment cannot compete with Belle limits and it would need an improvement of at least two orders of magnitude in order to explore a slightly larger parameter space.
\par
The outcomes of our analyses show that the experimental results on hadronic tau decays expected by Belle II could improve significantly the search for LFV in such processes. Although the search for $\ell$--$\tau$ conversion in nuclei cannot compete, at present, with the information coming from tau decays, it could be used to unveil the relative weights of different dimension-6 operators. Finally, we have explicitly demonstrated the necessity to perform a marginalized numerical analysis of the parameters under consideration (see Fig.~\ref{fig:6} for explicit comparison): In this way, one can avoid naively deducing stronger estimates obtained when considering only one non-vanishing WC at a time.
\par 
This work sets a useful setting in the search of physics beyond the Standard Model --- namely charged-lepton-flavour violation --- through a systematic analysis within the framework of the Standard Model Effective Field Theory, taking into account all presently available information from experiments involving charged-lepton-flavour violation and the third family. Moreover, our tool will also be of use for analysing results from upcoming experiments like Belle II. 

\section*{Acknowledgements}
We thank Vincenzo Cirigliano for providing us with the results of the scalar matrix elements of Ref.~\cite{Celis:2013xja}.
We wish to thank Rusa Mandal for useful discussions on the topic of this paper, and Ana Pe\~nuelas and V\'{\i}ctor Miralles for their help with using $\texttt{HEPfit}$.
This work has been supported in part by Grants No. FPA2017-84445-P and SEV-2014-0398 (AEI/ERDF, EU) and by PROMETEO/2017/053 (GV).

\appendix

\renewcommand{\theequation}{\Alph{section}.\arabic{equation}}
\renewcommand{\thetable}{\Alph{section}.\arabic{table}}
\renewcommand{\thefigure}{\Alph{section}.\arabic{figure}}

\let\appsect\section
\renewcommand{\section}{
\setcounter{equation}{0}
\setcounter{table}{0}
\setcounter{figure}{0}
\appsect}

\section*{Appendices}

\section{Amplitudes generated by \texorpdfstring{\boldmath $D=6$}{D=6} operators} 
\label{app:1}

The $D=6$ operators \cite{Buchmuller:1985jz} of the SMEFT Lagrangian that are noninvariant under $\text{U}(1)_e \times \text{U}(1)_\mu \times \text{U}(1)_\tau$ rotations of the lepton fields while keeping the diagonal $\text{U}(1)_L$ symmetry (conserving the overall lepton number) generate CLFV processes. Within this setting, operators listed in Table~\ref{tab:1} generate tree level and also some particular 1-loop amplitudes to those processes. The latter have been considered by other authors and we also include them in our study. All the relevant amplitudes are collected in this appendix.

\subsection{\texorpdfstring{\boldmath The tree-level amplitudes for $\tau^- \rightarrow \ell^- \,  \overline{q} q$ and $\ell^- q \rightarrow \tau^-  q$, with $\ell = e,\mu$}{The tree-level amplitudes for tau- -> l- qbar q and l- q -> tau- q, with l = e,mu}}
\label{apps:1}
The amplitudes for these processes with light quarks in the final state, namely $q=u,d,s$, can be divided into four structures:
\begin{equation} \label{eq:fourst}
{\cal M}_{{\text{tree}}}\, = \, {\cal M}_{{\text{loc}}} + {\cal M}_{Z} + {\cal M}_{\gamma} + {\cal M}_H \, . 
\end{equation}
${\cal M}_{{\text{loc}}}$ corresponds to the contributions of four-fermion local operators (like those shown in Figs.~\ref{fig:1}(a) or \ref{fig:3}(a)) and consists of the following matrix elements stemming from the respective operators (here we show the matrix elements for the $\ell_2 q \rightarrow \ell_1 q$ process; for different configurations, see the end of this section):
\begin{align} \label{eq:tauamp1}
\mathcal{M}_{LQ}^{(1)} & = \frac{ C_{LQ}^{(1)}}{\Lambda_{{\text{CLFV}}}^{2}}[\bar{u}_{\ell_1}\gamma_{\mu}P_\text{L}u_{\ell_2}][(\bar{u}_u \gamma^{\mu}P_\text{L}u_u)+(\bar{u}_{d_x}\gamma^{\mu}P_\text{L}u_{d_x})]\, , \nonumber \\
\mathcal{M}_{LQ}^{(3)} & = \frac{ C_{LQ}^{(3)}}{\Lambda_{{\text{CLFV}}}^{2}}[\bar{u}_{\ell_1}\gamma_{\mu}P_\text{L}u_{\ell_2}][-(\bar{u}_u\gamma^{\mu}P_\text{L}u_u)+(\bar{u}_{d_x}\gamma^{\mu}P_\text{L}u_{d_x})]\, , \nonumber \\
\mathcal{M}_{eu} & = \frac{ C_{eu}}{\Lambda_{{\text{CLFV}}}^{2}}[\bar{u}_{\ell_1}\gamma_{\mu}P_\text{R}u_{\ell_2}][\bar{u}_u\gamma^{\mu}P_\text{R}u_u]\, , \nonumber \\
\mathcal{M}_{ed} & = \frac{ C_{ed}}{\Lambda_{{\text{CLFV}}}^{2}}[\bar{u}_{\ell_1}\gamma_{\mu}P_\text{R}u_{\ell_2}][\bar{u}_{d_x}\gamma^{\mu}P_\text{R}u_{d_x}]\, , \nonumber \\
\mathcal{M}_{Lu} & = \frac{ C_{Lu}}{\Lambda_{{\text{CLFV}}}^{2}}[\bar{u}_{\ell_1}\gamma_{\mu}P_\text{L}u_{\ell_2}][\bar{u}_u\gamma^{\mu}P_\text{R}u_u]\, , \nonumber\\
\mathcal{M}_{Ld} & = \frac{ C_{Ld}}{\Lambda_{{\text{CLFV}}}^{2}}[\bar{u}_{\ell_1}\gamma_{\mu}P_\text{L}u_{\ell_2}][\bar{u}_{d_x}\gamma^{\mu}P_\text{R}u_{d_x}]\, ,  \\
\mathcal{M}_{Qe} & = \frac{ C_{Qe}}{\Lambda_{{\text{CLFV}}}^{2}}[\bar{u}_{\ell_1}\gamma_{\mu}P_\text{R}u_{\ell_2}][(\bar{u}_u\gamma^{\mu}P_\text{L}u_u)+(\bar{u}_{d_x}\gamma^{\mu}P_\text{L}u_{d_x})]\, , \nonumber \\
\mathcal{M}_{LedQ}& = \frac{ C_{LedQ}}{\Lambda_{{\text{CLFV}}}^{2}}\Big\{ [\bar{u}_{\ell_1}P_\text{R}u_{\ell_2}][\bar{u}_{d_x}P_\text{L}u_{d_x}]+[\bar{u}_{\ell_1}P_\text{L}u_{\ell_2}][\bar{u}_{d_x}P_\text{R}u_{d_x}]\Big\}\, , \nonumber \\
\mathcal{M}_{LeQu}^{(1)} & = -\frac{ C_{LeQu}^{(1)}}{\Lambda_{{\text{CLFV}}}^{2}}\Big\{ [\bar{u}_{\ell_1}P_\text{R}u_{\ell_2}][\bar{u}_uP_\text{R}u_u]+[\bar{u}_{\ell_1}P_\text{L}u_{\ell_2}][\bar{u}_uP_\text{L}u_u]\Big\}\, , \nonumber \\
\mathcal{M}_{LeQu}^{(3)} & = -\frac{ C_{LeQu}^{(3)}}{\Lambda_{{\text{CLFV}}}^{2}}\Big\{ [\bar{u}_{\ell_1}\sigma_{\mu \nu}P_\text{R}u_{\ell_2}][\bar{u}_u\sigma^{\mu \nu}P_\text{R}u_u]+[\bar{u}_{\ell_1}\sigma_{\mu \nu}P_\text{L}u_{\ell_2}][\bar{u}_u\sigma^{\mu \nu}P_\text{L}u_u]\Big\}\, . \notag
\end{align}
${\cal M}_Z$ and ${\cal M}_{\gamma}$ encode the contributions mediated by $Z$ and $\gamma$ bosons, respectively, i.e.\ the processes $\tau \rightarrow \ell \, \{Z, \gamma\}$, followed by  $ \{Z, \gamma\} \rightarrow \overline{q} q$ (Figs.~\ref{fig:1}(b) and \ref{fig:3}(b)):
\begin{align}
\label{eq:tauamp2}
\mathcal{M}_{\varphi e} & = \frac{ C_{\varphi e}M_{Z}^{2} }{\Lambda_{{\text{CLFV}}}^{2}} [\bar{u}_{\ell_1}\gamma^{\mu}P_\text{R}u_{\ell_2}]\,\frac{(-g_{\mu \nu}+q_{\mu}q_{\nu}/M_{Z}^{2})}{q^{2}-M_{Z}^{2}}\Big\{[\bar{u}_u\gamma^{\nu}(v_{u}-a_{u}\gamma_{5})u_u]+[\bar{u}_{d_x}\gamma^{\nu}(v_{d}-a_{d}\gamma_{5})u_{d_x}]\Big\}\, , \nonumber \\
\mathcal{M}_{\varphi L}^{(1)} & = \frac{ C_{\varphi L}^{(1)}M_{Z}^{2}}{\Lambda_{{\text{CLFV}}}^{2}}  [\bar{u}_{\ell_1}\gamma^{\mu}P_\text{L}u_{\ell_2}]\,\frac{(-g_{\mu \nu}+q_{\mu}q_{\nu}/M_{Z}^{2})}{q^{2}-M_{Z}^{2}}\Big\{[\bar{u}_u\gamma^{\nu}(v_{u}-a_{u}\gamma_{5})u_u]+[\bar{u}_{d_x}\gamma^{\nu}(v_{d}-a_{d}\gamma_{5})u_{d_x}]\Big\}\, , \nonumber \\
\mathcal{M}_{\varphi L}^{(3)} & = \frac{ C_{\varphi L}^{(3)}M_{Z}^{2} }{\Lambda_{{\text{CLFV}}}^{2}}  [\bar{u}_{\ell_1}\gamma^{\mu}P_\text{L}u_{\ell_2}]\,\frac{(-g_{\mu \nu}+q_{\mu}q_{\nu}/M_{Z}^{2})}{q^{2}-M_{Z}^{2}}\Big\{[\bar{u}_u\gamma^{\nu}(v_{u}-a_{u}\gamma_{5})u_u]+[\bar{u}_{d_x}\gamma^{\nu}(v_{d}-a_{d}\gamma_{5})u_{d_x}]\Big\}\, , \nonumber \\
\mathcal{M}_{eB}^{(Z)} & = \frac{i \,C_{e B} s_\text{W} M_{Z} }{\sqrt{2} \Lambda_{{\text{CLFV}}}^{2}}  [\bar{u}_{\ell_1}\sigma^{\mu \nu}u_{\ell_2}] \,\frac{\Omega_{\mu \nu \alpha}}{q^{2}-M_{Z}^{2}}
 \Big\{[\bar{u}_u\gamma^{\alpha}(v_{u}-a_{u}\gamma_{5})u_u]+[\bar{u}_{d_x}\gamma^{\alpha}(v_{d}-a_{d}\gamma_{5})u_{d_x}]\Big\}\, , \nonumber \\
\mathcal{M}_{eW}^{(Z)} & = \frac{i \,C_{e W} c_\text{W} M_{Z} }{\sqrt{2} \Lambda_{{\text{CLFV}}}^{2}}  [\bar{u}_{\ell_1}\sigma^{\mu \nu}u_{\ell_2}] \,\frac{\Omega_{\mu \nu \alpha}}{q^{2}-M_{Z}^{2}}
 \Big\{[\bar{u}_u\gamma^{\alpha}(v_{u}-a_{u}\gamma_{5})u_u]+[\bar{u}_{d_x}\gamma^{\alpha}(v_{d}-a_{d}\gamma_{5})u_{d_x}]\Big\}\, , \nonumber \\
\mathcal{M}_{eB}^{(\gamma)} & = -\frac{i \,C_{e B} }{ \Lambda_{{\text{CLFV}}}^{2}} \sqrt{2} s_\text{W}c_\text{W}^{2} M_{Z} Q_{q} [\bar{u}_{\ell_1}\sigma^{\mu \nu}u_{\ell_2}]\, \frac{\Omega_{\mu \nu \alpha}}{q^{2}}
\Big\{[\bar{u}_u\gamma^{\alpha}u_u]+[\bar{u}_{d_x}\gamma^{\alpha}u_{d_x}]\Big\}\, , \nonumber \\
\mathcal{M}_{eW}^{(\gamma)} & =  \frac{i \,C_{e W} }{ \Lambda_{{\text{CLFV}}}^{2}} \sqrt{2} s_\text{W}^{2}c_\text{W} M_{Z} Q_{q} [\bar{u}_{\ell_1}\sigma^{\mu \nu}u_{\ell_2}]\, \frac{\Omega_{\mu \nu \alpha}}{q^{2}}
\Big\{[\bar{u}_u\gamma^{\alpha}u_u]+[\bar{u}_{d_x}\gamma^{\alpha}u_{d_x}]\Big\}\, . 
\end{align}
Note that we have separated the contribution of the operators ${\cal O}_{eB}$ and ${\cal O}_{eW}$ into those governed by the photon and the $Z$ boson.
In Eqs.~(\ref{eq:tauamp1}) and (\ref{eq:tauamp2}), $c_\text{W} = \cos \theta_\text{W}$ and $s_\text{W} = \sin \theta_\text{W}$ are the trigonometric functions of the weak mixing (also called Weinberg) angle and the index $x$ at the $d$ spinors refers to the first or second family, i.e.\ $d_x\in\{ d,s \}$: Note that in ${\cal M}_{{\text{loc}}}$ we assume that there are no FCNCs in the quark bilinears. Further, we also used $P_\text{L,R}=\frac12(1\mp\gamma_5)$,
\begin{equation} \label{eq:omegas}
\Omega_{\mu \nu \alpha} \,  =   \, q_{\mu} \, g_{\nu \alpha}  - q_{\nu} \, g_{\mu \alpha} \, ,
\end{equation}
and the SM weak couplings are
\begin{alignat}{3}\label{eq:avs}
v_{u} &= \frac{1}{2}-\frac{4}{3} s_W^2\,, \qquad \qquad &v_{d}&=-\frac{1}{2}+\frac{2}{3} s_W^2\, , \nonumber  \\
a_{u} &= \frac{1}{2}\,, &a_{d}&=-\frac{1}{2} \, .
\end{alignat}
Finally, ${\cal M}_H$ corresponds to the intermediate-Higgs contribution: $\tau \rightarrow \ell H$, $H \rightarrow \overline{q} q$ (Figs.~\ref{fig:1}(c) and \ref{fig:3}(c)). This 
is driven by ${\cal O}_{e \varphi}$ and by the Higgs--quark--quark coupling in ${\cal L}_\text{eff}$ that we have obtained in 
Eq.~(\ref{eq:lefff}). As we are considering $m_u=m_d=0$ and $m_s \neq 0$, we only have contribution to $\tau \rightarrow \ell \bar{s} s$ given by
\begin{equation} \label{eq:tauamp3}
{\cal M}_H \, = \,  \frac{C_{e\varphi}}{\Lambda_{{\text{CLFV}}}^{2}} \frac{7 v}{6 \sqrt {2}} \frac{1}{(q^2-M_H^2)} \, 
[ \bar{u}_{\ell_1} u_{\ell_2}] \, m_s \bar{u}_s u_s \, ,
\end{equation} 
with $v= \langle 0 |\phi|0\rangle = (\sqrt{2} G_F)^{-1/2}$, which correspond to diagrams (c) in Figs.~\ref{fig:1} and \ref{fig:3}. 
Our results in Eqs.~(\ref{eq:tauamp1}), (\ref{eq:tauamp2}) and (\ref{eq:tauamp3}) are relevant for both $\tau \rightarrow \ell \overline{q} q$ and $\ell q \rightarrow \tau q$, changing the $u$ to $v$ spinors appropriately and applying the following choices:
\begin{itemize}
\item For $\tau (k) \rightarrow \ell(k') \overline{q}(p') q(p)$, $\ell_1 = \ell$ and $\ell_2=\tau$, with $q=k-k'=p+p'$. 
\item For $\ell(k) q(p) \rightarrow \tau(k') q(p')$, $\ell_1 = \tau$ and $\ell_2 = \ell$, with $q = k-k'=p'-p$\,.
\end{itemize}

\subsection{\texorpdfstring{\boldmath The one-loop amplitude for $\tau^- \rightarrow \ell^- g g$, with $\ell = e,\mu$}{The one-loop amplitude for tau- -> l- g g, with l = e,mu}} \label{apps:2}

We consider the gluon-involved contribution to the $\tau^- \rightarrow \ell^- \bar PP$ process ($P$ stands for a pseudoscalar meson) upon hadronization of the two gluons from the $\tau^- \rightarrow \ell^- g g$ amplitude that, as pointed out in Ref.~\cite{Celis:2013xja}, can be represented via the dominant Higgs-exchange contribution shown in Fig.~\ref{fig:2}. The associated matrix element is generated by operator ${\cal O}_{e \varphi}$ from Table~\ref{tab:1}, together with the part of Eq.~(\ref{eq:lefff}) related to the energy--momentum tensor that arises, essentially, from the gluon final state through the trace anomaly of QCD, as explained
in Section~\ref{apps:4}. The matrix element for the hadronization into a $\overline{P} P$ pair of pseudoscalar mesons reads
\begin{equation} \label{eq:emuu}
\mathcal{M}_{\tau gg} \, = \, \frac{C_{e \varphi}}{\Lambda_{{\text{CLFV}}}^{2}} \, \frac{v}{3 \sqrt{2}} \, \frac{1}{(q^2-M_H^{2})}\,[ \bar{u}_{\ell} \,  u_{\tau} ] \, \theta_{P}(q^2)\, ,
\end{equation}
where $\theta_{P}(q^2)\equiv \langle \,\overline{P}(p') \,  P(p) \, | \, \theta_{\mu}^{\mu} \, | \, 0 \, \rangle$ and $q=p+p'$.

\subsection{\texorpdfstring{\boldmath The one-loop amplitude for $\ell^- g \rightarrow \tau^- g$, with $\ell = e,\mu$}{The one-loop amplitude for l- g -> tau- g, with l = e,mu}} \label{apps:3}

We include two one-loop diagrams contributing to the $\ell$--$\tau$ conversion process; see Fig.~\ref{fig:4}. The Higgs contribution was already considered in Ref.~\cite{Takeuchi:2017btl}, where it was claimed to represent the dominant Higgs amplitude to this process; in addition, we consider the $Z$ contribution. The peculiarities of the loop part of those diagrams are discussed in detail in Appendix~\ref{app:2}. The matrix element for the Higgs contribution to the $\ell g(p) \rightarrow \tau g(p')$ amplitude is
\begin{equation} \label{eq:mhgg1}
{\cal M}_{Hl} =  \, \frac{C_{e \varphi}}{\Lambda_{{\text{CLFV}}}^{2}} \, \frac{3 \, v}{\sqrt{2}} \,
[ \bar{u}_{\tau}  u_{\ell}   ] \, \frac{g_{Hgg}}{q^2 - M_H^2}  \left[ q^2 g_{\mu \nu} - 2 p'_{\mu} p_{\nu} \right] 
\varepsilon_a^{\mu}(p) \, \varepsilon_a^{*\nu}(p') \,,
\end{equation}
where 
\begin{equation} \label{eq:mhgg2}
g_{Hgg} = \sum_{Q=c,b,t} \, \frac{\alpha_\text{S}}{2 \, \pi} \, \frac{ m_q^2}{q^2} \left[ 1 - \frac{q^2}{2}
\left( 1  -  \frac{4 \, m_q^2}{q^2} \right) C_0(q^2,m_q^2) \right] .
\end{equation}
Above, the sum runs over the heavy quarks only (namely $Q=c,b,t$), $q=p'-p$ and $C_0(q^2,m_q^2)$ is given by Eq.~(\ref{eq:c000}). Notice that in Eq.~(\ref{eq:mhgg1}) there is a sum over the color ($a$) in the gluon polarizations.
The matrix element for the $Z$ contribution is
\begin{equation} \label{eq:mzgg1}
\begin{split}
{\cal M}_{Zl} \, & = \,  \left( \, \frac{C_{\varphi e}}{\Lambda_{{\text{CLFV}}}^{2}}  \left[ \overline{u}_{\tau} \gamma_{\sigma} P_\text{R} \, u_{\ell}
\right] \, + \, \left( \frac{C_{\varphi L}^{(1)}}{\Lambda_{{\text{CLFV}}}^{2}} +  \frac{C_{\varphi L}^{(3)}} {\Lambda_{{\text{CLFV}}}^{2}} \right) \left[ \overline{u}_{\tau} \gamma_{\sigma} P_\text{L} \, u_{\ell} \right] \right) \\
& \times \frac{\alpha_\text{S}}{\pi} \,  \frac{q^{\sigma}}{q^2} \, \varepsilon_{\alpha \beta \mu \nu} \, q^{\alpha} \, (p+p')^{\beta} \, \varepsilon_a^{\mu}(p) \,  \varepsilon_a^{*\nu}(p') \, \sum_q \, I^3_{\text{w},q} \, m_q^2 \, C_0(q^2,m_q^2) \, ,
\end{split}
\end{equation}
where the sum now runs over all quark flavours and $I^3_{\text{w},q}=\pm\frac12$ is the quark weak isospin (eigenvalue of the $\sigma_3/2$ generator) same for each quark family. For completeness, we use the $\epsilon_{0123}=-1$ convention for the Levi-Civita tensor, even though the phase of the last equation has no physical effect on the resulting cross section.

\section{Triangle diagrams}
\label{app:2}

The computation of diagrams involving gluons in Figs.~\ref{fig:2} and \ref{fig:4} imply several features that we intend to explain in this appendix, and are due to the trace anomaly of QCD \cite{Crewther:1972kn,Chanowitz:1972vd,Collins:1976yq,Shifman:1978zn} and the Landau--Yang theorem \cite{Landau:1948kw,Yang:1950rg}.

\subsection{SVV Green function} \label{apps:4}

The $Hgg$ vertex at one loop contributes both to the $H \rightarrow gg$ decay in Fig.~\ref{fig:2} and the $g H \rightarrow g$ dynamical vertex
in diagram $(a)$ of Fig.~\ref{fig:4}. In the latter case, it is a part of the computation of the $\mu$--$\tau$ conversion in nuclei, and the gluon hadronization at $E \gg m_{\tau}$ will then be carried out through the nucleon PDFs. We are interested here in the hadronization mechanism that involves `$ g g \rightarrow \text{hadrons}$' in the contribution to tau decays in Fig.~\ref{fig:2}, in particular, into a pseudoscalar pair. 
\par
\begin{figure}[!t]
\capstart
\begin{center}
\includegraphics[scale=0.75]{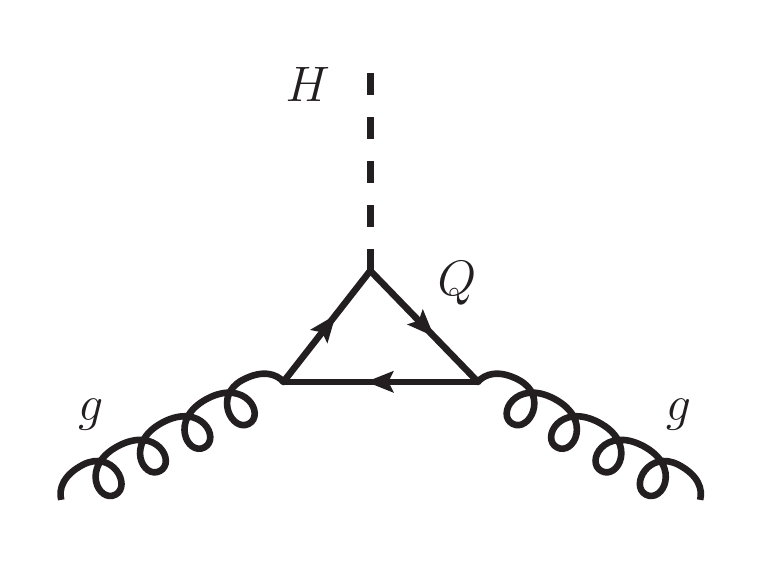}
\caption[]{\label{fig:b1} Triangle diagram contributing to $H \rightarrow g g$. $Q$ indicates a heavy quark, $Q=c,b,t$. For the final contribution one needs to add an analogous (cross) contribution with the gluons interchanged (or, equivalently, reversed quark momenta in the loop).}
\end{center}
\end{figure}

 
The Higgs interaction with quarks is given, after spontaneous breaking of the electroweak symmetry, by the Standard Model Lagrangian
\begin{equation} \label{eq:smhqq}
{\cal L} = - \sum_q \, \frac{m_q}{v} \, h \, \bar{\psi}_q \psi_q \, , 
\end{equation}
where $v = (\sqrt{2} G_F )^{-1/2} \approx 246 \, \mbox{GeV}$ and the sum extends on light $q_\ell = u,d,s$ and heavy quarks $Q=c,b,t$.
With the quark-gluon vertices of the QCD Lagrangian, we can now compute the diagram in Fig.~\ref{fig:b1} for an off-shell Higgs field with
$q^2 \lesssim m_{\tau}^2$ by including only the (dominant) heavy quarks $Q$ in the loop. 
For large quark masses $m_Q \gg m_{\tau}$ we have a low-energy local effective Lagrangian independent of the heavy quark mass
\cite{Donoghue:1990xh}:
\begin{equation} \label{eq:eftQ}
{\cal L}_\text{eff} = \frac{\alpha_\text{S} \, n_Q}{12 \pi \, v} \, h \, G_{\mu \nu}^a G^{\mu \nu \, a} - \sum_{q=u,d,s} \frac{m_q}{v} \, h \, \bar{\psi}_q \psi_q \, ,
\end{equation}
where $n_Q$ is the number of heavy quarks in the loop and $G_{\mu \nu}^a$ is the strength field tensor of the QCD gluon. In order to get the matrix element of the gluon operator in ${\cal L}_\text{eff}$ for a two-pseudoscalar-mesons final state, we use the relation of that operator with the trace of the energy--momentum tensor of QCD. The latter has an anomaly and reads \cite{Crewther:1972kn,Chanowitz:1972vd,Collins:1976yq,Shifman:1978zn}
\begin{equation} \label{eq:tracean}
\theta_{\mu}^{\mu} = \frac{\beta(\alpha_\text{S})}{4 \, \alpha_\text{S}} \, G_{\mu \nu}^a G^{\mu \nu \, a} + 
\sum_q \, m_q \, (1+ \gamma_{m_q}) \, \bar{\psi}_q  \psi_q \, ,
\end{equation}
where $q=u,d,c,s,b,t$. Above,
\begin{equation} \label{eq:betaa}
\gamma_{m_q}(\mu) =  \mu\,\frac{\diff \ln m_q}{\diff\mu} \; ,\qquad 
\beta(\alpha_\text{S}) = - \left(9 - \frac{2}{3} \, n_Q\right) \frac{\alpha_\text{S}^2}{2 \pi} + {\cal O}(\alpha_\text{S}^3) \, . 
\end{equation}
Note that $\theta_{\mu}^{\mu}$ is a scale-independent composite operator \cite{Ji:1995sv}. 
\par 
The gluon part of the effective action in Eq.~(\ref{eq:eftQ}) arises from the contribution of the heavy quarks $Q$ in the loop shown in 
Fig.~\ref{fig:b1}, using for the Higgs--quark--quark vertex the interaction term from Eq.~(\ref{eq:smhqq}). Hence, neglecting the higher-order $\gamma_m$ 
terms in $\theta_{\mu}^{\mu}$, we can integrate out the heavy quarks obtaining
\begin{equation} \label{eq:thetas}
\theta_{\mu}^{\mu} = \frac{\beta(\alpha_\text{S})}{4 \, \alpha_\text{S}} \, G_{\mu \nu}^a G^{\mu \nu \, a} -
\frac{\alpha_\text{S}}{12 \pi} \, n_Q G_{\mu \nu}^a G^{\mu \nu \, a}+ 
\sum_{q=u,d,s} m_q \overline{\psi}_q  \psi_q \, ,
\end{equation}
and the $n_Q$ dependence cancels. 
Finally, we can rewrite our effective action as
\begin{equation} \label{eq:lefff}
{\cal L}_\text{eff}  =  -  \frac{h}{9 \, v} \left( 2 \, \theta_{\mu}^{\mu} + 7 \sum_{q=u,d,s} m_q \overline{\psi}_q \psi_q \right). 
\end{equation}

\subsection{AVV Green function} \label{apps:5}

\begin{figure}[!ht]
\capstart
\begin{center}
\includegraphics[scale=0.75]{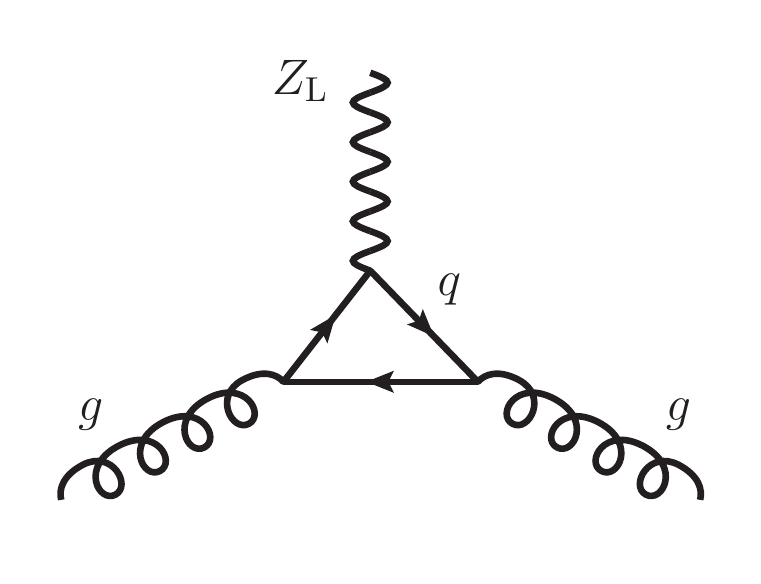}
\caption[]{\label{fig:b2} Triangle diagram contributing to $Z_\text{L} \rightarrow g g$. Here, $q$ is a generic quark, $q=u,d,c,s,t,b$. The final contribution comes from adding to this diagram the analogous one with the quark momenta in the loop reversed.  }
\end{center}
\end{figure}

In order to compute the diagram $(b)$ in Fig.~\ref{fig:4} contributing to the $\mu$--$\tau$ conversion in nuclei, we need to consider the subdiagram in Fig.~\ref{fig:b2}. 
Because the $V-A$ structure of the $Z$--quark--quark vertex, it contributes both to the $VVV$ and $AVV$ Green functions. The Landau--Yang theorem  \cite{Landau:1948kw,Yang:1950rg} states that a massive vector ($J=1$) cannot decay into two on-shell identical massless vectors; 
hence, we cannot have $Z \rightarrow \gamma \gamma$ or $Z \rightarrow g g$ (as the gluons have identical colour in this process).
For any off-shell vector, the theorem does not apply. In our case we notice that the $VVV$ contribution vanishes identically and independently of the on- or off-shellness of the $Z$ boson: surely, a consequence of Furry's theorem. For the $AVV$ component, we observe that the two-gluon system catches the scalar part ($J=0$) of the off-shell $Z$ boson, i.e.\ its longitudinal component $Z_\text{L}$ that, accordingly, does not give a pole. Hence, the only non-vanishing contribution is given by $Z_\text{L} \rightarrow g g$. 
\par
We recall that the $AVV$ Green function carries the axial (Adler--Bell--Jackiw) anomaly \cite{Bell:1969ts,Horejsi:1992tw,Horejsi:1992tx}. Using the 
diagram in Fig.~\ref{fig:b2} to compute $T_{\alpha \mu \nu}(p,p') \equiv i\mathcal{M}\big( g_{\mu}(p) \, Z_{\alpha}(q) \to g_{\nu}(p') \big)$ for on-shell gluons we obtain
\begin{equation} \label{eq:tamunu}
T_{\alpha \mu \nu} =  \frac{\alpha_\text{S}}{2 \pi} \, \frac{e}{\sin 2 \theta_W } \,   \frac{q_{\alpha}}{q^2} \, 
\varepsilon_{\mu \nu \kappa \lambda} \,  q^{\kappa} \, (p+p')^{\lambda} \, \sum_{q} I^3_{\text{w},q} \,[ 1 \,+ 2 m_q^2 C_0(q^2,m_q^2)]  ,
\end{equation}
where the sum extends to all quarks, $I^3_{\text{w},q}$ is the weak isospin of a quark of flavour $q$ and $C_0(q^2,m_q^2)$ is the Passarino--Veltman scalar triangle function \cite{Passarino:1978jh}
\begin{equation} \label{eq:c000}
C_0(q^2,m_q^2)
\equiv C_0 (0,0,q^2,m_q^2,m_q^2,m_q^2) = \frac{1}{2 q^2} \, \ln^2\bigg[ \frac{\sqrt{1-\tau}-1}{\sqrt{1-\tau} + 1} +  i \varepsilon \bigg],
\end{equation}
with $\tau \equiv 4 m_q^2/q^2$. The first term in $[1+2m_q^2 C_0(q^2,m_q^2)]$ in Eq.~(\ref{eq:tamunu}) is the contribution of the axial anomaly. Note, however,
that as this term is multiplied by the $I^3_{\text{w},q}=\pm\frac12$ factor, the anomalous term cancels when adding the two members of each family of quarks, which results in the anomalous-free amplitude, as is desirable.
In addition, and as commented above, $T_{\alpha \mu \nu} \propto q_{\alpha}$, where $q_{\alpha}$ is the 4-momentum of the $Z$ boson. Accordingly, when contracted with the gauge-boson propagator, the pole in the latter cancels, as corresponds to the fact that the longitudinal component (scalar part) of a spin-1 boson is the only one contributing here.

\section{Resonance Chiral Theory}
\label{app:3}

R$\chi$T is a phenomenological framework based on the dynamics driven by effective field theories and the chiral symmetry of QCD  \cite{Ecker:1988te,Ecker:1989yg,Cirigliano:2006hb}. It extends the model-independent $\chi$PT scheme by adding to the theory the fields of octets of hadron resonances that:
\begin{itemize}
\item i) lie in the $E \lesssim 2 \, \mbox{GeV}$ region although, in practice, only the lightest multiplets are included;
\item ii) cannot be generated by loops of the pseudoscalar mesons, i.e.\ remain in the limit of large number of colours ($N_\text{C} \rightarrow \infty$).
\end{itemize}
A summary of its features is given in Ref.~\cite{Portoles:2010yt}. Here, we follow the notation and definitions of Ref.~\cite{Cirigliano:2006hb,Cata:2007ns} with $N_\text{F}=3$ flavours. 
\par 
The basic structure of the R$\chi$T Lagrangian is generically given by chiral and $\text{SU}(3)$ symmetry as:
\begin{equation} \label{eq:rchtl}
\begin{split}
{\cal L}_\text{R$\chi$T} \, &= \, {\cal L}_\text{GB} \, + \, \sum_i {\cal L}_\text{kin}(R_i) \, + \, \sum_i \, \left\langle \,  R_i \, \left( \alpha_a \, \chi_2^a + \beta_a \, \chi_4^a + \dots\right) \, \right\rangle  \\
& + \,\sum_{i,j} \, \left\langle \, R_i \, R_j \,( \gamma_a \, \chi_2^a +\dots) \, \right\rangle \, + \dots \, ,
\end{split}
\end{equation}
where
\begin{equation} \label{eq:gbr}
{\cal L}_\text{GB} \, = \, \frac{F^2}{4} \, \langle \, u_{\mu} \, u^{\mu} \, + \,  \chi_+ \, \rangle \, + \, \sum_{i=1}^{12} \, L_i^\text{SD} \, O_i^{(4)} \, + \, \sum_{j=1}^{94} \, C_j^\text{SD} \, O_j^{(6)} \, + \dots
\end{equation}
is the chiral Lagrangian involving only the octet of pseudoscalar Goldstone fields and the external auxiliary fields. Here, $F$ is the decay constant of the pion. The first term in Eq.~\eqref{eq:gbr} is the ${\cal O}(p^2)$ Lagrangian of $\chi$PT, while the higher-order operators have the same structure as those of the chiral Lagrangian, but with different couplings. The couplings in Eq.~\eqref{eq:gbr} (of ${\cal O}(p^n)$ for $n>2$) labelled `SD' do not have contributions that could be obtained upon integration of the resonance fields in ${\cal L}_\text{R$\chi$T}$ --- because the latter are explicit in the theory --- and are, a priori, unknown. In this way, double counting is avoided. 
In the above equations, $\langle A \rangle$ indicates the $\text{SU}(3)$ trace of the matrix $A$, $R_i$ are the
hadron resonance fields, and $\chi_i^a$ are chiral operators of order ${\cal O}(p^i)$ that transform as the resonance fields under $\text{SU}(3)$ and contain again only the pseudoscalar Goldstone and the external fields. In addition, note that, for instance, $\alpha_a \,\chi_2^a \, = \, \alpha_1 \, \chi_2^1 \, + \, \alpha_2 \, \chi_2^2 \, + \dots$ and the sum extends to all possible operators that the symmetry allows.  The couplings $\alpha_a$, $\beta_a$, $\dots$ are not given by the symmetry alone. We will only consider the simplest (leading ${\cal O}(p^2)$) structure $\langle \, R_i \, \chi_2 \, \rangle$ in our resonance Lagrangian. 
\par 
A priori, the R$\chi$T Lagrangian does not know anything about the short-distance structure of QCD. Hence, it must always be implemented, as much as possible, with short-distance constraints \cite{Ecker:1989yg,Knecht:2001xc,Pich:2002xy,Bijnens:2003rc} that will provide the information on the (a priori) unknown couplings, namely $L_i^\text{SD}$, $C_i^\text{SD}$, $\alpha_a$, $\beta_a$, $\dots$ and so on.
\par 
As explained in the main text, we intend to use the R$\chi$T framework to provide the hadronization of the quark bilinears of our perturbative results. We are interested in the $\tau$ lepton decaying into one or two pseudoscalars, or a vector resonance. Hence, we only need to include the resonances that can contribute to the dynamics of those final states. These will be $\text{SU}(3)$ octets of scalars, pseudoscalars, vectors and spin-2 tensors.%
\footnote{We use the antisymmetric representation for spin-1 fields \cite{Ecker:1988te}. In this realization, there is no mixing of axial-vector resonances and pseudoscalar fields.}
It is important to point out that we should include only those that remain in the $N_\text{C} \rightarrow \infty$ limit. As it is well known, the identification of those multiplets with the experimentally determined resonances in the PDG \cite{Zyla:2020zbs} is clear in all cases except for the scalars; see Ref.~\cite{Dai:2019lmj} and references therein.
\par 
Our R$\chi$T Lagrangian is, finally,
\begin{equation} \label{eq:ourrl}
{\cal L}_\text{R$\chi$T} \, = \, {\cal L}_\text{GB} \, + \, {\cal L}_\text{S} \, + \, {\cal L}_\text{P} \, +\, {\cal L}_\text{V} \, + \, {\cal L}_\text{T} \, ,
\end{equation}
where in our case
\begin{equation} \label{eq:ourgb}
 {\cal L}_\text{GB} \,  = \,  \frac{F^2}{4} \, \langle \, u_{\mu} \, u^{\mu} \, + \,  \chi_+ \, \rangle \, + \, \sum_{i=1}^{12} \, L_i^\text{SD} \, O_i^{(4)}  \, 
 + \, \Lambda_1^\text{SD} \, \langle \, t_+^{\mu \nu} \, f_{ + \, \mu \nu} \, \rangle \, - \, i \, \Lambda_2^\text{SD} \, \langle \, t_+^{\mu \nu} \, u_{\mu} \, u_{\nu} \, \rangle \, . 
\end{equation}
Above, $O_i^{(4)}$ are the operators of the Gasser and Leutwyler Lagrangian \cite{Gasser:1984gg} and the tensor-involved operators have been recalled from Ref.~\cite{Cata:2007ns}. The resonance terms \cite{Ecker:1988te,Ecker:2007us,Mateu:2007tr} are (including their kinetic terms)
\begin{equation} \label{eq:resoo}
\begin{split}
{\cal L}_\text{S} & = \frac{1}{2} \, \langle \, \nabla^{\mu} \, S \, \nabla_{\mu} \, S \, - \, M_S^2 \, S^2 \, \rangle \, + \, \langle \, S \, \chi_S \, \rangle \, , \\
{\cal L}_\text{P} & = \frac{1}{2} \, \langle \, \nabla^{\mu} \, P \, \nabla_{\mu} \, P \, - \, M_P^2 \, P^2 \, \rangle \, + \, \langle \, P \, \chi_P \, \rangle \, , \\
{\cal L}_\text{V} & = - \frac{1}{2} \, \langle \, \nabla^{\lambda} \, V_{\lambda \mu} \, \nabla_{\nu} \, V^{\nu \mu} \,- \, \frac{M_V^2}{2} \, V_{\mu \nu} \, V^{\mu \nu} \, \rangle \, + \, \langle \, V_{\mu \nu} \, \chi_V^{\mu \nu} \, \rangle \, , \\
{\cal L}_\text{T} & = - \frac{1}{2} \, \langle \, T_{\mu \nu} \, D_T^{\mu \nu , \rho \sigma} \, T_{\rho \sigma} \, \rangle  \, + \, \langle \, T_{\mu \nu} \, \chi_T^{\mu \nu} \, \rangle \, .
\end{split}
\end{equation}
Here, the interaction is provided by the following currents:
\begin{equation} \label{eq:currin}
\begin{split}
\chi_S & = c_d \, u_{\mu} \, u^{\mu} \, + \, c_m \, \chi_{+} \, , \\
\chi_P & = i \, d_m \, \chi_- \, , \\
\chi_V^{\mu \nu} & = \frac{F_V}{2 \sqrt{2}} \, f_{+}^{\mu \nu} \, + \, i \, \frac{G_V}{\sqrt{2}} \, u^{\mu} \, u^{\nu} \, + \, T_V \, t_+^{\mu \nu} \,, \\
\chi_T^{\mu \nu} & = g_T \, \{ \, u^{\mu}, u^{\nu} \, \}\, + \, \beta \, g^{\mu \nu} \, u^{\alpha} \, u_{\alpha} \, + \, \gamma \, g^{\mu \nu} \, \chi_+ \, .
\end{split}
\end{equation}
Apart from the $L_i^\text{SD}$ couplings in ${\cal L}_\text{GB}$, we have several couplings involving the resonances, namely $c_d$, $c_m$, $d_m$, $F_V$, $G_V$, $T_V$, $g_T$, $\beta$ and $\gamma$. Some of these couplings could be fixed from the phenomenology: For instance, $F_V$ could be determined from $\rho \rightarrow e^+ e^-$. However, the real strength of R$\chi$T resides in obtaining as much information from the QCD structure as possible via the implementation of short-distance constraints. 
\par 
Most of this work has already been done \cite{Pich:2002xy,Cirigliano:2006hb,Ecker:2007us}. We get
\begin{alignat}{3} \label{eq:sdc}
F_V \, G_V \,& =\, F^2 \, , \qquad \qquad\qquad & 4 \, c_d \, c_m \, = \, F^2 \, ,  \nonumber \\
\beta \, & = -g_T \, , \qquad \qquad & 8 \, (c_m^2 - d_m^2) \, = \, F^2 \, .
\end{alignat}
The interacting term for the pseudoscalar resonance (proportional to $d_m$) from Eqs.~(\ref{eq:resoo}) and \eqref{eq:currin} produces a mixing between the resonance and the pseudoscalar Goldstone bosons. We can avoid this mixing through a redefinition of the pseudoscalar resonance: $P \rightarrow P + \, i \, \frac{d_m}{M_P^2} \, \chi_-$. The $d_m$ term in $\chi_P$ from Eq.~(\ref{eq:currin}) is cancelled, but the local contribution that we have to consider is generated:%
\footnote{The $L_{12}$ $\chi$PT coupling corresponds to $H_2$ in the Gasser and Leutwyler Lagrangian \cite{Gasser:1984gg}.}
\begin{equation} \label{eq:l8h2}
{\cal L} \, = \, L_8^P \, O^{(4)}_8 \, + \, L_{12}^P \, O^{(4)}_{12} \; = \; - \, \frac{d_m^2}{2 M_P^2} \, \left( \, O^{(4)}_8 \, - \, 2 \, O^{(4)}_{12} \right) \; = \;
\, - \,  \frac{d_m^2}{2 M_P^2} \,\langle \chi_-^2 \rangle \, .
\end{equation}
We notice that our redefinition of the pseudoscalar resonance field implies that it is being integrated out from our Lagrangian. Accordingly, we recover the pseudoscalar resonance contributions to $L_8$ and $L_{12}$ from Ref.~\cite{Ecker:1988te}.
\par 
We have noticed that in the hadronization of the scalar current the contribution of the spin-2 resonances spoils its high-energy behaviour \cite{Lepage:1980fj}. To solve this problem, we fix
\begin{equation} \label{eq:l5sd}
L_5^\text{SD} \, = \, - \, \frac{2}{3} \, \frac{\beta \, \gamma}{M_T^2} \, ,
\end{equation}
where $M_T$ is the octet mass of the spin-2 resonances. In addition,
\begin{equation} \label{eq:lisd}
L_i^\text{SD} \, = \, 0 \, , \qquad \, i \, \neq \, 5 \, .
\end{equation}
\par 
In the hadronization of the tensor current (\ref{eq:finalh}), the $\Lambda_2^\text{SD}$ coupling from Eq.~(\ref{eq:ourgb}) appears. There is no a priori knowledge on the short-distance component of this coupling. However, the same hadronized tensor current in Eq.~(\ref{eq:finalh}) gives us an answer. Requiring the appropriate high energy behaviour of this current \cite{Lepage:1980fj} we obtain
\begin{equation} \label{eq:lambda2sd}
\Lambda_2^\text{SD} \, = \, 0 \, .
\end{equation}
In fact, we can also determine the vector-resonance contribution to the $\chi$PT coupling $\Lambda_2$, namely $\Lambda_2^\text{R}$, upon its integration between the $G_V$ and $T_V$ terms from Eq.~\eqref{eq:currin}. We get
\begin{equation} \label{eq:lambda2res}
\Lambda_2^\text{R} \, = \, \sqrt{2} \, \frac{T_V \, G_V}{M_V^2} \, . 
\end{equation}
The value of $\Lambda_2$ has been determined in Ref.~\cite{Gonzalez-Solis:2019lze}: $\Lambda_2 = (11.1 \pm 0.4) \, \mbox{MeV}$. If we assume resonance saturation, we can, in fact, use this relation to get a value for the coupling $T_V$:%
\footnote{For numerical inputs, we use the values from Appendix~\ref{app:7}.}
\begin{equation} \label{eq:tvvalue}
T_V \, \approx \, 0.1147 \, \mbox{GeV}^2 \, . 
\end{equation}
This gives $f_V^\perp \approx 0.148 \, \mbox{GeV}$ to be compared with the result from Ref.~\cite{Mateu:2007tr}, 
$f_V^\perp (1 \, \mbox{GeV}) \, = \, 0.165 \pm 0.031 \, \mbox{GeV}$.


\section{\texorpdfstring{\boldmath $\Omega$}{Omega} coefficients in Eq.~(\ref{eq:finalh})} 
\label{app:4}

In this appendix, we collect the values for the $\Omega$ coefficients used in Eq.~(\ref{eq:finalh}) for every final- and intermediate-state contribution. We define $\sin\theta_{P}\equiv s_{P}$, $\cos\theta_{P}\equiv c_{P}$, $\sin\theta_{V}\equiv s_{V}$, $\cos\theta_{V}\equiv c_{V}$, with $\sin 2\theta_{P} \equiv s_{2P}$ and so on, and $m_{\pi/K}\equiv m_{\pi}/m_{K}$. Furthermore, $\theta_{P}$ has been defined in Eq.~\eqref{eq:mixinga} and $\theta_{V}$ in Eq.~\eqref{eq:mixingv}.
\begin{table}[htb]
\begin{center}
\renewcommand{\arraystretch}{1.5}
\begin{tabular}{c||c|c|c|c|c|}
\cline{2-6}
& \multicolumn{5}{c|}{$\Omega_P^{(1)}(ij)$}  \\
\hline
$P$ & $uu$ & $dd$ & $ss$ & $ds$ & $sd$ \\
\hline
\hline
$\pi^0$ & $\frac{1}{2}$ & -$\frac{1}{2}$ &  0 & 0  &  0  \\
\hline
$K^0$ & 0  & 0 & 0 & $\frac{1}{\sqrt{2}}$ & 0 \\
\hline
$\overline{K}^0$ & 0 & 0 & 0 &  0 & $\frac{1}{\sqrt{2}}$ \\
\hline
$\eta$ &  $\frac{c_{P}-\sqrt{2}s_{P}}{2 \sqrt{3}}$  & $\frac{c_{P}-\sqrt{2} s_{P}}{2 \sqrt{3}}$  & $-\frac{2 \sqrt{3}c_{P} +\sqrt{6}s_{P}}{6}$ & 0 & 0 \\
\hline
$\eta'$ & $\frac{ s_{P}+\sqrt{2}c_{P}}{2 \sqrt{3}} $ &  $\frac{ s_{P}+\sqrt{2}c_{P}}{2 \sqrt{3}} $   & $\frac{\sqrt{6} c_{P} -2 \sqrt{3} s_{P} }{6}$ & 0 & 0 \\
\hline
\end{tabular}
\end{center}
\vspace*{-0.5cm}
\caption{\label{tab:omega1a} Factor $\Omega_P^{(1)}(ij)$.
}
\end{table}
\begin{table}[htb]
\begin{center}
\renewcommand{\arraystretch}{1.5}
\begin{tabular}{c||c|c|c|c|c|}
\cline{2-6}
& \multicolumn{5}{c|}{$\Omega_P^{(2)}(ij)$}  \\
\hline
$P$ & $uu$ & $dd$ & $ss$ & $ds$ & $sd$ \\
\hline
\hline
$\pi^0$&  $-4  m_{\pi/K}^{2}$ & $4  m_{\pi/K}^{2}$ & 0  & 0 & 0 \\
\hline
$K^0$ & 0 & 0 & 0 & $-4\sqrt{2}$ & 0 \\
\hline
$\overline{K}^0$ & 0 & 0 & 0 &  0 & $-4\sqrt{2}$\\
\hline
$\eta$ & $-\frac{4 m_{\pi/K}^{2} \left(c_{P}-\sqrt{2} s_{P}\right)}{ \sqrt{3}}$ & $-\frac{4 m_{\pi/K}^{2} \left(c_{P}-\sqrt{2} s_{P}\right)}{ \sqrt{3}}$ &  $ \frac{4 (2-m_{\pi/K}^{2}) \left(2c_{P}+\sqrt{2} s_{P}\right)}{ \sqrt{3}}$ & 0 & 0 \\
\hline
$\eta'$ &  $-\frac{ 4 m_{\pi/K}^{2} \left(s_{P}+c_{P} \sqrt{2}\right)}{ \sqrt{3}}$  &  $-\frac{ 4 m_{\pi/K}^{2} \left(s_{P}+c_{P} \sqrt{2}\right)}{ \sqrt{3}}$  & $\frac{4 (2-m_{\pi/K}^{2}) \left(2 s_{P}-\sqrt{2}c_{P}\right)}{ \sqrt{3}}$ & 0 & 0\\
\hline
\end{tabular}
\end{center}
\vspace*{-0.5cm}
\caption{\label{tab:omega2a} Factor $\Omega_P^{(2)}(ij)$.
}
\end{table}
\begin{table}[htb]
\begin{center}
\renewcommand{\arraystretch}{1.5}
\begin{tabular}{c||c|c|c|c|c|}
\cline{2-6}
& \multicolumn{5}{c|}{$\Omega_{A}^{(1)}(ij)$}  \\
\hline
$P$ & $uu$ & $dd$ & $ss$ & $ds$ & $sd$ \\
\hline
\hline
$\pi^0$ & $\frac{1}{2}$ &  -$\frac{1}{2}$& 0 & 0 & 0 \\
\hline
$K^0$ & 0 & 0 & 0 & $\frac{1}{\sqrt{2}}$ & 0\\
\hline
$\overline{K}^0$ & 0 & 0 & 0 & 0 & $\frac{1}{\sqrt{2}}$\\
\hline
$\eta$ & $\frac{1}{6} \left(\sqrt{3} c_{P}-\sqrt{6} s_{P}\right)$ &  $\frac{1}{6} \left(\sqrt{3} c_{P}-\sqrt{6} s_{P}\right)$  & $-\frac{\sqrt{2} c_{P}+s_{P}}{\sqrt{6}} $  & 0 & 0 \\
\hline
$\eta'$ & $\frac{s_{P}+ \sqrt{2}c_{P}}{2 \sqrt{3}}$ & $\frac{s_{P}+\sqrt{2}c_{P} }{2 \sqrt{3}}$ & $\frac{ c_{P}-\sqrt{2}s_{P}}{\sqrt{6}}$  & 0 & 0 \\
\hline
\end{tabular}
\end{center}
\vspace*{-0.5cm}
\caption{\label{tab:omega3a} Factor $\Omega_A^{(1)}(ij)$.
}
\end{table}
\begin{table}[htb]
\begin{center}
\renewcommand{\arraystretch}{1.5}
\begin{tabular}{c||c|c|c|c|c||c|c|c|c|c|}
\cline{2-11}
& \multicolumn{5}{c||}{$\Omega_V^{(1)}(ij)$} & \multicolumn{5}{c|}{$\Omega_T^{(1)}(ij)$} \\
\hline
$V$ & $uu$ & $dd$ & $ss$ & $ds$ & $sd$ & $uu$ & $dd$ & $ss$ & $ds$ & $sd$\\
\hline
\hline
$\rho^0$ & $-\frac{1}{2}$ &  $\frac{1}{2} $ & 0 & 0 & 0 & $\frac{1}{\sqrt{2}}$ & 0 & 0 & 0 & 0\\
\hline
$\phi$ &  $\frac{ \sqrt{6} s_{V}-\sqrt{3}c_{V}}{6}$  & $\frac{\sqrt{6} s_{V}-\sqrt{3} c_{V}}{6} $ & $\frac{\sqrt{2}c_{V}+s_{V}}{\sqrt{6}}$  & 0 & 0 &  $\frac{c_{V}-\sqrt{2}s_{V}}{\sqrt{6}}$ & 0 & 0 & 0 & 0  \\
\hline
$\omega$ &  $-\frac{s_{V}+\sqrt{2}c_{V} }{2 \sqrt{3}} $  & $-\frac{s_{V}+\sqrt{2}c_{V} }{2 \sqrt{3}}$ & $\frac{\sqrt{2} s_{V}-c_{V}}{\sqrt{6}}$ & 0 & 0 & $\frac{\sqrt{2}c_{V}+s_{V}}{\sqrt{6}}$ & 0 & 0 & 0 & 0 \\
\hline
$K^{0 \, *}$ & 0 & 0 & 0 & $-\frac{1}{\sqrt{2}}$  & 0 & 0 & 0 & 0 & 0 & 0 \\
\hline
$\overline{K}^{0 \, *}$ & 0 & 0 & 0 & 0 & $-\frac{1}{\sqrt{2}}$ & 0 & 0 & 0 & 0 & 0 \\
\hline
\end{tabular}
\end{center}
\vspace*{-0.5cm}
\caption{\label{tab:omega1vt} Factors $\Omega_V^{(1)}(ij)$ and $\Omega_T^{(1)}(ij)$.
}
\end{table}
\begin{table}[htb]
\begin{center}
\renewcommand{\arraystretch}{1.5}
\begin{tabular}{c||c|c|c|c|c|}
\cline{2-6}
& \multicolumn{5}{c|}{$\Omega_S^{(1)}(ij)$} \\
\hline
$P_1 P_2$ & $uu$ & $dd$ & $ss$ & $ds$ & $sd$ \\
\hline
\hline
$\pi^0 \pi^0$ & $\frac{1}{4}$  & $\frac{1}{4}$ & 0 & 0 & 0  \\
\hline
$\pi^+ \pi^-$ & $\frac{1}{2}$ & $\frac{1}{2}$ & 0 & 0 & 0 \\
\hline
$K^0 \overline{K}^0$ & 0 &  $\frac{1}{2}$ & $\frac{1}{2}$ & 0 & 0\\
\hline
$K^+ K^-$ & $\frac{1}{2}$ & 0 & $\frac{1}{2}$ & 0 & 0  \\
\hline
$\eta \eta$ & $ \frac{-c_{2P}-2 \sqrt{2} s_{2P}+3}{24}$ & $ \frac{-c_{2P}-2 \sqrt{2} s_{2P}+3}{24}$ &  $\frac{ c_{2P}+2 \sqrt{2} s_{2P}+3}{12}$ & 0 & 0 \\
\hline
$\pi^0 K^0$ & 0 & 0 & 0 & $-\frac{1}{2\sqrt{2}}$ & 0 \\
\hline
$\pi^0 \overline{K}^0$ & 0 & 0 & 0 & 0 & $-\frac{1}{2\sqrt{2}}$ \\
\hline
$\pi^+ K^-$ & 0 & 0 & 0 & 0 & $\frac{1}{2}$ \\
\hline
$K^+ \pi^-$ & 0 & 0 & 0 & $\frac{1}{2}$ & 0 \\
\hline
$\pi^0 \eta$ & $\frac{\sqrt{3} c_{P}-\sqrt{6} s_{P}}{6}$ &  $\frac{\sqrt{6} s_{P}-\sqrt{3} c_{P}}{6}$ & 0 & 0 & 0 \\
\hline
$\pi^0 \eta'$ & $\frac{s_{P}+ \sqrt{2}c_{P}}{2 \sqrt{3}}$ & $-\frac{s_{P}+ \sqrt{2}c_{P}}{2 \sqrt{3}}$ & 0 & 0 & 0 \\
\hline
$K^0 \eta$ & 0 & 0 & 0 & $-\frac{4 s_{P}+\sqrt{2} c_{P} }{4 \sqrt{3}} $ & 0 \\
\hline
$K^0 \eta'$ & 0 & 0 & 0 & $\frac{2 \sqrt{2}c_{P}-s_{P}}{2\sqrt{6}}$ & 0 \\
\hline
$\overline{K}^0  \eta$ & 0 & 0 & 0 & 0 & $-\frac{4 s_{P}+\sqrt{2} c_{P} }{4 \sqrt{3}} $ \\
\hline
$\overline{K}^0 \eta'$ & 0 & 0 & 0 & 0 & $\frac{2 \sqrt{2}c_{P}-s_{P}}{2\sqrt{6}}$\\
\hline
$\eta \eta'$ &  $\frac{2 \sqrt{2} c_{2P}-s_{2P}}{12} $ &  $\frac{2 \sqrt{2} c_{2P}-s_{2P}}{12}$ & $\frac{s_{2P}-2 \sqrt{2} c_{2P}}{6}$ & 0 & 0 \\
\hline
\end{tabular}
\end{center}
\vspace*{-0.5cm}
\caption{\label{tab:omega12s} Factors $\Omega_S^{(1)}(ij)$.
}
\end{table}
\begin{landscape}
\begin{table}[htb]
\begin{center}
\renewcommand{\arraystretch}{1.5}
\begin{tabular}{c||c|c|c|c|c|}
\cline{2-6}
& \multicolumn{5}{c|}{$\Omega_S^{(2)}(ij)$} \\
\hline
$P_1 P_2$ & $uu$ & $dd$ & $ss$ & $ds$ & $sd$ \\
\hline
\hline
$\pi^0 \pi^0$ & $-4 m_{\pi/K}^{2}$ & $-4 m_{\pi/K}^{2}$ & 0 & 0 & 0 \\
\hline
$\pi^+ \pi^-$ & $-8 m_{\pi/K}^{2}$ & $-8 m_{\pi/K}^{2}$ & 0 & 0 & 0 \\
\hline
$K^0 \overline{K}^0$ & 0 & $-8 $ & $-8 $ & 0 & 0 \\
\hline
$K^+ K^-$ &  $-8 $ & 0 & $-8 $ & 0 & 0 \\
\hline
$\eta \eta$  &  $\frac{m_{\pi/K}^{2}(4  \sqrt{2}s_{2P}+2c_{2P}-6)}{3}$ & $\frac{m_{\pi/K}^{2}(4  \sqrt{2}s_{2P}+2c_{2P}-6)}{3}$ & $\frac{(4m_{\pi/K}^{2}-8)(2 \sqrt{2}s_{2P}+c_{2P}+3)}{3}$ & 0 & 0 \\
\hline
$\pi^0 K^0$  & 0 & 0 & 0 & $2\sqrt{2}(1+m_{\pi/K}^{2})$ & 0 \\
\hline
$\pi^0 \overline{K}^0$ & 0 & 0 & 0 & 0 & $2\sqrt{2}(1+m_{\pi/K}^{2})$ \\
\hline
$\pi^+ K^-$  & 0 & 0 & 0 & 0 & $-4(1+m_{\pi/K}^{2})$ \\
\hline
$K^+ \pi^-$   & 0 & 0 & 0 & $-4(1+m_{\pi/K}^{2})$ & 0 \\
\hline
$\pi^0 \eta$  & $\frac{8 m_{\pi/K}^{2}(\sqrt{2} s_{P}-c_{P})}{\sqrt{3}}$ & -$\frac{8 m_{\pi/K}^{2}(\sqrt{2} s_{P}-c_{P})}{\sqrt{3}}$ & 0 & 0 & 0 \\
\hline
$\pi^0 \eta'$ &  $-\frac{8 m_{\pi/K}^{2}( s_{P}+\sqrt{2}c_{P})}{\sqrt{3}}$ & $\frac{8 m_{\pi/K}^{2}( s_{P}+\sqrt{2}c_{P})}{\sqrt{3}}$ & 0 & 0 & 0 \\
\hline
$K^0 \eta$ & 0 & 0 & 0 & $\frac{2(\sqrt{2}c_{P}(5-3m_{\pi/K}^{2})+8  s_{P})}{\sqrt{3}}$ & 0 \\
\hline
$K^0 \eta'$ & 0 & 0 & 0 & $\frac{2(\sqrt{2}s_{P}(5-3m_{\pi/K}^{2})-8 c_{P})}{\sqrt{3}} $ & 0\\
\hline
$\overline{K}^0  \eta$ & 0 & 0 & 0 & 0 & $\frac{2(8  s_{P}+\sqrt{2}c_{P}(5-3m_{\pi/K}^{2}))}{\sqrt{3}}$ \\
\hline
$\overline{K}^0 \eta'$  & 0 & 0 & 0 & 0 & $\frac{2(\sqrt{2}s_{P}(5-3m_{\pi/K}^{2})-8 c_{P})}{\sqrt{3}} $ \\
\hline
$\eta \eta'$  & $\frac{4 m_{\pi/K}^{2}(s_{2P}-2 \sqrt{2}c_{2P})}{3}$ & $\frac{4 m_{\pi/K}^{2}(s_{2P}-2 \sqrt{2}c_{2P})}{3}$ & $\frac{(8m_{\pi/K}^{2}-16)(s_{2P}-2 \sqrt{2}c_{2P})}{3}$ & 0 & 0 \\
\hline
\end{tabular}
\end{center}
\vspace*{-0.5cm}
\caption{\label{tab:omega13s} Factors  $\Omega_S^{(2)}(ij)$.
}
\end{table}
\end{landscape}
\begin{table}[t]
\begin{center}
\renewcommand{\arraystretch}{1.5}
\begin{tabular}{c||c|c|c|c|c|}
\cline{2-6}
& \multicolumn{5}{c|}{$\Omega_{S}^{(3)}(ij)$}  \\
\hline
$P$ & $uu$ & $dd$ & $ss$ & $ds$ & $sd$ \\
\hline
\hline
$a_0^0$ & $-\sqrt{2}$  &$\sqrt{2}$ & 0 & 0 & 0 \\
\hline
$f_0^8$ & $-\sqrt{\frac{2}{3}}$ & $-\sqrt{\frac{2}{3}}$ & $2\sqrt{\frac{2}{3}}$& 0 &  0\\
\hline
$f_0^0$ & $ -\frac{2}{\sqrt{3}} $ & $ -\frac{2}{\sqrt{3}} $ &  $ -\frac{2}{\sqrt{3}} $ & 0  &  0\\
\hline 
$K_0^{0 \, *}$ & 0 &  0 & 0 & $-2$ & 0\\
\hline
$\overline{K}_0^{0 \, *}$ & 0 & 0 & 0 & 0 & $-2$\\
\hline
\end{tabular}
\end{center}
\vspace*{-0.5cm}
\caption{\label{tab:omega1sk} Factor $\Omega_{S}^{(3)}(ij)$.
}
\end{table}
\begin{table}[b]
\begin{center}
\renewcommand{\arraystretch}{1.5}
\begin{tabular}{c||c|c|c|c|c|}
\cline{2-6}
& \multicolumn{5}{c|}{$\Omega_{S}^{(4)}$,\; $\Omega_{T}^{(3)}/2$,\; $\Omega_{T}^{(4)}$}  \\
\hline
$P_1 P_2$ &
\hfill$a_0^0$ \hfill\Big|\hfill $a_2^0$\hfill\hfill &
\hfill$f_0^8$ \hfill\Big|\hfill $f_2^8$\hfill\hfill &
\hfill$f_0^0$ \hfill\Big|\hfill $f_2^0$\hfill\hfill &
\hfill$K_0^{0 \, *}$ \hfill\Big|\hfill $K_2^{0 \, *}$\hfill\hfill &
\hfill$\overline{K}_0^{0 \, *}$ \hfill\Big|\hfill $\overline{K}_2^{0 \, *}$\hfill\hfill \\
\hline
\hline
$\pi^0 \pi^0$ & 0 & $\sqrt{\frac{2}{3}}$ & $\frac{2}{\sqrt{3}}$ & 0 & 0	 \\
\hline
$\pi^+ \pi^-$ & 0 & 2$\sqrt{\frac{2}{3}}$ & $\frac{4}{\sqrt{3}}$  & 0 & 0	 \\
\hline
$K^0 \overline{K}^0$ & $-\sqrt{2}$ & $-\sqrt{\frac{2}{3}}$ & $\frac{4}{\sqrt{3}}$  & 0 & 0 \\
\hline
$K^+ K^-$ & $\sqrt{2}$ & $-\sqrt{\frac{2}{3}}$& $\frac{4}{\sqrt{3}}$  & 0 & 0 	 \\
\hline
$\eta \eta$ & 0 & $-\frac{c_{P}(4 s_{P}+\sqrt{2}c_{P})}{\sqrt{3}}$ & $\frac{2}{\sqrt{3}}$  & 0 & 0  \\
\hline
$\pi^0 K^0$ & 0 & 0 & 0 & $-\sqrt{2}$ & 0 \\
\hline
$\pi^0 \overline{K}^0$ & 0 & 0 & 0 & 0 & $-\sqrt{2}$  \\
\hline
$\pi^+ K^-$ & 0 & 0 & 0 & 0 & $2$ \\
\hline
$K^+ \pi^-$ & 0 & 0 & 0 & $2$ & 0 \\
\hline
$\pi^0 \eta$ & $\frac{2(\sqrt{2}c_{P}-2 s_{P})}{\sqrt{3}}$ & 0 & 0 & 0 & 0 \\
\hline
$\pi^0 \eta'$ &$\frac{2(2c_{P}+\sqrt{2}s_{P})}{\sqrt{3}}$ & 0 & 0 & 0 & 0	\\
\hline
$K^0 \eta$ & 0 & 0 & 0 & $-\frac{4s_{P}+\sqrt{2}c_{P}}{\sqrt{3}}$ & 0 \\
\hline
$K^0 \eta'$ & 0 & 0 & 0 & $\frac{4c_{P}-\sqrt{2}s_{P}}{\sqrt{3}}$ & 0 \\
\hline
$\overline{K}^0  \eta$ & 0 & 0 & 0 & 0 & $-\frac{4s_{P}+\sqrt{2}c_{P}}{\sqrt{3}}$	 \\
\hline
$\overline{K}^0 \eta'$ & 0 & 0 & 0 & 0 &  $\frac{4c_{P}-\sqrt{2}s_{P}}{\sqrt{3}}$ \\
\hline
$\eta \eta'$ & 0 & $\frac{4c_{2P}-\sqrt{2}s_{2P}}{\sqrt{3}}$ & 0 & 0 & 0 	 \\
\hline
\end{tabular}
\end{center}
\vspace*{-0.5cm}
\caption{\label{tab:omega23s} Factors $\Omega_{S}^{(4)}$, $\Omega_{T}^{(3)}$ and $\Omega_{T}^{(4)}$.
}
\end{table}
\begin{landscape}
\begin{table}[htb]
\begin{center}
\renewcommand{\arraystretch}{1.5}
\begin{tabular}{c||c|c|c|c|c|}
\cline{2-6}
& \multicolumn{5}{c|}{$\Omega_{S}^{(5)}$,\; $\Omega_{T}^{(5)}$} \\
\hline
$P_1 P_2$ &
\hfill$a_0^0$\hfill\Big|\hfill $a_2^0$\hfill\hfill & \hfill$f_0^8$\hfill\Big|\hfill$f_2^8$\hfill\hfill &
\hfill$f_0^0$\hfill\Big|\hfill$f_2^0$\hfill\hfill &
\hfill$K_0^{0 \, *}$\hfill\Big|\hfill$K_2^{0 \, *}$\hfill\hfill &
\hfill$\overline{K}_0^{0 \, *}$\hfill\Big|\hfill$\overline{K}_2^{0 \, *}$\hfill\hfill\\
\hline
\hline
$\pi^0 \pi^0$ & 0	& $\sqrt{\frac{2}{3}} m_{\pi/K}^{2}$ & $\frac{2 m_{\pi/K}^{2}}{\sqrt{3}}$ & 0 & 0 \\
\hline
$\pi^+ \pi^-$ 	 &	0 & $2\sqrt{\frac{2}{3}} m_{\pi/K}^{2}$ & $\frac{4 m_{\pi/K}^{2}}{\sqrt{3}}$ & 0 & 0 \\
\hline
$K^0 \overline{K}^0$ 	& $-\sqrt{2}$ & $-\sqrt{\frac{2}{3}} $ & $\frac{4 }{\sqrt{3}}$ & 0 & 0 \\
\hline
$K^+ K^-$ 	& $\sqrt{2}$ & $-\sqrt{\frac{2}{3}}$ &  $\frac{4 }{\sqrt{3}}$ & 0 & 0 \\
\hline
$\eta \eta$  	&	0 &  $\frac{(m_{\pi/K}^{2}-4)(2\sqrt{2}s_{2P}+c_{2P})+9m_{\pi/K}^{2}-12}{3 \sqrt{6}}$ &  $\frac{(1-m_{\pi/K}^{2})(4\sqrt{2}s_{2P}+2c_{2P})+6}{3 \sqrt{3}}$ & 0 & 0 \\
\hline
$\pi^0 K^0$ &	0 & 0 & 0 & $-\frac{1+m_{\pi/K}^{2}}{\sqrt{2}}$ & 0 \\
\hline
$\pi^0 \overline{K}^0$ 	&	0 & 0  & 0 & 0 & $-\frac{1+m_{\pi/K}^{2}}{\sqrt{2}}$ \\
\hline
$\pi^+ K^-$ &		0 & 0 & 0 & 0 & $1+m_{\pi/K}^{2}$  \\
\hline
$K^+ \pi^-$   &	0 & 0 & 0 & $1+m_{\pi/K}^{2}$ & 0 \\
\hline
$\pi^0 \eta$  &	$\frac{2 m_{\pi/K}^{2} (\sqrt{2} c_{P}-2s_{P})}{\sqrt{3}}$ & 0 & 0 & 0 & 0  \\
\hline
$\pi^0 \eta'$  &	$\frac{2 m_{\pi/K}^{2} ( 2 c_{P}+\sqrt{2}s_{P})}{\sqrt{3}}$ & 0 & 0 & 0 & 0 \\
\hline
$K^0 \eta$ 	&	0 & 0 & 0 & $\frac{c_{P}(3m_{\pi/K}^{2}-5)-4\sqrt{2}s_{P}}{\sqrt{6}}$ & 0 \\
\hline
$K^0 \eta'$ &	0 & 0 & 0 & $\frac{s_{P}(3m_{\pi/K}^{2}-5)+4\sqrt{2}c_{P}}{\sqrt{6}}$ & 0\\
\hline
$\overline{K}^0  \eta$  &	0 & 0 & 0 & 0 & $\frac{c_{P}(3m_{\pi/K}^{2}-5)-4\sqrt{2}s_{P}}{\sqrt{6}}$ \\
\hline
$\overline{K}^0 \eta'$ & 0 & 0 & 0 & 0 & $\frac{s_{P}(3m_{\pi/K}^{2}-5)+4\sqrt{2}c_{P}}{\sqrt{6}}$  \\
\hline
$\eta \eta'$  &	 0 & $\frac{(4-m_{\pi/K}^{2})(4c_{2P}-\sqrt{2}s_{2P})}{3\sqrt{3}}$ & $\frac{4(1-m_{\pi/K}^{2})(s_{2P}-2\sqrt{2}c_{2P})}{3\sqrt{3}}$  & 0 & 0 \\
\hline
\end{tabular}
\end{center}
\vspace*{-0.5cm}
\caption{\label{tab:omega14s} Factors  $\Omega_{S}^{(5)}$ and $\Omega_{T}^{(5)}$.
}
\end{table}
\end{landscape}
\vfill
\begin{table}[htb]
\begin{center}
\renewcommand{\arraystretch}{1.5}
\begin{tabular}{c||c|c|c|c|c|}
\cline{2-6}
& \multicolumn{5}{c||}{$\Omega_{T}^{(2)}(ij)$}  \\
\hline
$T$ & $uu$ & $dd$ & $ss$ & $ds$ & $sd$ \\
\hline
\hline
$a_2^0$ & $-\sqrt{2}$ & $\sqrt{2}$ & 0 & 0 & 0 \\
\hline
$f_2^8$ & $-\sqrt{\frac{2}{3}}$ & $-\sqrt{\frac{2}{3}}$ & $2\sqrt{\frac{2}{3}}$ & 0 & 0 \\
\hline
$f_2^0$ & $-\frac{2}{\sqrt{3}}$ & $-\frac{2}{\sqrt{3}}$ & $-\frac{2}{\sqrt{3}}$ &  0& 0\\
\hline
$K_2^{0 \, *}$ & 0 & 0 & 0 & $-2$ & 0 \\
\hline
$\overline{K}_2^{0 \, *}$ & 0 & 0 & 0 & 0 & $-2$ \\
\hline
\end{tabular}
\end{center}
\vspace*{-0.5cm}
\caption{\label{tab:omega1kT} Factor $\Omega_{T}^{(2)}(ij)$.
}
\end{table}
\vfill
\begin{table}[!htb]
\begin{center}
\renewcommand{\arraystretch}{1.5}
\begin{tabular}{c||c|c|c|c|c||c|c|c|c|c|}
\cline{2-11}
& \multicolumn{5}{c||}{$\Omega_V^{(2)}(ij)$} & \multicolumn{5}{c|}{$\Omega_T^{(6)}(ij)$} \\
\hline
$P_1 P_2$ & $uu$ & $dd$ & $ss$ & $ds$ & $sd$ & $uu$ & $dd$ & $ss$ & $ds$ & $sd$\\
\hline
\hline
$\pi^+ \pi^-$ & $\frac{1}{2}$ & $-\frac{1}{2}$ & 0 & 0 & 0 & $2$ & 0 & 0 & 0 & 0 \\
\hline
$K^0 \overline{K}^0$  & 0 & $\frac{1}{2}$ & $-\frac{1}{2}$ & 0 & 0 & 0 & 0 & 0 & 0 & 0 \\
\hline
$K^+ K^-$ & $\frac{1}{2}$ & 0 & $-\frac{1}{2}$ & 0 & 0 & $2$ & 0 & 0 & 0 & 0 \\
\hline
$\pi^0 K^0$   & 0 & 0 & 0 & $\frac{1}{2\sqrt{2}}$ & 0 & 0 & 0 & 0 & 0 & 0 \\
\hline
$\pi^0 \overline{K}^0$  & 0 & 0 & 0 & 0 & $-\frac{1}{2\sqrt{2}}$ & 0 & 0 & 0 & 0 & 0\\
\hline
$\pi^+ K^-$ & 0 & 0 & 0 & 0 & $\frac{1}{2}$ & 0 & 0 & 0 & 0 & 0 \\
\hline
$K^+ \pi^-$ & 0 & 0 & 0 & $\frac{1}{2}$ & 0 & 0 & 0 & 0 & 0 & 0 \\
\hline
$K^0 \eta $ & 0 & 0 & 0 & $-\sqrt{\frac{3}{2}}\, \frac{c_{P}}{2}$ & 0 & 0 & 0 & 0 & 0 & 0 \\
\hline
$K^0 \eta' $  & 0 & 0 & 0 & $-\sqrt{\frac{3}{2}}\, \frac{s_{P}}{2}$ & 0 & 0 & 0 & 0 & 0 & 0 \\
\hline
$\overline{K}^0 \eta $  & 0 & 0 & 0 & 0 & $\sqrt{\frac{3}{2}}\, \frac{c_{P}}{2}$ & 0 & 0 & 0 & 0 & 0 \\
\hline
$\overline{K}^0 \eta' $  & 0 & 0 & 0 & 0 & $\sqrt{\frac{3}{2}}\, \frac{s_{P}}{2}$ & 0 & 0 & 0 & 0 & 0 \\
\hline
\end{tabular}
\end{center}
\vspace*{-0.5cm}
\caption{\label{tab:omega22vt} Factors $\Omega_V^{(2)}(ij)$ and $\Omega_T^{(6)}(ij)$.
}
\end{table}
\vfill
\begin{table}[!htb]
\begin{center}
\renewcommand{\arraystretch}{1.5}
\begin{tabular}{c||c|c|c|c|c|c|}
\cline{2-6}
& \multicolumn{5}{c||}{$\Omega_{V}^{(3)}$} \\
\hline
$P_1 P_2$ & $\rho^0$ & $\phi$ & $\omega$ & $K^{0 \, *}$ & $\overline{K}^{0 \, *}$ \\
\hline
\hline
$\pi^+ \pi^-$ & $-\sqrt{2} $ & 0 & 0 & 0 & 0 \\
\hline
$K^0 \overline{K}^0$  & $\frac{1}{\sqrt{2}}$ & $-\sqrt{\frac{3}{2}}\, c_{V}$ & $-\sqrt{\frac{3}{2}}\, s_{V}$ & 0 & 0 \\
\hline
$K^+ K^-$  & $-\frac{1}{\sqrt{2}}$ & $-\sqrt{\frac{3}{2}}\, c_{V}$ & $-\sqrt{\frac{3}{2}}\, s_{V}$ & 0 & 0  \\
\hline
$\pi^0 K^0$  & 0 & 0 & 0 & $-\frac{1}{\sqrt{2}}$ & 0 \\
\hline
$\pi^0 \overline{K}^0$ & 0 & 0 & 0 & 0 & $\frac{1}{\sqrt{2}}$ \\
\hline
$\pi^+ K^-$ & 0 & 0 & 0 & 0 & $-1$ \\
\hline
$K^+ \pi^-$ & 0 & 0 & 0 & $-1$ & 0 \\
\hline
$K^0 \eta $ & 0 & 0 & 0 & $\sqrt{\frac{3}{2}}\, c_{P}$ & 0 \\
\hline
$K^0 \eta' $   & 0 & 0 & 0 & $\sqrt{\frac{3}{2}}\, s_{P}$ & 0 \\
\hline
$\overline{K}^0 \eta $ & 0 & 0 & 0 & 0 & $-\sqrt{\frac{3}{2}}\, c_{P}$ \\
\hline
$\overline{K}^0 \eta' $ & 0 & 0 & 0 & 0 & $-\sqrt{\frac{3}{2}}\, s_{P}$ \\
\hline
\end{tabular}
\end{center}
\vspace*{-0.5cm}
\caption{\label{tab:omega4vk} Factors $\Omega_{V}^{(3)}$.
}
\end{table}

\vfill

\section{Kinematics of \texorpdfstring{\boldmath $\ell$--$\tau$}{l--tau} conversion in nuclei} \label{app:5}

$\ell$--$\tau$ conversion in nuclei is a two-body to two-body process described at tree level within the SMEFT framework by the perturbative diagrams in Section~\ref{sss:2.2.1}. Hence, the squared unpolarized amplitudes as well as the phase space can be described by just two invariant variables. In our case, we choose $\xi$ and $Q^{2}$ (see Section~\ref{sss:2.2.2}).
The perturbative cross sections are then given (in terms of these invariant variables) by Eqs.~(\ref{eq:pertampl1}--\ref{eq:pertampl3}), where the phase-space factor is written in terms of the K\"all\'en's triangle function $\lambda$.

The total cross section of the process is given by Eq.~\eqref{eq:totcrosssect}, where the integration limits for $\xi$ and $Q^{2}$ are as follows:
As usual, we consider that the parton cannot (or it is very unlikely to) have a momentum larger than the nucleus in which it is confined, which leads to 
\begin{equation}
\xi_\text{max}=1\, .
\end{equation}
Considering massive quarks and leptons modifies the typically assumed vanishing lower limit of $\xi$ to
\begin{equation}
\xi_\text{min}=\frac{\sqrt{E_{\ell}^{2}-m_{\ell}^{2}+(m_{\tau}+m_{j})^{2}}-E_{\ell}}{M}\, .
\end{equation}
For the variable $Q^{2}$, we have
\begin{equation}
 Q^{2}_{\pm}=\frac{\xi M E_{\ell}(m_{\ell}^{2}-m_{\tau}^{2}+\xi^{2} M^{2})+2E_{\ell}^{2}\xi^{2} M^{2}-m_{j}^{2}(E_{\ell} \xi M +m_{\ell}^{2})-\xi^{2} M^{2}m_{\tau}^{2}\pm \xi M \sqrt{X Y}}{M \xi (2 E_{\ell}+M\xi)+m_{\ell}^{2}}\, ,
\end{equation}
where
\begin{equation}
X=E_{\ell}^{2}-m_{\ell}^{2}\, ,
\end{equation}
and
\begin{equation}
Y=m_{j}^{4}+[m_{\ell}^{2}-m_{\tau}^{2}+\xi M(2E_{\ell}+\xi M)]^{2}-2m_{j}^{2}[m_{\ell}^{2}+m_{\tau}^{2}+\xi M(2E_{\ell}+\xi M)]\, .
\end{equation}
However, since the parton distribution functions provided by the nCTEQ15 group are expected not to be reliable below $Q=1.3$ GeV, we take the square of this value as the lower limit of our integral~\eqref{eq:totcrosssect}. This leads to a small underestimation of the total cross section and thus more conservative resulting constraints on the Wilson coefficients.

\section{The tau decay widths} 
\label{app:8}

In this appendix, we collect expressions for the branching ratios of the decays of the tau lepton into pseudoscalars:
\begin{equation} \label{eq:brlp}
\mathcal{B}( \tau \rightarrow \ell P ) \, = \, \frac{\lambda^{1/2}(m_{\tau}^2,m_{\ell}^2,m_P^2)}{16 \, \pi \, m_{\tau}^3 \, \Gamma_{\pi}} \, 
\frac{1}{2} \sum_{i,f} \, | {\cal M}(P) |^2 \, , 
\end{equation}
\begin{equation} \label{eq:brlpp}
\mathcal{B}( \tau \rightarrow \ell P_1 P_2 ) \, = \, \frac{1}{256 \, \pi^3 \, m_{\tau}^3 \, \Gamma_{\tau}} \, \int_{s_{\text{min}}}^{s_{\text{max}}}
\diff s \ \int_{t_{\text{min}}}^{t_{\text{max}}} \diff t \, \frac{1}{2} \sum_{i,f} \, | {\cal M}(P_1, P_2) |^2 \, ,
\end{equation}
with
\begin{equation} \label{eq:ssttlimits}
\begin{split}
t_{\text{min}}^{\text{max}}  & =  \frac{1}{4 \, s} \, \left[ (m_{\tau}^2 - m_{\ell}^2 + m_{P_1}^2 - m_{P_2}^2)^2 \, - \, 
\left( \lambda^{1/2}(s,m_{P_1}^2,m_{P_2}^2) \, \mp \, \lambda^{1/2}(m_{\tau}^2,s,m_{\ell}^2) \right)^2 \right],\\
s_{\text{min}} & = (m_{P_1} + m_{P_2})^2\,, \\
s_{\text{max}} & = (m_{\tau} - m_{\ell})^2\,,
\end{split}
\end{equation}
where $\lambda(a,b,c)$ is the K\"all\'en's triangle function.
\par
The calculation of observables involving hadron resonances as external states is not properly defined within quantum field theory because hadron resonances decay strongly and are not proper asymptotic states, as is required in that framework. Hence, in order to describe the $\tau \rightarrow \ell V$ decays, we need to provide an appropriate definition. We intend to study the processes with $V = \rho^{0}(770),\omega(782),\phi(1020),K^{*0}(892),\bar{K}^{*0}(892)$. All but the $\omega(782)$ decay strongly into two pseudoscalars. For these cases we can use the definition that has already been employed in Refs.~\cite{Arganda:2008jj} and \cite{Lami:2016vrs}:
\begin{equation}
\label{eq:brconst2}
\mathcal{B}(\tau \rightarrow \ell V) =    \sum_{P_1 P_2} \, \mathcal{B}(\tau \rightarrow \ell P_1 P_2)\Big|_V \,.
\end{equation}
In the above equation, the branching ratios for the $P_1 P_2$ decays from Eq.~\eqref{eq:brlpp} have the same $t$ limits as shown in Eq.~\eqref{eq:ssttlimits}, but the $s$ limits are now restricted to
\begin{equation}
s_{\text{min}} \, = \, M_V^2 - \frac{1}{2} M_V \Gamma_V \, , \qquad \qquad s_{\text{max}} \, = \, M_V^2 + \frac{1}{2} M_V \Gamma_V \, .
\end{equation}
In Eq.~\eqref{eq:brconst2}, $P_1 P_2$, from a chiral point of view, are indistinguishable from the $V$ resonance, i.e.\ the pair of pseudoscalar mesons have the same $J$ and $I$ quantum numbers. Accordingly, they are the dominant strong-decay channel of the resonance $V$. This definition is based on the fact that, experimentally, no $V$ resonance is observed, only its decay products (pairs $P_1 P_2$). The correspondence is 
$\left\{ \rho, \phi, K^* \right\} \leftrightarrow \left\{ \pi \pi, K \overline{K}, K \pi \right\}$, where a sum of contributions is understood: For instance, for the $\phi$ resonance we have to sum over the $K^+ K^-$ and $K^0 \overline{K}^0$ decay modes.
\par
The $\omega(782)$ decays dominantly into three pions. Hence, the procedure above does not work for this decay. As the ratio between its width and mass is around $1~\%$, we will consider the $\omega(782)$ as an asymptotic state and proceed as in the case of one pseudoscalar. An analogous check with the $\phi(1020)$ case shows that, within this approach, we should get the right order of magnitude for the $\tau \rightarrow \ell \omega$ decay width.

\section{Correlation matrices of the marginalized numerical analysis} 
\label{app:6}

\begin{figure}[!t]
\capstart
\begin{center}
\includegraphics[width=\textwidth]{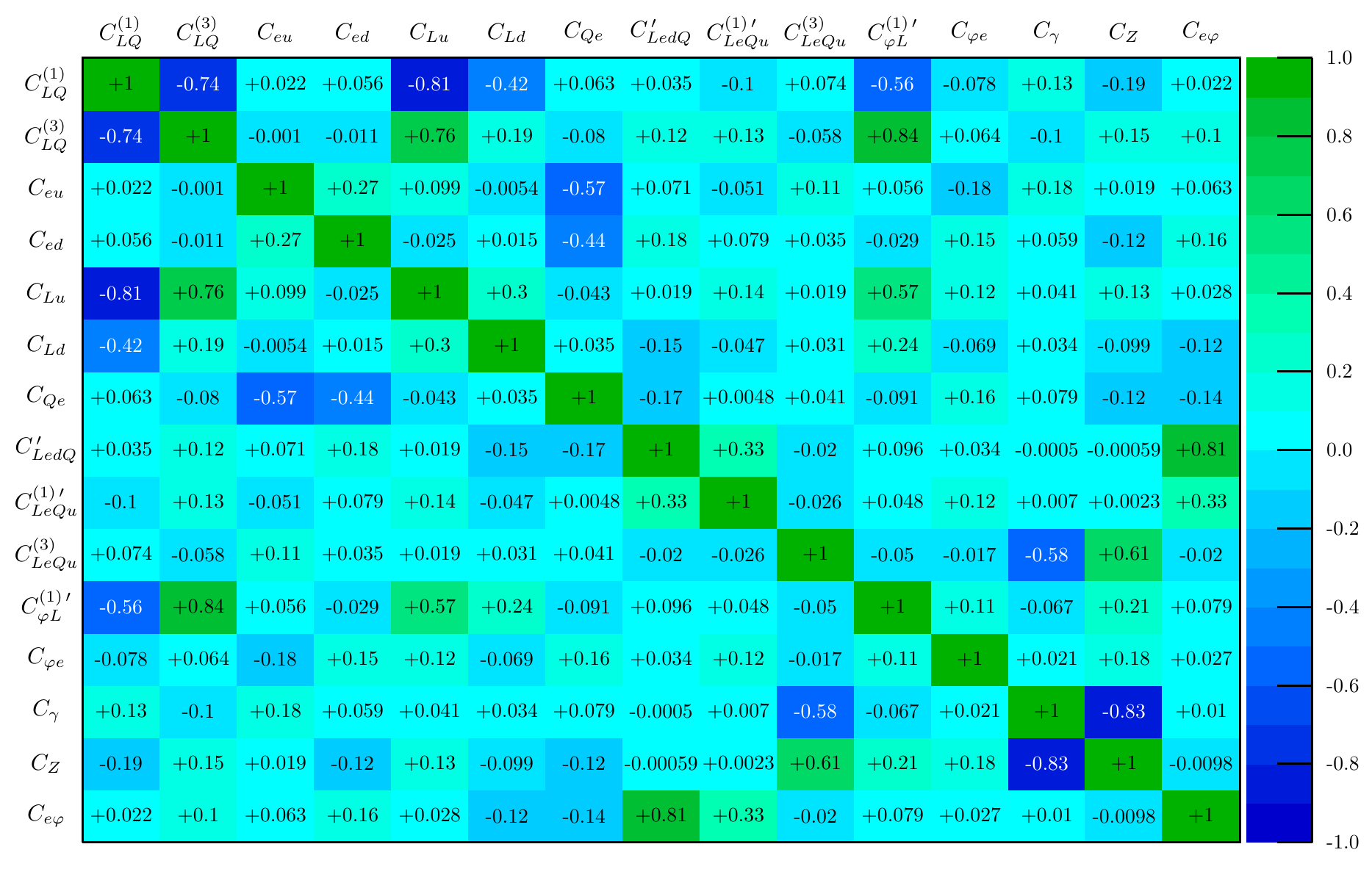}
\caption[]{\label{fig:G1} Correlations among all the Wilson coefficients from the numerical analysis considering Belle limits and {\em including} FCNCs.}
\end{center}
\end{figure}

\begin{figure}[!t]
\capstart
\begin{center}
\includegraphics[width=\textwidth]{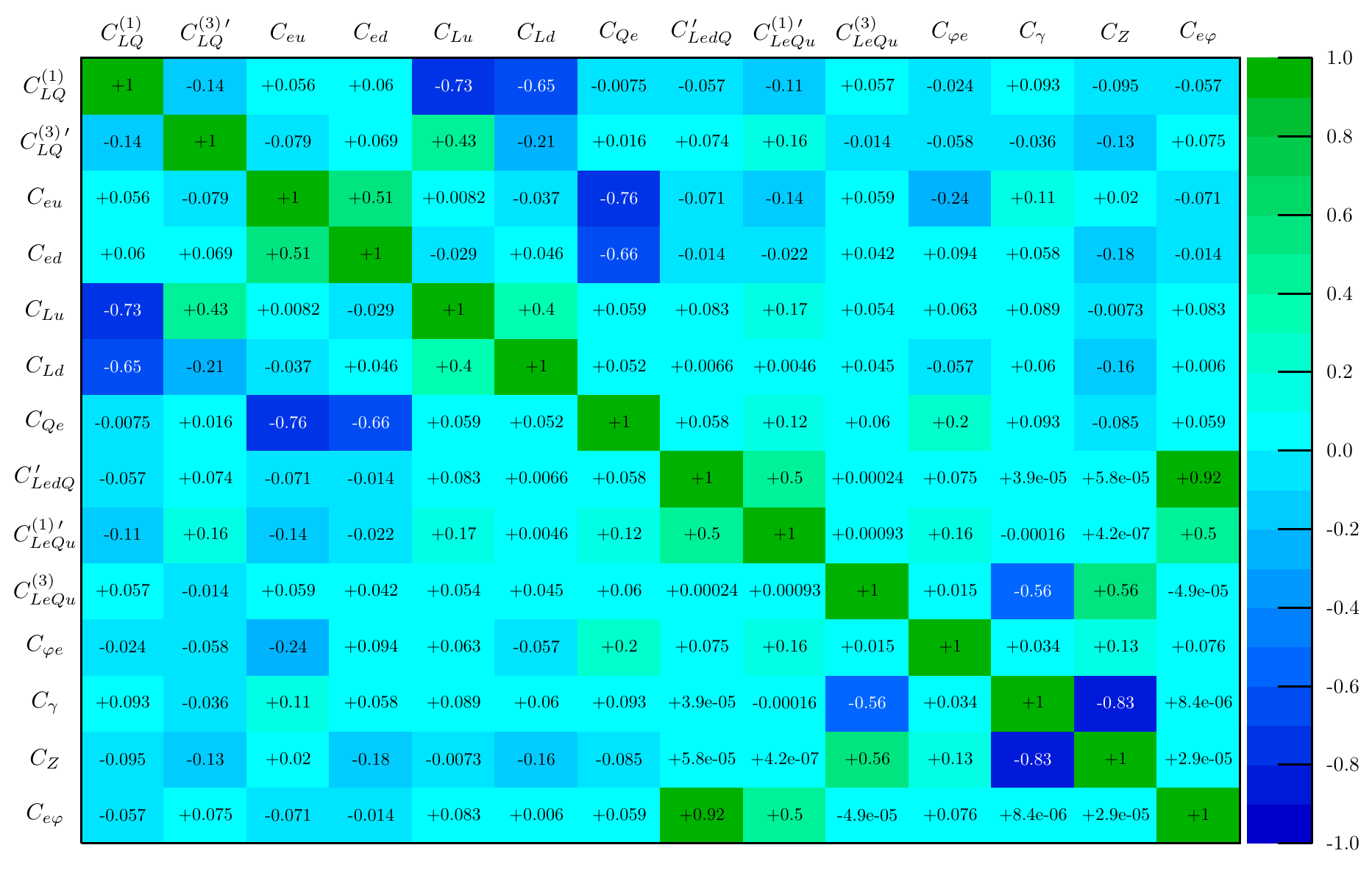}
\caption[]{\label{fig:G2} Correlations among all the Wilson coefficients from the numerical analysis considering Belle limits and {\em excluding} FCNCs.}
\end{center}
\end{figure}

\begin{figure}[!t]
\capstart
\begin{center}
\includegraphics[width=\textwidth]{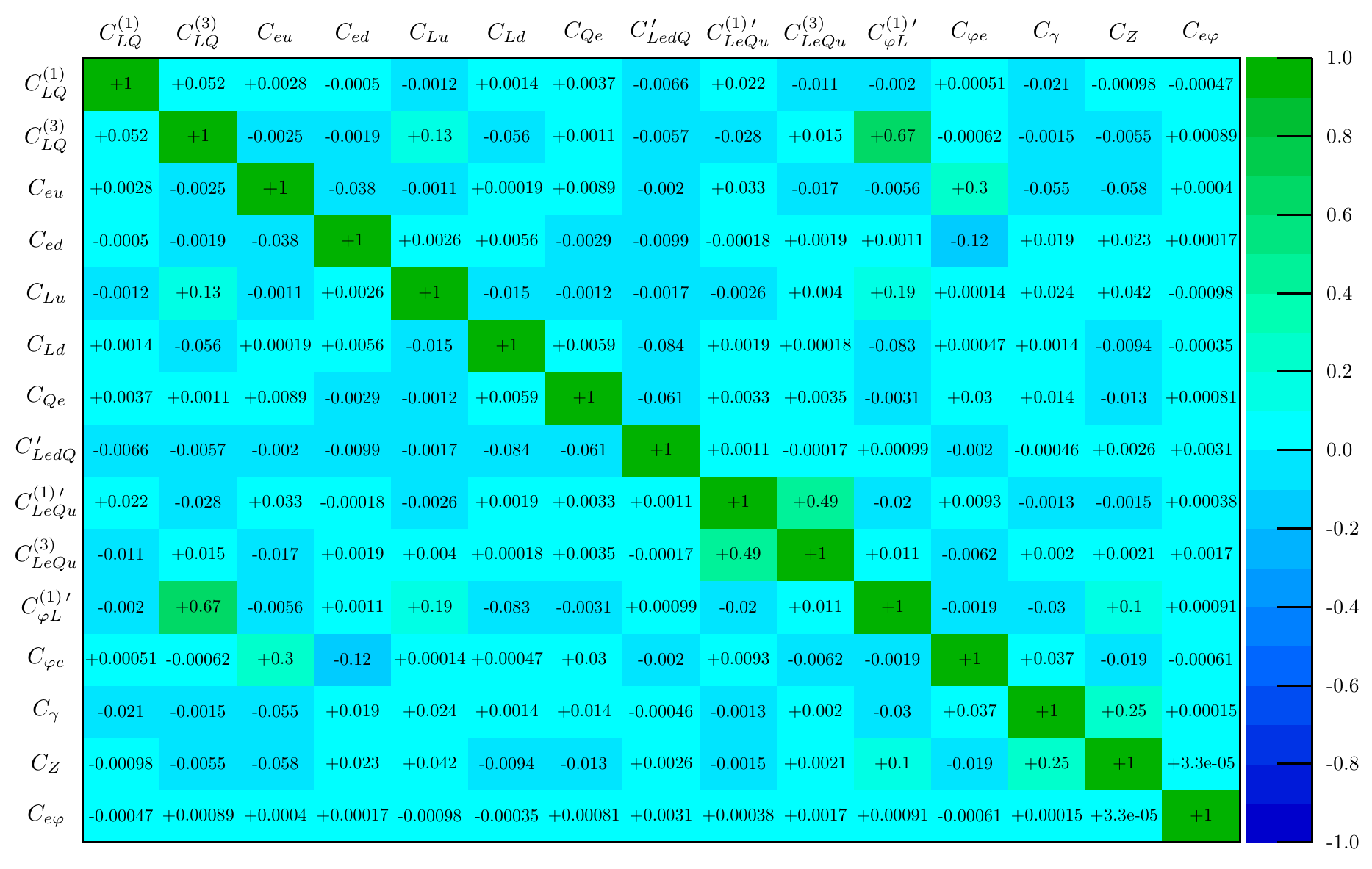}
\caption[]{\label{fig:G3} Correlations among all the Wilson coefficients from the numerical analysis considering $\ell$--$\tau$ limits and including FCNCs.}
\end{center}
\end{figure}

In this section, we present the main correlation matrices of our numerical analysis using \texttt{HEPfit}. 
For the hadronic $\tau$ decays, the correlation matrix of all the Wilson coefficients obtained from the numerical analysis considering only the Belle limits and including (excluding) FCNCs is shown in Fig.~\ref{fig:G1} (\ref{fig:G2}). We regard it interesting to compare these two matrices, as is described in detail in Section~\ref{sss:3.2.1}.
The correlation matrix obtained from the numerical analysis considering only the limits of $\ell$--$\tau$ conversion in nuclei and including FCNCs is presented in Fig.~\ref{fig:G3}.

\section{Numerical inputs}
\label{app:7}

In this appendix, we collect the numerical inputs for our calculations: Due to the hadronic incertitudes, we explain our choices for the related parameters; for the rest we take the PDG values \cite{Zyla:2020zbs}.
\par 
For the masses of the hadrons, we take the values listed in Tab.~\ref{tab:massid}: For the pseudoscalar mesons, we take the isospin-averaged values. For the vectorial resonances, we take masses from Ref.~\cite{Zyla:2020zbs}. For the rest of the resonances, we then consider a single mass for the whole multiplet chosen as the mass of the associated isotriplet.
\begin{table}[!h]
\capstart
\begin{center}
\renewcommand{\arraystretch}{1.5}
\begin{tabular}{|c|c|c|c|c|c|c|c|}
\hline
 $m_{\pi}$ & $m_K$ & $m_{\eta }$ & $m_{\eta'}$ & $M_S$ & $M_P$ & $M_T$   \\
\hline
 $0.138$ & $0.496$ & $0.548 $ & $0.958$ & $1.450$  & $1.3 $ & $1.320$ \\
\hline
\end{tabular}
\end{center}
\vspace*{-0.5cm}
\caption{\label{tab:massid} Masses (given in GeV) for the pseudoscalars and resonances.
}
\end{table}
\par 
Our knowledge of the hadron couplings in the R$\chi$T Lagrangian is rather sketchy. This is due to our poor insight about the final-state interactions, so relevant in strong processes. We use the values from Tab.~\ref{tab:coupid} together with the relations (\ref{eq:sdc}). It remains to comment on the $\gamma$ coupling in the spin-2 resonance Lagrangian (\ref{eq:currin}): There is no information on this coupling. However, we notice that its numerical relevance is rather suppressed since it accompanies the masses of the pseudoscalar mesons. Therefore, its specific value is not relevant in the numerical computations. For definiteness, we take $\gamma = \beta$. 
\begin{table}[!h]
\capstart
\begin{center}
\renewcommand{\arraystretch}{1.5}
\begin{tabular}{|c|c|c|c|c|}
\hline
 $F$\,[GeV] \cite{Zyla:2020zbs}& $F_V$\,[GeV] \cite{Dumm:2009va} & $c_d$\,[GeV] \cite{Sanz-Cillero:2017fvr} & $g_T$\,[GeV] \cite{Ecker:2007us,Sanz-Cillero:2017fvr} & $T_V$ ($\mbox{GeV}^2$) \\
\hline
 $0.092$ & $0.206$ & $0.030$ & $0.028$ & $0.115$ \\
\hline
\end{tabular}
\end{center}
\vspace*{-0.5cm}
\caption{\label{tab:coupid} Couplings involving hadron resonances. Their justification is based on the quoted references. For the value of $T_V$ see the discussion at the end of appendix~\ref{app:3}.
}
\end{table}
\par 
We consider now the mixing angle between the octet ($\eta_8$) and singlet ($\eta_0$) strong-interaction eigenstates of the pseudoscalar meson multiplet giving the $\eta$ and $\eta'$ physical states. We define this angle via the following relation:
\begin{equation} \label{eq:mixinga}
\left( \begin{array}{c} \eta \\ \eta' \end{array} \right) \, = \, 
\left( \begin{array}{cc} \cos \theta_P & - \sin \theta_P \\
                         \sin \theta_P & \cos \theta_P
       \end{array} \right) \, 
\left( \begin{array}{c} \eta_8 \\ \eta_0 \end{array} \right) \, ,
\end{equation}
and take $\theta_P = - 20^{\circ}$ arising in the large-$N_\text{C}$ analyses \cite{HerreraSiklody:1997kd,Kaiser:1998ds}.
Finally, we define the analogous mixing angle for the vector resonances as
\begin{equation} \label{eq:mixingv}
\left( \begin{array}{c} \phi(1020) \\ \omega(782) \end{array} \right) = 
\left( \begin{array}{cc} \cos \theta_V & - \sin \theta_V \\
                         \sin \theta_V & \cos \theta_V 
       \end{array} \right)
\left( \begin{array}{c} V_8 \\ V_0 \end{array} \right) .
\end{equation}
We consider ideal mixing $\theta_V =  35^{\circ}$.


\end{document}